\documentclass[preprint, 10pt]{elsarticle}

\usepackage[latin1]{inputenc}
\usepackage[T1]{fontenc}

\usepackage{lineno,hyperref}
\usepackage{amsmath,amssymb,amsthm,mathrsfs,graphicx}
\usepackage{subfigure}
\usepackage{geometry}
\usepackage{fancyhdr}
\usepackage{psfrag}
\usepackage{color} % necessario para usar eps_tex !!! 
\usepackage{subfigure}

\modulolinenumbers[5] % TO BE USED WITH lineno PACKAGE

\journal{Computer Methods in Applied Mechanics and Engineering}

\begin{document}

\begin{frontmatter}

\title{A non-ordinary peridynamics implementation for anisotropic materials}
%\tnotetext[mytitlenote]{Fully documented templates are available in the elsarticle package on \href{http://www.ctan.org/tex-archive/macros/latex/contrib/elsarticle}{CTAN}.}

%% Group authors per affiliation:
%\author{Gabriel Hattori}
%\address{Radarweg 29, Amsterdam}
%\fntext[myfootnote]{Since 1880.}

%% or include affiliations in footnotes:
\author[durham]{Gabriel Hattori\corref{mycorrespondingauthor}}
\cortext[mycorrespondingauthor]{Corresponding author}
\ead{gabriel.hattori@durham.ac.uk}

\author[durham]{Jon Trevelyan}
\ead{jon.trevelyan@durham.ac.uk}

\author[durham]{William M. Coombs}
\ead{w.m.coombs@durham.ac.uk}

\address[durham]{Department of Engineering, Durham University, DH1 3LE, Durham, UK}
%\address[mysecondaryaddress]{360 Park Avenue South, New York}

\begin{abstract}
Peridynamics (PD) represents a new approach for modelling fracture mechanics, where a continuum domain is modelled through particles connected via physical bonds. This formulation allows us to model crack initiation, propagation, branching and coalescence without special assumptions. Up to date, anisotropic materials were modelled in the PD framework as different isotropic materials (for instance, fibre and matrix of a composite laminate), where the stiffness of the bond depends on its orientation. A non-ordinary state-based formulation will enable the modelling of generally anisotropic materials, where the material properties are directly embedded in the formulation. Other material models include rocks, concrete and biomaterials such as bones. In this paper, we implemented this model and validated it for anisotropic composite materials. A composite damage criterion has been employed to model the crack propagation behaviour. Several numerical examples have been used to validate the approach, and compared to other benchmark solution from the finite element method (FEM) and experimental results when available. 
\end{abstract}


\end{frontmatter}

%\linenumbers % TO BE USED WITH lineno PACKAGE

\section{Introduction}

%Introduction anisotropic materials, works that have been done recently and problems
Anisotropic materials have been studied since the first works of Sih et al. \cite{Sih65}, and ever since have attracted attention of fracture mechanics researchers. Anisotropic materials are usually brittle, almost ensuring that cracks will appear during their lifetime. The effect of cracks in an anisotropic material is more complicated than the equivalent problem with an isotropic material. For instance, the crack propagation path is also a function of the material properties, rather than depending only on the orientation of the applied load and the specimen geometry.

These materials are widely used in composites in the aerospace and automobile industries \cite{Nobile05,motamedi2014stochastic}, as sensors and actuators (piezoelectric \cite{garcia2005anisotropic,Wunsche11} and magnetoelectroelastic \cite{milazzo2012equivalent,Wu00} materials) and more recently have applications in biomechanics \cite{giordano2017anisotropic,santiuste2014influence} and even in hydraulic fracturing \cite{giraud2008effective,hattori2016numerical}, just to mention some of the works. Cracks in composite materials can be responsible for complete failure of the component, resulting in economic losses or even loss of life.
%Damaged smart materials exhibit different electric and magnetic fields compared to the pristine material, leading to wrong interpretations of the measurements.
Damaged smart materials exhibit different electric and magnetic fields compared to the pristine material, incurring in errors of the response of a sensor for instance.
Therefore, it is important to accurately quantify the effect of discontinuities in anisotropic materials.

Fracture mechanics have been studied for nearly a century, from the first work of Griffith \cite{griffith1921phenomena} for brittle materials. 
Over the years a number of researchers have modelled fracture mechanics analytically \cite{Muskhelishvili53,sih1991mechanics} for simple problems and geometries, and more commonly using numerical frameworks, such as the extended finite element method (XFEM) \cite{bouhala2013modelling,motamedi2012fracture} and the extended boundary element method (XBEM) \cite{hattori2016extended}, among many others. Nevertheless, these methods suffer when crack branching or coalescence are involved. 

The phase-field method has been shown to model crack branching behaviour \cite{biner2009simulation,henry2008study}. The method consists in describing the crack as an interface directly in the formulation and is used conjointly to the finite element method (FEM) \cite{schluter2014phase}. Nevertheless, the method requires a fine mesh around the crack to model the interface correctly. Another drawback of the method is that it can provide unrealistic results. 
%Crack initiation is also difficult to model, since the linear elastic fracture mechanics (LEFM) assumes the presence of an initial crack. 
A novel numerical method entitled peridynamics (PD) \cite{silling2000reformulation} has been recently developed, and has shown great potential in fracture mechanics problems involving initiating, propagating, branching and coalescing cracks.

%INTRODUCTION PERIDYNAMICS, bond-based formulation
The original peridynamics (PD) formulation was proposed by Silling \cite{silling2000reformulation}, where he redefined the classical approach for continuum mechanics using an integral framework instead of partial derivatives. The main reason for using this approach is that the partial derivatives pose a challenge when dealing with fracture mechanics problems, since the governing partial differential equations in elasticity imply that singularities will appear due to the presence of discontinuities, which is not desirable. Due to the integral form of the formulation, no special assumptions are needed to deal with singularities, such as a crack in the domain.

The first PD formulation described a continuum medium through discrete particles, interacting between each other through physical connections entitled bonds. Each bond has a stiffness associated with it, being analogous to a spring in continuum mechanics theory. However, each particle has an area of influence, interacting with all other particles contained within a perimeter called the horizon of the particle, and is a characteristic of non-local formulations (see \cite{eringen1972nonlocal,lorentz2003analysis} for other types of non-local approaches). The material properties in PD are calculated using the material parameters of the classical continuum mechanics, and also parameters from PD such as the horizon size. The tractions between different particles are in the same direction as the bond, have opposite sense and the same magnitude. This first formulation obtained by Silling was denoted bond-based PD. 

The crack is formed when the bonds between particles are broken, a key feature of the PD formulation. This characteristic also enables the modelling of crack initiation without further assumptions. Additionally, crack branching can appear if elastic wave reflections generate instabilities at the crack tip, which is very difficult to model in standard numerical techniques. However this theory presented limitations with respect to the material properties. Silling has stated that the so called bond-based PD limits the Poisson's ratio of the material (1/3 for 2D and 1/4 for 3D) \cite{silling2000reformulation,silling2007peridynamic}. A more generalised framework called state-based PD has been  developed so any material properties can be assumed without restrictions \cite{silling2007peridynamic}. 

The state-based PD is divided into two main approaches: the ordinary state-based PD, which represents a generalisation of the bond-based theory, and the non-ordinary state-based PD. Some of the main differences lie in the orientation of tractions between particles and how material properties are obtained. The tractions in ordinary state-based PD are still in the same direction of the bond, but they are not constrained to have the same magnitude. A relation more similar to continuum mechanics is present due to the use of state vectors. The balance of linear and angular momentum is automatically satisfied in bond-based and ordinary state-based theories, but the same is not valid for non-ordinary framework \cite{silling2007peridynamic}. 

The non-ordinary theory uses non-local approximations of some of the continuum mechanics variables. This permits a more general representation of the continuum mechanics using stress-strain relations, so that the material constitutive matrix can be used, instead of calculating equivalent properties for the stiffness in the bonds as in bond-based and ordinary state-based PD. Additionally, the tractions acting on the particles are no longer constrained to the direction of the bonds nor have the same magnitude. A drawback of these properties is that the balance of linear and angular momentum are not implicitly satisfied and have to be proved.
Silling et al. \cite{silling2007peridynamic,silling2010peridynamic} have demonstrated how these criteria can be satisfied for a specific non-ordinary state-based formulation.

%TALK ABOUT STATE-BASED PLUS EXAMPLES!
Madenci and Oterkus \cite{madenci2014peridynamic} detail extensively the use of ordinary state-based PD. Warren et al. \cite{warren2009non} and  Breitenfeld et al. \cite{breitenfeld2014non} have performed some of the first works in non-ordinary state-based PD for explicit and implicit implementations, respectively. Yaghoobi and Chorzepa \cite{yaghoobi2015meshless} have modelled fibre reinforced concrete problems for non-ordinary state-based PD. Wu et al. \cite{wu2015stabilized} have implemented a non-ordinary state-based for the metal machining process in ductile materials, where the loss of material usually leads to instabilities due to the strain localisation problem. A simple stabilisation technique was implemented to eliminate these instabilities. Wang et al. \cite{wang2016numerical} have studied crack propagation problems in rock type materials. 

Some authors have investigated different anisotropic materials using PD. Hu et al. \cite{hu2012peridynamic} and Oterkus et al. \cite{oterkus2012peridynamic,oterkus2012combined} have implemented bond-based models for composite materials, where the fibre and the matrix have different material properties, and are defined by the orientation of the bond. However, these PD models assume only two material constants (stiffness at the bonds and in the matrix). Hu et al. \cite{hu2014peridynamic} have also analysed delamination as well as damage in the fibre and the bonds. 

Ghajari et al. \cite{ghajari2014peridynamic} have implemented a bond-based model for orthotropic materials. The anisotropy is generated by changing the stiffness of the bonds with their orientation. A limitation of this model is to use only two constants to define the material properties, instead of the four used in the continuum mechanics model. Another limitation imposed by the bond-based formulation is that mode II behaviour is dependent on mode I, which is not desirable. To the best of the authors' knowledge, a general formulation for anisotropic materials in PD where the material properties are considered in the formulation rather than due to the orientation of the bonds has not been found in the literature.

In this paper we propose a non-ordinary state-based formulation for generally anisotropic materials for dynamic crack problems. We implement an anisotropic damage criterion in order to capture the appropriate crack propagation path. We validate our model using benchmark problems and comparison against other numerical methods or experimental results. The remainder of the paper is organised as follows: the continuum mechanics theory is briefly introduced in Section 2. We describe the state-based PD formulation in Section 3, with emphasis on the non-ordinary formulation. In Section 4, the anisotropic damage criterion is explained, while the explicit integration scheme is briefly explained in Section 5. The numerical results are presented in Section 6. The main conclusions are summarised in Section 7.

\section{Classical continuum mechanics}

In this section we introduce some of the parameters of continuum mechanics employed in PD. The deformation gradient characterises the behaviour of motion in the neighbourhood of a point, and it is defined as \cite{holzapfel2000nonlinear}
\begin{equation}
\mathbf F(\mathbf x) = \frac{\partial \mathbf y}{\partial \mathbf x}
\label{eq:def_grad_cont}
\end{equation}
where $\mathbf x$ denotes an arbitrary point in the reference configuration, and 
\begin{equation}
\mathbf y = \mathbf x + \mathbf u 
\end{equation}
denotes a point in the deformed configuration, while $\mathbf u$ correspond to the displacements. The deformation gradient is in principle not symmetric. If the displacements $\mathbf u$ are zero, then the deformation gradient is the identity matrix.

The determinant of the deformation gradient is defined as $J = det(\mathbf F(\mathbf x))$, and this gives the ratio of the volumes between the reference and deformed configurations. Since $J > 0$, the volume of the material in the deformed configuration will never be zero. Moreover, the inverse of the deformation gradient can always be calculated.

The first Piola-Kirchhoff stress is given by \cite{holzapfel2000nonlinear}
\begin{equation}
\mathbf P(\mathbf x) = J \boldsymbol \sigma \mathbf F(\mathbf x)^T
\label{eq:piola_cont}
\end{equation}
where $\boldsymbol \sigma$ stands for the Cauchy stress and the superscript $T$ denotes the transpose of a matrix.

% The Green-Lagrange strain tensor is defined as 
% \begin{equation}
% %\mathbf E(\mathbf x) = \frac{1}{2}\left( \mathbf F(\mathbf x)^T \mathbf F(\mathbf x) - \mathbf I\right)
% \mathbf E(\mathbf x) = \frac{1}{2}\left( \mathbf F(\mathbf x)^T \mathbf F(\mathbf x) - \mathbf I \right) 
% \end{equation}
% where the superscript $T$ denotes the transpose of a matrix and $\mathbf I$ is the identity matrix. Let us remark that the Green-Lagrange strain tensor is symmetric.

In the small strain assumption, the first Piola-Kirchhoff stress can be approximated by the Cauchy stress, i.e.,
\begin{equation}
\boldsymbol \sigma \approx \mathbf P(\mathbf x) \\
\end{equation}
and the infinitesimal strain can be defined in terms of the deformation gradient such as 
\begin{equation}
\boldsymbol \varepsilon \approx \frac{1}{2}\left( \mathbf F(\mathbf x)^T + \mathbf F(\mathbf x) \right) - \mathbf I
\label{eq:inf_strain_def}
\end{equation}
where $\mathbf I$ is the identity matrix.

In the classical continuum mechanics, the equation of motion is defined as
\begin{equation}
%\nabla \cdot \boldsymbol{\sigma} + \mathbf b(\mathbf x,t)= \rho \ddot{\mathbf u}
\sigma_{ij,j} + b_i(x_i,t) = \rho \ddot{u_i}
\label{eq:motion1}
\end{equation}
where $\mathbf b(\mathbf x,t)$ denotes the body forces per unit volume, $\rho$ is the density and $\ddot{\mathbf u}$ is the acceleration. 

The generalised Hooke's law is the relation between stresses and strains for an elastic material and is expressed as
\begin{equation}
\sigma_{ij} = C_{ijkl} \varepsilon_{kl}
\label{comportamiento}    
\end{equation}
where $\varepsilon_{kl}$ are the strains, and $C_{ijkl}$ denotes the material $4^{th}$ order constitutive tensor. The most general form of anisotropy consists of 81 independent components in the constitutive tensor. Nevertheless, most anisotropic materials present symmetry properties that admit the following relations
\begin{equation}\label{simetria}
    C_{ijkl} = C_{jikl} = C_{ijlk} = C_{klij}
\end{equation}
leading to a tensor with only 21 independent components for the 3D case, and 6 components in the 2D case.

Classical linear strain-displacement relations are assumed as
\begin{equation}\label{cinemat}
\varepsilon_{ij} = \frac{1}{2}(u_{i,j}+ u_{j,i})
\end{equation}
%where $u_i$ are the displacements

\section{State-based peridynamics}

The equation of motion in state-based peridynamics (PD) is defined as \cite{silling2007peridynamic}
\begin{equation}
\int_{\mathscr H} \{ \underline{\mathbf T}[\mathbf x,t] \langle \mathbf x' - \mathbf x \rangle  - \underline{\mathbf T}[\mathbf x',t] \langle \mathbf x - \mathbf x' \rangle  \} dV_{\mathbf x'} + \mathbf b(\mathbf x,t) = \rho \ddot{\mathbf u}(\mathbf x,t)
\label{eq:state_eq1}
\end{equation}
where $\mathscr{H}$ delimits the area of influence of a particle $\mathbf x$ and is also denominated as the horizon of $\mathbf x$, $\mathbf x$ is the particle of interest, $\mathbf x'$ represents the particles inside the horizon of $\mathbf x$, $\underline{\mathbf T}$ is the force vector state field, and square brackets denote that the variables are taken in the state vector framework. In order for Eq. (\ref{eq:state_eq1}) to be valid, it must satisfy both balance of linear and angular momentum. The proofs can be found in \cite{silling2007peridynamic,silling2010peridynamic}.

Silling et al. \cite{silling2007peridynamic} have proposed a generalised peridynamics (PD) formulation, where the PD variables are expressed in terms of vector states. 
For instance, the reference state vector $\underline{\mathbf X}\langle \boldsymbol \xi \rangle $ is defined as
\begin{equation}
\underline{\mathbf X}\langle \boldsymbol \xi \rangle = \boldsymbol \xi, \quad \forall \boldsymbol \xi \in \mathscr{H}
\label{eq:mapping_X}
\end{equation}
and 
\begin{equation}
\boldsymbol \xi =\mathbf x' - \mathbf x
\end{equation}

Eq. (\ref{eq:mapping_X}) represents the mapping between all particles $\mathbf x'$ with respect to the particle $\mathbf x$. The state vector nomenclature includes an underline to make a clear distinction from matrices, while the angle brackets relate to other variables the state vector relies upon.

Another way to visualise the vector state concept is to assume a matrix form as stated in \cite{madenci2014peridynamic}. The deformation state vector $\underline{\mathbf Y}\langle  \mathbf x' - \mathbf x \rangle$ is then defined as
\begin{equation}
\underline{\mathbf Y}\langle  \mathbf x - \mathbf x' \rangle = 
\left[
\begin{array}{c}
\mathbf y_1 - \mathbf y \\
\mathbf y_2 - \mathbf y \\
\vdots \\
\mathbf y_N - \mathbf y \\
\end{array} \right]
\end{equation}
where $\mathbf y = \mathbf y(\mathbf x,t)$ represents the deformed position of particle $\mathbf x$ at time $t$, and $\mathbf y_i = \mathbf y(\mathbf x',t)$, $i=1, \cdots N$ denotes all the particles $\mathbf x'$ contained within the horizon of $\mathbf x$.

\begin{figure}[!htb]
\centering
\def\svgwidth{0.5\linewidth}
\begingroup%
  \makeatletter%
  \providecommand\color[2][]{%
    \errmessage{(Inkscape) Color is used for the text in Inkscape, but the package 'color.sty' is not loaded}%
    \renewcommand\color[2][]{}%
  }%
  \providecommand\transparent[1]{%
    \errmessage{(Inkscape) Transparency is used (non-zero) for the text in Inkscape, but the package 'transparent.sty' is not loaded}%
    \renewcommand\transparent[1]{}%
  }%
  \providecommand\rotatebox[2]{#2}%
  \ifx\svgwidth\undefined%
    \setlength{\unitlength}{685.74325676bp}%
    \ifx\svgscale\undefined%
      \relax%
    \else%
      \setlength{\unitlength}{\unitlength * \real{\svgscale}}%
    \fi%
  \else%
    \setlength{\unitlength}{\svgwidth}%
  \fi%
  \global\let\svgwidth\undefined%
  \global\let\svgscale\undefined%
  \makeatother%
  \begin{picture}(1,0.70832591)%
    %\put(0,0){\includegraphics[width=\unitlength]{pd_state_based.eps}}%
    \put(0,0){\includegraphics[width=\unitlength,page=1]{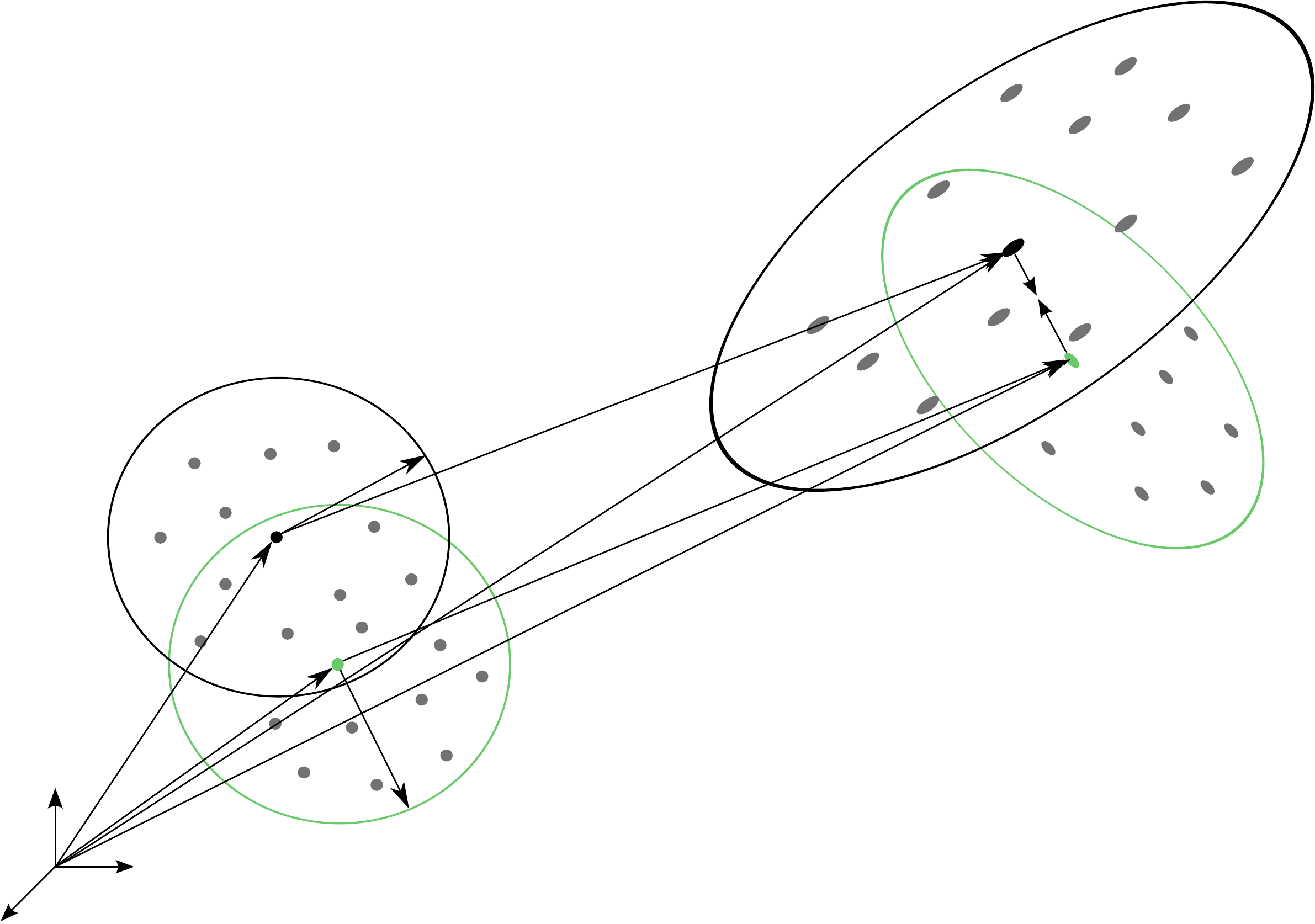}}%
    \put(0.82,0.41){\color[rgb]{0,0,0}\makebox(0,0)[lb]{\footnotesize \color{green}{$\mathbf y'$}}}%
    \put(0.785,0.52){\color[rgb]{0,0,0}\makebox(0,0)[lb]{\footnotesize $\mathbf y$}}%
    \put(0.785,0.495){\color[rgb]{0,0,0}\makebox(0,0)[lb]{\footnotesize $\mathbf T$}}%
    \put(0.75,0.43){\color[rgb]{0,0,0}\makebox(0,0)[lb]{\footnotesize \color{green}{$\mathbf T'$}}}%
    \put(0.215,0.265){\color[rgb]{0,0,0}\makebox(0,0)[lb]{\footnotesize $\mathbf x$}}%
    \put(0.24,0.205){\color[rgb]{0,0,0}\makebox(0,0)[lb]{\footnotesize \color{green}{$\mathbf x'$}}}%
    \put(0.29,0.14){\color[rgb]{0,0,0}\makebox(0,0)[lb]{\footnotesize \color{green}{$\delta'$}}}%
    \put(0.245,0.325){\color[rgb]{0,0,0}\makebox(0,0)[lb]{\footnotesize $\delta$}}%
    \put(0.02,0.0){\color[rgb]{0,0,0}\makebox(0,0)[lb]{$z$}}%
    \put(0.10,0.02){\color[rgb]{0,0,0}\makebox(0,0)[lb]{$x$}}%
    \put(0.05,0.09){\color[rgb]{0,0,0}\makebox(0,0)[lb]{$y$}}%
    \put(0.13,0.48){\color[rgb]{0,0,0}\makebox(0,0)[lb]{reference state}}%
    \put(0.68,0.245){\color[rgb]{0,0,0}\makebox(0,0)[lb]{deformed state}}%
    \put(0.43,0.40){\color[rgb]{0,0,0}\makebox(0,0)[lb]{\footnotesize $\mathbf u$}}%
    \put(0.49,0.31){\color[rgb]{0,0,0}\makebox(0,0)[lb]{\footnotesize $\mathbf u'$}}%
  \end{picture}%
\endgroup%
\caption{Reference and deformed configuration in state-based PD.}
\label{fig:deformed}
\end{figure}
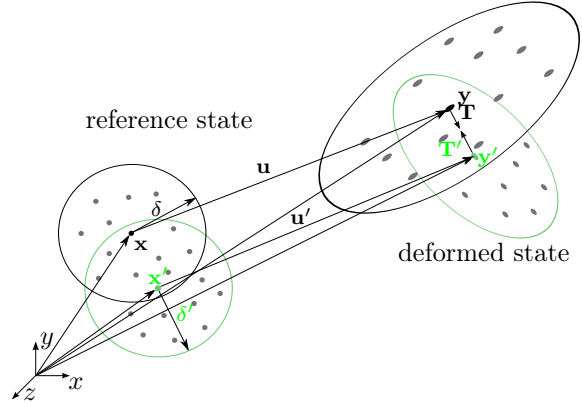

Figure \ref{fig:deformed} illustrates the reference (or undeformed) configuration, and the deformed configuration after a displacement $\mathbf u$ and $\mathbf u'$ has been imposed on particles $\mathbf x$ and $\mathbf x'$, respectively. Similarly, $\delta$ and $\delta'$ correspond to the horizon of particles $\mathbf x$ and $\mathbf x'$, respectively.

In the original bond-based formulation \cite{silling2000reformulation}, the particles within the distance $\delta$ of $\mathbf x$ are said to be inside the horizon of that particle, thus making a contribution to the displacement solution. The bonds possess stiffness, and can be considered as springs or trusses since the bonds only have tractions in the direction of the bond. Macek and Silling \cite{macek2007peridynamics} have used this idea to implement a bond-based formulation where the bonds were modelled by truss elements. They concluded that this formulation provided similar results to the ones found with a commercial finite element software.

However, the bond-based PD formulation presents limitations with respect to the material properties. The Poisson's ratio is restricted to 1/3 for 2D problems and to 1/4 for 3D problems \cite{silling2000reformulation}, since the particles within the horizon $\delta'$ of particle $\mathbf x'$ are not all included during the analysis of $\mathbf x$. The state-based theory \cite{silling2007peridynamic} removes this limitation, allowing the modelling of any material properties. Other authors have developed different approaches to overcome this limitation in bond-based models (see \cite{gerstle2007peridynamic} for instance). In this work, we focus on the state-based PD theory.

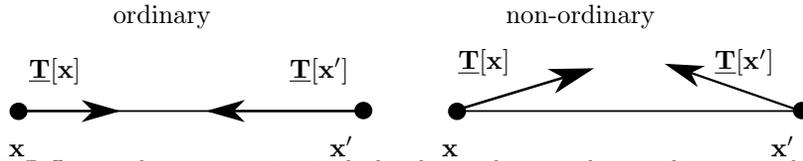
\begin{figure}[!htb]
\centering
\begingroup%
  \makeatletter%
  \providecommand\color[2][]{%
    \errmessage{(Inkscape) Color is used for the text in Inkscape, but the package 'color.sty' is not loaded}%
    \renewcommand\color[2][]{}%
  }%
  \providecommand\transparent[1]{%
    \errmessage{(Inkscape) Transparency is used (non-zero) for the text in Inkscape, but the package 'transparent.sty' is not loaded}%
    \renewcommand\transparent[1]{}%
  }%
  \providecommand\rotatebox[2]{#2}%
  \ifx\svgwidth\undefined%
    \setlength{\unitlength}{302.29375bp}%
    \ifx\svgscale\undefined%
      \relax%
    \else%
      \setlength{\unitlength}{\unitlength * \real{\svgscale}}%
    \fi%
  \else%
    \setlength{\unitlength}{\svgwidth}%
  \fi%
  \global\let\svgwidth\undefined%
  \global\let\svgscale\undefined%
  \makeatother%
  \begin{picture}(1,0.18653214)%
    %\put(0,0){\includegraphics[width=\unitlength]{non-ord.eps}}%
    \put(0,0){\includegraphics[width=\unitlength,page=1]{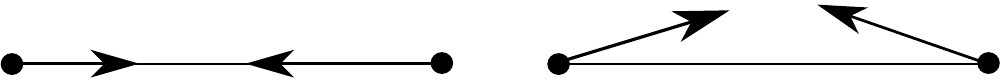}}%
    \put(0.13,0.12){\color[rgb]{0,0,0}\makebox(0,0)[lb]{ordinary}}%
    \put(0.62,0.12){\color[rgb]{0,0,0}\makebox(0,0)[lb]{non-ordinary}}%
    \put(-0.00,-0.04){\color[rgb]{0,0,0}\makebox(0,0)[lb]{$\mathbf x$}}%
    \put(0.40,-0.04){\color[rgb]{0,0,0}\makebox(0,0)[lb]{$\mathbf x'$}}%
    \put(0.54,-0.04){\color[rgb]{0,0,0}\makebox(0,0)[lb]{$\mathbf x$}}%
    \put(0.95,-0.04){\color[rgb]{0,0,0}\makebox(0,0)[lb]{$\mathbf x'$}}%
    \put(0.025,0.05){\color[rgb]{0,0,0}\makebox(0,0)[lb]{$\mathbf{\underline T}[\mathbf x]$}}%
    \put(0.35,0.05){\color[rgb]{0,0,0}\makebox(0,0)[lb]{$\mathbf{\underline T}[\mathbf x']$}}%
    \put(0.56,0.06){\color[rgb]{0,0,0}\makebox(0,0)[lb]{$\mathbf{\underline T}[\mathbf x]$}}%
    \put(0.88,0.06){\color[rgb]{0,0,0}\makebox(0,0)[lb]{$\mathbf{\underline T}[\mathbf x']$}}%
  \end{picture}%
\endgroup%
\caption{Differences between traction in the bonds in ordinary and non-ordinary state-based PD.}
\label{fig:non_ord}
\end{figure}

There are two types of state-based formulation: ordinary and non-ordinary. In the ordinary theory, the forces in the bonds are defined in the direction of the bonds, in the same way as in the bond-based formulation. However, the forces do not need to have the same magnitude, as in the bond-based approach. On the other hand, the non-ordinary formulation presents no restriction with respect to the direction of the bond and the magnitude of the tractions. These differences are illustrated in Figure \ref{fig:non_ord}.

The process of obtaining the equivalent material properties in the ordinary state-based formulation is not straightforward. There are no direct equivalences of stresses and strains in the ordinary framework. In this sense, a typical approach is to draw an equivalence between the strain energy for the continuum mechanics theory and the strain energy density in the PD framework. Then the material properties are obtained by solving a series of simple problems with known deformation, such as pure shear, uni-axial deformation, bi-axial deformation \cite{madenci2014peridynamic}. For isotropic materials or some specific anisotropic (e.g. orthotropic) materials, analytical solutions can be obtained. However, analytical solutions are not possible for generally anisotropic materials. 

The equivalence between strain energy densities from continuum mechanics and PD framework also poses another inconvenience. Particles closer to the boundaries of the analysed domain will share bonds with fewer particles than those in the middle of the domain, for example. However, it is assumed that the strain energy density for the particles, regardless of the number of particles in the horizon, leading to an overestimation of the material properties at the boundaries \cite{madenci2014peridynamic,silling2015variable}. Correction factors have been proposed to modify the strain energy for a particle whose horizon contains a reduced number of particles (see \cite{madenci2014peridynamic} for a detailed explanation).

% A potential issue of the ordinary state based theory is how to obtain the equivalent material properties from the classical continuum mechanics. A typical approach is to draw an equivalence between the strain energy for the continuum mechanics theory and the strain energy density in the PD framework. There are 2 drawbacks for this scheme: firstly, an initial problem for a specific deformation has to be solved in order to obtain the material properties, then the problem of interest can be solved. For isotropic materials or some specific anisotropic (e.g. orthotropic) materials, analytical solutions can be obtained. However, analytical solutions are not possible for generally anisotropic materials. Secondly, since the strain energy is compared to its equivalent in continuum mechanics, an issue arises for particles close to boundaries \cite{madenci2014peridynamic,silling2015variable}. In this case, the horizon of the particle will be smaller than the other particles, but it is still assumed that they have the same strain energy density, leading to an overestimation of the material properties at the boundaries. Correction factors have been proposed to adjust the strain energy for a reduced horizon (see \cite{madenci2014peridynamic} for a detailed explanation).

\subsection{Non-ordinary state-based PD}

The non-ordinary PD framework is a more generalised approach, where some of the main parameters in continuum mechanics, such as the deformation gradient, are expressed in terms of the PD formulation. Since these parameters are not constrained by the physical bonds connecting the particles, the tractions in the bonds are not enforced to be in the same direction as the bonds themselves, as is the case in ordinary PD models.
Another advantage is that the material properties in the non-ordinary state-based formulation come from the material constitutive matrix, which is not the case for the ordinary state-based PD.

The deformation gradient $\mathbf F(\mathbf x)$ in the classical continuum mechanics is given by Eq. (\ref{eq:def_grad_cont}). The non-local deformation gradient $\mathbf F(\mathbf x)$ for each particle is given by \cite{silling2007peridynamic}
\begin{align}
\mathbf F(\mathbf x) &= \left[ \int_{\mathscr H} \omega(|\xi|)(\underline{\mathbf Y}(\xi) \otimes \xi)dV_{\mathbf x} \right] . \mathbf B(\mathbf x) \label{eq:def_grad1} \\
\mathbf B(\mathbf x) &= \left[ \int_{\mathscr H} \omega(|\xi|)(\xi \otimes \xi) dV_{\mathbf x} \right]^{-1} \label{eq:def_grad2}
\end{align}
where $\mathbf B(\mathbf x)$ is the shape tensor, $\otimes$ denotes the dyadic product of two vectors, and $\omega(|\boldsymbol \xi|)$ is a dimensionless weight function, used to increase the influence of the nodes closes to $\mathbf x$. 
%The use of this factor is still under study \cite{warren2009non}, but the assumption of $\omega(|\xi|)=1$ has been seen to provide good results.
In this work, we assumed that the influence function has a triangular shape, and is given by
\begin{equation}
\omega(|\boldsymbol \xi|) = 1 - \frac{\xi}{\delta}
\label{eq:influence_func}
\end{equation}
where $\xi=|\boldsymbol \xi|$. The reason for using this influence function is to make the behaviour of particles closer to $\mathbf x$ more dominant than the ones more distant. Warren et al. \cite{warren2009non} have used a unitary influence function, while Queiruga and Moridis \cite{queiruga2017numerical} have investigated several influence functions for simple problems in 2D and concluded that the triangular function given by Eq. (\ref{eq:influence_func}) leads to lower errors for the non-ordinary state-based PD. % CHECK THIS !!!
%$\omega(|\xi|)=1$.

The discretisation of Eqs. (\ref{eq:def_grad1}) and (\ref{eq:def_grad2}) can be expressed as a Riemann sum as \cite{warren2009non}
\begin{align}
\mathbf F(\mathbf x_j) &= \left[ \sum_{n=1}^m \omega(|\mathbf x_n- \mathbf x_j|)(\underline{\mathbf Y}\langle \mathbf x_n- \mathbf x_j \rangle \otimes (\mathbf x_n- \mathbf x_j))V_n \right] \\
\mathbf B(\mathbf x_j) &= \left[ \sum_{n=1}^m \omega(|\mathbf x_n- \mathbf x_j|)( (\mathbf x_n- \mathbf x_j) \otimes (\mathbf x_n- \mathbf x_j))V_n \right]^{-1}
\end{align}
where $m$ is the number of particles with the horizon of node $j$ and $V_n$ is the volume of particle $n$. Let us remark that each particle $\mathbf x_j$ must be connected to at least two other particles in different orientations to ensure that $\mathbf B(\mathbf x_j)$ will not be singular for 2D problems. For 3D problems, at least three particles in different planes must be connected to avoid singularity of $\mathbf B(\mathbf x_j)$.

%The traction in the bonds is a parameter that depends only on the deformation state vector. In order to obtain an expression for the traction, it can be shown that the strain energy density is a function of the deformation state vector, i.e.,

According to \cite{silling2007peridynamic}, a material is denominated simple if the traction state depends only on the deformation state, i.e., $\underline T = \underline {\mathbf T}(\underline {\mathbf Y})$. A material is simple and elastic if the traction state can be expressed as 
\begin{equation}
\underline {\mathbf T} = \nabla W(\underline {\mathbf Y})
\label{eq:T_state}
\end{equation}
where $W$ is the strain energy density and $\nabla$ represents the Fr\'echet derivative. 

The force and deformation can be related in a state vector framework by using a stress-strain model as an intermediate step \cite{silling2008convergence}. For a strain energy density $W(\mathbf F)$, the stress tensor can be expressed as \cite{warren2009non}
\begin{equation}
\underline {\mathbf T} = \nabla W = \frac{\partial W}{\partial \mathbf F}\nabla \mathbf F
\label{eq:T_state2}
\end{equation}

% To incorporate the kinematic stress into the PD model, the transpose of the first Piola-Kirchhoff stress is equivalent to \cite{silling2008convergence}
% \begin{equation}
% \mathbf P(\mathbf x)^T = \frac{\partial W}{\partial \mathbf F}
% \end{equation}
% with $W$ being the strain energy density function.

The transpose of the first Piola-Kirchhoff stress is a measure of the derivative of the strain energy density $W$ with respect to the deformation gradient, i.e.
\begin{equation}
\mathbf P(\mathbf x)^T = \frac{\partial W}{\partial \mathbf F}
\label{eq:transpose_Piola}
\end{equation}

Substituting Eq. (\ref{eq:transpose_Piola}) into Eq. (\ref{eq:T_state}) and evaluating the Fr\'echet derivative, the traction state is defined explicitly as
%The traction state at time $t$ is finally stated as 
\begin{equation}
\underline{\mathbf T}[\mathbf x,t]\langle \mathbf x' - \mathbf x \rangle = \omega(|\mathbf x' - \mathbf x|)\mathbf P(\mathbf x)^T . \mathbf B(\mathbf x) . (\mathbf x' - \mathbf x)
\label{eq:traction_state}
\end{equation}

Let us remark that there is no dependence on time for $\mathbf P(\mathbf x)$ and $\mathbf B(\mathbf x)$, however these parameters are modified when bonds are broken. 

The processing of mapping a stress tensor as a peridynamic force state is the inverse of the process of approximating the deformation state by a deformation gradient tensor. A peridynamic constitutive model that uses stress as an intermediate quantity results in general in bond forces which are not parallel to the deformed bonds \cite{silling2007peridynamic}.

\section{Damage criterion for anisotropic materials}

In PD, damage is modelled through the bond breakage between pairs of particles. Once a bond is broken, the interaction between particles provided by that bond will not be used during the rest of the analysis. A damage index $\varphi(\mathbf x,t)$ is used to measure the relation of damaged bonds and active bonds for any given particles and is given by
\begin{equation}
\varphi(\mathbf x,t) = 1 - \frac{\int_{\mathscr{H}}\mu(\boldsymbol \xi,t)dV_\xi }{\int_{\mathscr{H}} dV_\xi}
\label{eq:damage_index}
\end{equation}
and 
\begin{equation}
\mu(\boldsymbol \xi,t) = \left \{
\begin{array}{cc}
1 & \text{if the bond is active}\\
0 & \text{if the bond is broken}
\end{array} \right.
\label{eq:damage_index2}
\end{equation}

From Eq. (\ref{eq:damage_index}), $0 \leq \varphi(\mathbf x,t) \leq 1$, where 0 represents the undamaged state and 1 represents the breakage of all the bonds of a given particle. The parameter $\mu(\boldsymbol \xi,t)$ is used only to specify if a particular bond is active or broken. The broken bonds will eventually lead to a softening material response, since failed bonds cannot sustain any load.

There are different damage criteria for anisotropic materials. For instance, Pens\'ee et al. \cite{pensee2002micromechanical} have considered a micromechanical approach for modelling damage in anisotropic brittle materials such as rocks and concrete, based on energy and a multiscale approach. The damage criterion is related to the type of anisotropic material analysed. In this work, we have employed the Tsai-Hill criterion to define damage in the bonds. This criterion is used for composite laminates and can take into account failure between different modes and is given by the following expression
\begin{equation}
\left( \frac{\sigma_L}{\sigma_{Lu}} \right)^2 + \left( \frac{\sigma_T}{\sigma_{Tu}} \right)^2 - \frac{\sigma_L}{\sigma_{Lu}} \frac{\sigma_T}{\sigma_{Lu}}  + \left( \frac{\tau_{LT}}{\tau_{LTu}} \right)^2 = 1
\label{eq:tsai-hill}
\end{equation}
where $\sigma_L$, $\sigma_T$ and $\tau_{LT}$ stand for the longitudinal stress (in the direction of the fibre), transversal stress (perpendicular to the fibre) and shear stress, respectively. $\sigma_{Lu}$, $\sigma_{Tu}$ and $\tau_{LTu}$ are the respective tensile strength of the composite material for different loading.

In order to use this criterion in PD, the stress in a bond is defined as the average stress between the interacting particles, such as
\begin{equation}
\sigma(\mathbf x, \mathbf x') = \frac{1}{2}\left( \sigma(\mathbf x) + \sigma(\mathbf x') \right)
\end{equation}

In this case, we can employ the Cauchy stresses instead of the first Piola-Kirchhoff stress since we are using the small strain assumption.

Next, the stress at the bond is expressed in terms of a local coordinate system using the rotation matrix $\mathbf R(\theta)$, which depends on the fibre orientation $\theta$ and is defined as
\begin{equation}
\mathbf R(\theta) = \left[ 
\begin{array}{ccc}
\cos^2{\theta} & \sin^2{\theta} & 2\cos{\theta}\sin{\theta} \\
\sin^2{\theta} & \cos^2{\theta} & -2\cos{\theta}\sin{\theta} \\
-\cos{\theta}\sin{\theta} & \cos{\theta}\sin{\theta} & \cos^2{\theta}-\sin^2{\theta}
\end{array} \right]
\end{equation}

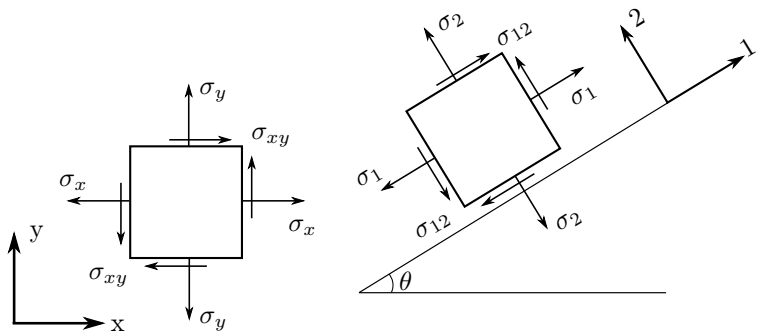
\begin{figure}[!htb]
\centering
\def\svgwidth{0.65\linewidth}
\begingroup%
  \makeatletter%
  \providecommand\color[2][]{%
    \errmessage{(Inkscape) Color is used for the text in Inkscape, but the package 'color.sty' is not loaded}%
    \renewcommand\color[2][]{}%
  }%
  \providecommand\transparent[1]{%
    \errmessage{(Inkscape) Transparency is used (non-zero) for the text in Inkscape, but the package 'transparent.sty' is not loaded}%
    \renewcommand\transparent[1]{}%
  }%
  \providecommand\rotatebox[2]{#2}%
  \ifx\svgwidth\undefined%
    \setlength{\unitlength}{576.42101231bp}%
    \ifx\svgscale\undefined%
      \relax%
    \else%
      \setlength{\unitlength}{\unitlength * \real{\svgscale}}%
    \fi%
  \else%
    \setlength{\unitlength}{\svgwidth}%
  \fi%
  \global\let\svgwidth\undefined%
  \global\let\svgscale\undefined%
  \makeatother%
  \begin{picture}(1,0.43236642)%
    %\put(0,0){\includegraphics[width=\unitlength]{fibre_rotation.eps}}%
    \put(0,0){\includegraphics[width=\unitlength,page=1]{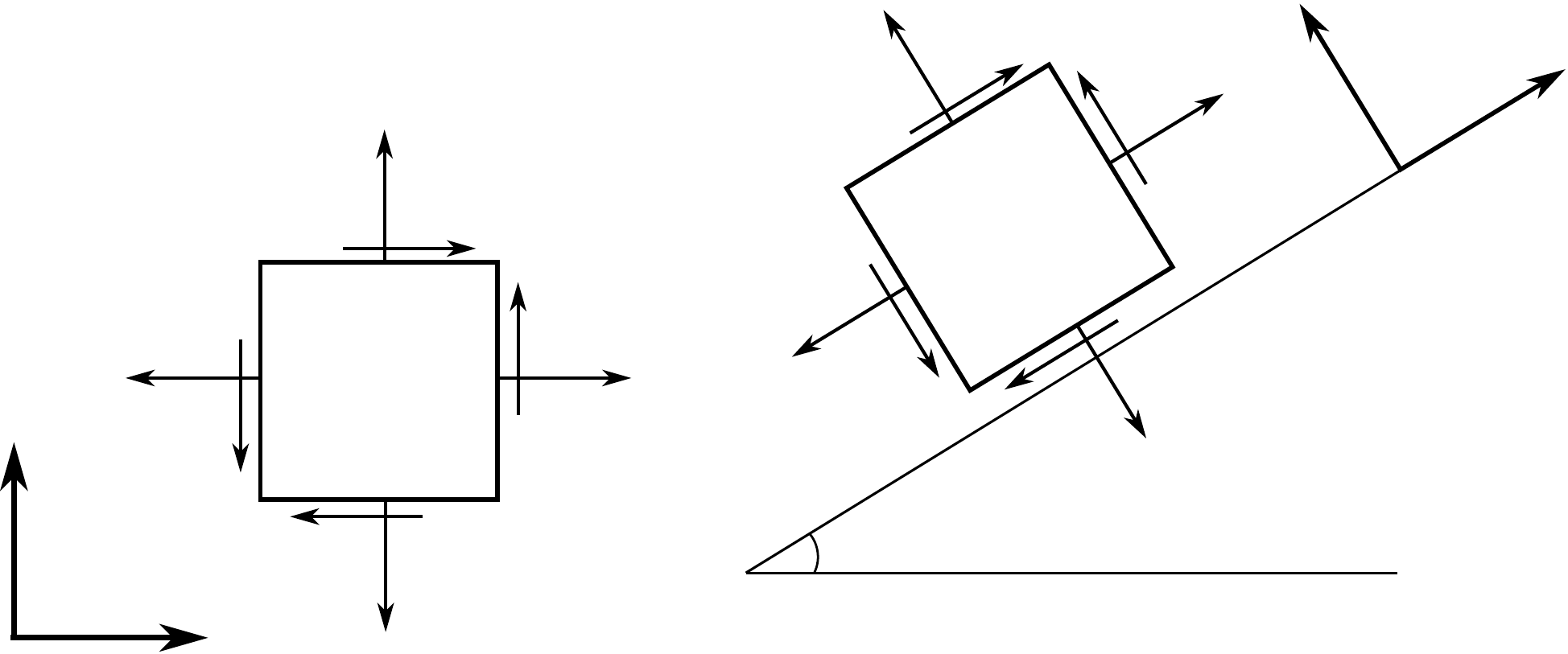}}%
    \put(0.33,0.25){\color[rgb]{0,0,0}\makebox(0,0)[lb]{$\sigma_{xy}$}}%
    \put(0.11,0.06){\color[rgb]{0,0,0}\makebox(0,0)[lb]{$\sigma_{xy}$}}%
    \put(0.26,-0.0){\color[rgb]{0,0,0}\makebox(0,0)[lb]{$\sigma_y$}}%
    \put(0.26,0.31){\color[rgb]{0,0,0}\makebox(0,0)[lb]{$\sigma_y$}}%
    \put(0.38,0.13){\color[rgb]{0,0,0}\makebox(0,0)[lb]{$\sigma_x$}}%
    \put(0.07,0.19){\color[rgb]{0,0,0}\makebox(0,0)[lb]{$\sigma_x$}}%
    \put(0.14,0.00){\color[rgb]{0,0,0}\makebox(0,0)[lb]{x}}%
    \put(0.03,0.12){\color[rgb]{0,0,0}\makebox(0,0)[lb]{y}}%
    \put(0.67,0.38){\color[rgb]{0,0,0}\rotatebox{31.38}{\makebox(0,0)[lb]{$\sigma_{12}$}}}%
    \put(0.56,0.12){\color[rgb]{0,0,0}\rotatebox{31.38}{\makebox(0,0)[lb]{$\sigma_{12}$}}}%
    \put(0.75,0.13){\color[rgb]{0,0,0}\rotatebox{31.38}{\makebox(0,0)[lb]{$\sigma_2$}}}%
    \put(0.59,0.40){\color[rgb]{0,0,0}\rotatebox{31.38}{\makebox(0,0)[lb]{$\sigma_2$}}}%
    \put(0.48,0.19){\color[rgb]{0,0,0}\rotatebox{31.38}{\makebox(0,0)[lb]{$\sigma_1$}}}%
    \put(0.77,0.30){\color[rgb]{0,0,0}\rotatebox{31.38}{\makebox(0,0)[lb]{$\sigma_1$}}}%
    \put(0.53,0.055){\color[rgb]{0,0,0}\makebox(0,0)[lb]{$\theta$}}%
    \put(1.00,0.37){\color[rgb]{0,0,0}\rotatebox{31.38}{\makebox(0,0)[lb]{1}}}%
    \put(0.85,0.41){\color[rgb]{0,0,0}\rotatebox{31.38}{\makebox(0,0)[lb]{2}}}%
  \end{picture}%
\endgroup%
\caption{Rotation of the stresses from global (x,y) to local (1,2) coordinates.}
\label{fig:fibre_rotation}
\end{figure}

Figure \ref{fig:fibre_rotation} illustrates the rotation with the global and local stresses, with respect to the fibre orientation.

Finally, the local stresses are given by
\begin{equation}
\left[
\begin{array}{c}
\sigma_L \\ \sigma_T \\ \tau_{LT}
\end{array} \right] = \mathbf R(\theta) \left[
\begin{array}{c}
\sigma_{xx} \\ \sigma_{yy} \\ \tau_{xy}
\end{array} \right]
\label{eq:loc_stress}
\end{equation}
where $\sigma_{xx}$, $\sigma_{yy}$ and $\tau_{xy}$ are the stresses in the global coordinate system. The local stresses of Eq. (\ref{eq:loc_stress}) are combined  into Eq. (\ref{eq:tsai-hill}), and if the result is higher than 1, then the current bond breaks.

Let us remark that different criteria could be used for non-ordinary state-based PD. For instance, Wang et al. \cite{wang2016numerical} and Zhou et al. \cite{zhou2016numerical} have used a stress criterion to model damage in rocks using a linear Mohr-Coulomb failure criterion, but they consider the rock to be an isotropic material.

\section{Numerical discretisation}

In this work, an explicit integration scheme was employed to calculate the displacements, velocities and accelerations in the PD framework, in a similar way as in the work of \cite{warren2009non}. A drawback of the PD formulation is the requirement for a large computational power, since a large number of particles are typically used. Moreover, each particle interacts with a number of other particles, which contributes for the method to be computationally expensive. However, a parallel implementation of a PD explicit formulation is straightforward with OpenMP or MPI for instance.

For the small strain assumption, the infinitesimal strain state can be approximated by Eq. (\ref{eq:inf_strain_def}) and the Cauchy stresses $\boldsymbol \sigma$ are evaluated using Eq. (\ref{comportamiento}).

The values of acceleration are calculated directly from Eq. (\ref{eq:state_eq1}). The velocities are integrated using a forward difference approach, while the displacements are obtained through a backward scheme. The numerical integration is summarised by
\begin{align}
\ddot{\mathbf u}(\mathbf x,t) &= \frac{1}{\rho} \left( \int_{\mathscr H} \{ \underline{\mathbf T}[\mathbf x,t] \langle \mathbf x' - \mathbf x \rangle  - \underline{\mathbf T}[\mathbf x',t] \langle \mathbf x - \mathbf x' \rangle  \} dV_{\mathbf x'} + \mathbf b(\mathbf x,t)  \right) \\
\dot{\mathbf u}(\mathbf x,\Delta t + t) &= \dot{\mathbf u}(\mathbf x,t) + \ddot{\mathbf u}(\mathbf x,t)\Delta t \\
\mathbf u(\mathbf x,\Delta t + t) &= \mathbf u(\mathbf x,t) + \dot{\mathbf u}(\mathbf x,t)\Delta t
\end{align}
where $\dot{\mathbf u}(\mathbf x,t)$ are the velocities and $\mathbf u(\mathbf x,t)$ are the displacements. 

Due to the use of an explicit approach, the time step must be smaller than a certain critical value in order for the analysis to be valid. Silling and Askari \cite{silling2005meshfree} and Madenci and Oterkus \cite{madenci2014peridynamic} have obtained the critical time step for bond-based and ordinary state-based theories, respectively. Warren et al. \cite{warren2009non} have used the Courant-Friedrichs-Lewy condition  \cite{courant1928uber} to estimate the critical time step for a non-ordinary state-based PD. In this case, the critical time step is proportional $\delta/c_p$, where $c_p = \sqrt{C_{22}/\rho}$ is the dilatational wave speed and $C_{22} = C_{2222}$. We used a conservative approach to guarantee a time step size smaller than the critical value. In this work we assumed $\Delta t = 0.01 \frac{\delta}{c_p}$.

\section{Numerical simulations}

In this section we investigate several applications for the PD formulation in anisotropic materials for 2D problems. The dynamic stress intensity factors (DSIF) are calculated and compared with converged FEM solutions. We have employed the extrapolation method in order to calculate the DSIF as follows \cite{hattori2016extended,Wunsche11}
\begin{equation}
\left( \begin{array}{c}
K_{II}(t) \\ K_I(t)
\end{array} \right) = \sqrt{\frac{\pi}{8 \overline r}} (\Re{(i \mathbf A \mathbf B^{-1})})^{-1}
\left( \begin{array}{cc}
\Delta u_{1}(t) \\ \Delta u_{2}(t)
\end{array} \right)
\end{equation}
where $K_I(t)$ and $K_{II}(t)$ are the dynamic mode I and mode II at time $t$, respectively; $\Delta u_1(t)$ and $\Delta u_2(t)$ are the crack opening displacement at time $t$ in the x and y-direction, respectively; $\mathbf A$, $\mathbf B$ come from the material properties and are obtained from the Stroh formalism \cite{hattori2016extended,Ting96}; $\Re(\cdot)$ represents the real part of $(\cdot)$ while  $i$ is the imaginary component; and $\overline r$ is the distance where the crack opening displacements are measured to the crack tip.

\subsection{Edge crack in an anisotropic 2D plate}

\begin{figure}[!htb]
\centering
\def\svgwidth{0.35\linewidth}
\begingroup%
  \makeatletter%
  \providecommand\color[2][]{%
    \errmessage{(Inkscape) Color is used for the text in Inkscape, but the package 'color.sty' is not loaded}%
    \renewcommand\color[2][]{}%
  }%
  \providecommand\transparent[1]{%
    \errmessage{(Inkscape) Transparency is used (non-zero) for the text in Inkscape, but the package 'transparent.sty' is not loaded}%
    \renewcommand\transparent[1]{}%
  }%
  \providecommand\rotatebox[2]{#2}%
  \ifx\svgwidth\undefined%
    \setlength{\unitlength}{342.60016122bp}%
    \ifx\svgscale\undefined%
      \relax%
    \else%
      \setlength{\unitlength}{\unitlength * \real{\svgscale}}%
    \fi%
  \else%
    \setlength{\unitlength}{\svgwidth}%
  \fi%
  \global\let\svgwidth\undefined%
  \global\let\svgscale\undefined%
  \makeatother%
  \begin{picture}(1,1.21536082)%
    %\put(0,0){\includegraphics[width=\unitlength]{edge_crack_anisot.eps}}%
    \put(0,0){\includegraphics[width=\unitlength,page=1]{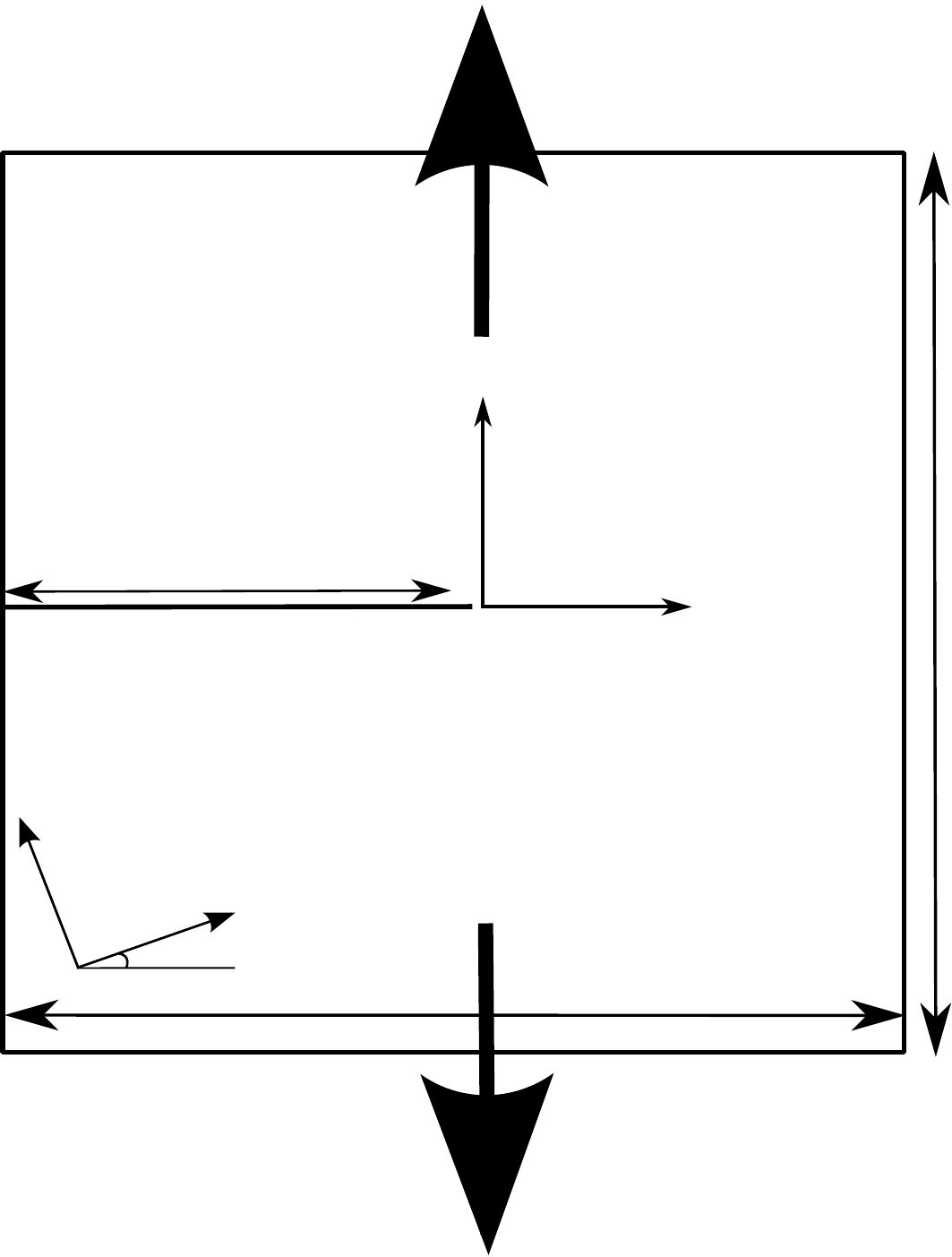}}%
    \put(0.57,0.27){\color[rgb]{0,0,0}\makebox(0,0)[lb]{w}}%
    \put(0.985,0.64){\color[rgb]{0,0,0}\makebox(0,0)[lb]{2h}}%
    \put(0.24,0.72){\color[rgb]{0,0,0}\makebox(0,0)[lb]{a}}%
    \put(0.55,-0.01){\color[rgb]{0,0,0}\makebox(0,0)[lb]{$\dot u(\mathbf x)$}}%
    \put(0.55,1.22){\color[rgb]{0,0,0}\makebox(0,0)[lb]{$\dot u(\mathbf x)$}}%
    \put(0.0479693,0.40935508){\color[rgb]{0,0,0}\makebox(0,0)[lb]{\footnotesize $E_2$}}%
    \put(0.26,0.36){\color[rgb]{0,0,0}\makebox(0,0)[lb]{\footnotesize  $E_1$}}%
    \put(0.22,0.30){\color[rgb]{0,0,0}\makebox(0,0)[lb]{\footnotesize $\theta$}}%
    \put(0.72,0.695){\color[rgb]{0,0,0}\makebox(0,0)[lb]{$x$}}%
    \put(0.52,0.85){\color[rgb]{0,0,0}\makebox(0,0)[lb]{$y$}}%
  \end{picture}%
\endgroup%
\caption{Anisotropic edge crack.}
\label{fig:edge2d_noprop}
\end{figure}
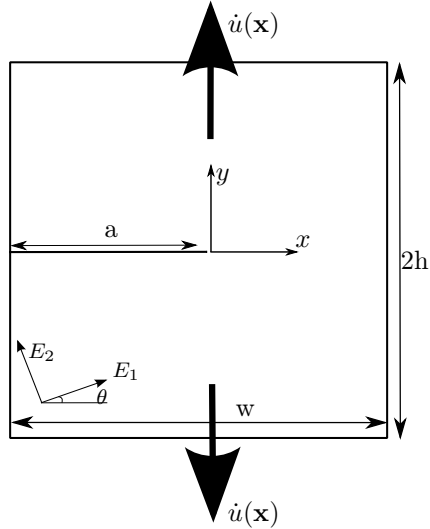

A square plate containing an edge crack is analysed in this section. The plate has dimensions $h=w=0.1$ m, and the length of the crack is $a=0.05$ m. Figure \ref{fig:edge2d_noprop} illustrates this example. The plate is a symmetric angle ply composite laminate of four graphite-epoxy laminae, with the following material properties: $E_1 = 144.8$ GPa, $E_2 = 11.7$ GPa, $G_{12} = 9.66$ GPa and $\nu_{12} = 0.21$. The density was assumed to be $\rho = 2710$ kg/m$^3$. The material constitutive matrix $\mathbf C_{IJ}$ in Voigt notation is calculated as
\begin{equation}
\mathbf C_{IJ} = \left( 
\begin{array}{ccc}
1/E_1 & -\nu_{12}/E_1 & 0\\
-\nu_{12}/E_1 & 1/E_2 & 0\\
0 & 0 & 1/G_{12}
\end{array} \right)^{-1} 
\end{equation}

The material properties have been rotated by an angle $\theta$ ranging from $0^\circ$ to $90^\circ$ in order to evaluate the effect in the corresponding SIFs. The rotation of the constitutive matrix is given by 
\begin{equation}
C_{ijkl} =r_{im}(\theta) r_{jn}(\theta) r_{ko}(\theta) r_{lm}(\theta) C_{mnop}
\label{eq:Cijkl_rot}
\end{equation}
where $C_{mnop}$ is the unrotated material properties, $C_{ijkl}$ is the rotated one and 
\begin{equation}
r_{ij}(\theta) = \left( 
\begin{array}{cc}
\cos{\theta} & \sin{\theta}\\
-\sin{\theta} & \cos{\theta}
\end{array} \right)
\label{eq:rotation1}
\end{equation}
is the rotation matrix.

The plate is subjected to an initial velocity field in the y-direction and it is defined as
\begin{equation}
\dot u(\mathbf x,t) = \frac{\partial u(\mathbf x,0)}{\partial t} = 50\frac{y}{2h} \text{m/s}
\end{equation}

Bobaru et al. \cite{bobaru2009convergence} studied the convergence in PD for 1D problems, and concluded that there are three main different approaches: 
\begin{enumerate}
 \item $m-$convergence: the number of particles $m$ increases as the horizon remains fixed. The PD converges to the non-local solution for that particular horizon;
 \item $\delta-$convergence: $\delta \rightarrow 0$ while $m$ is fixed. In this case the PD formulation converges to the local solution, i.e., the solution obtained with FEM for instance;
 \item $\delta m-$convergence: the number of particles $m$ increases as $\delta \rightarrow 0$, with $m$ increasing faster than $\delta$ decreases. The solution converges uniformly to the local solution and faster than using the $m-$ convergence alone. 
\end{enumerate}

\begin{figure}[!htb]
\centering
\subfigure[$200 \times 200$ particles]{\includegraphics[scale=0.55]{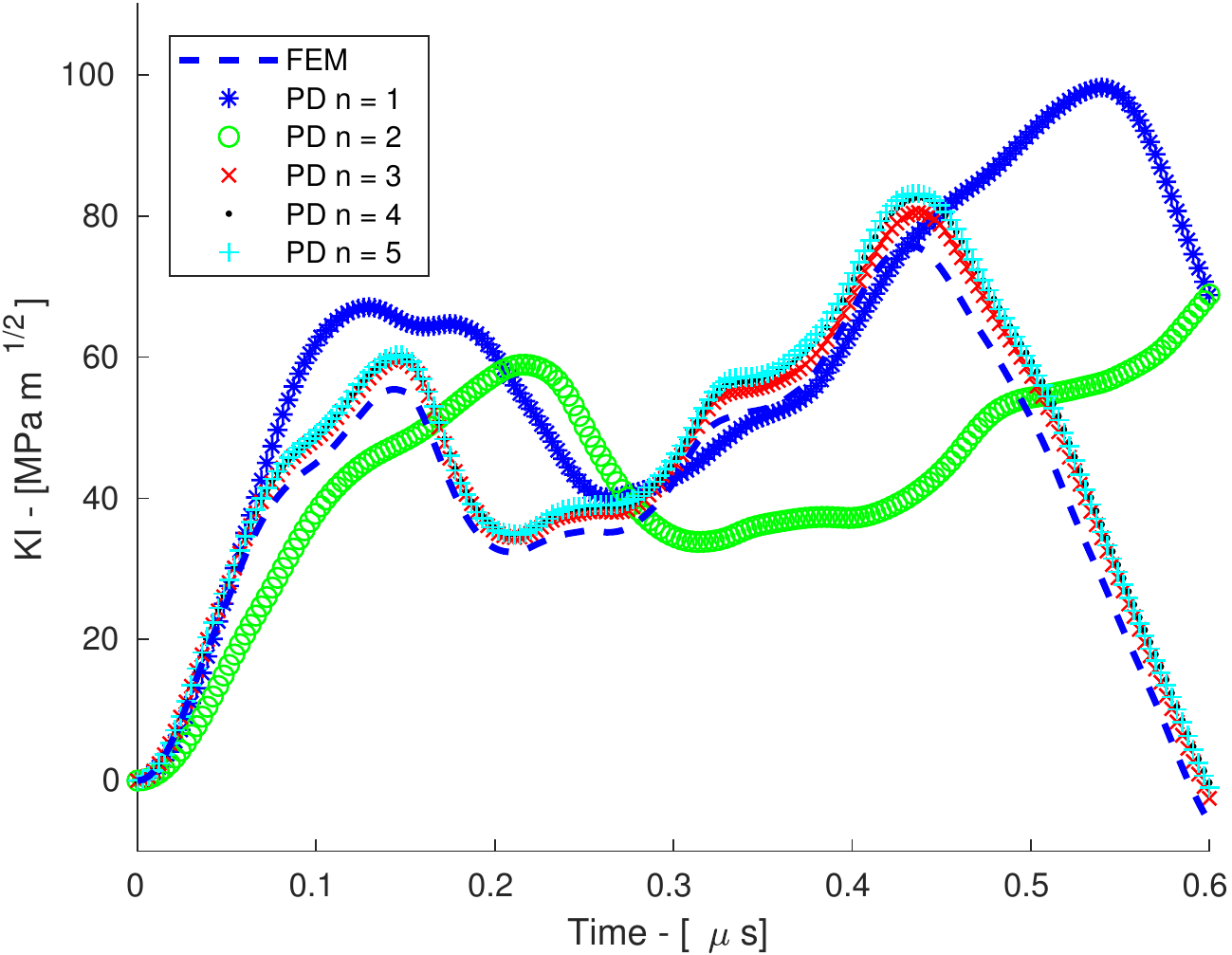} \label{fig:horizon_KI_201}}
\subfigure[$300 \times 300$ particles]{\includegraphics[scale=0.55]{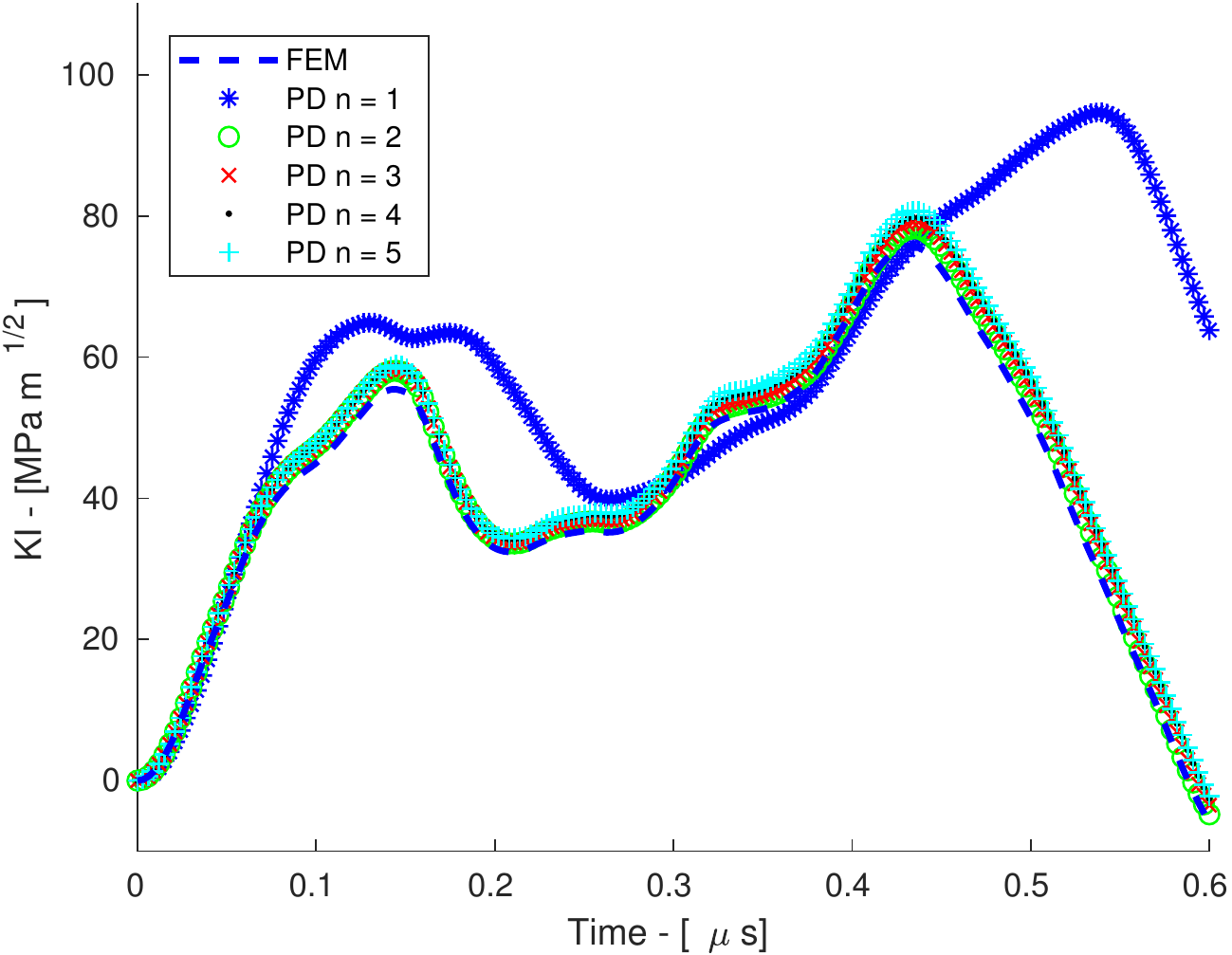} \label{fig:horizon_KI_301}}
\subfigure[$400 \times 400$ particles]{\includegraphics[scale=0.55]{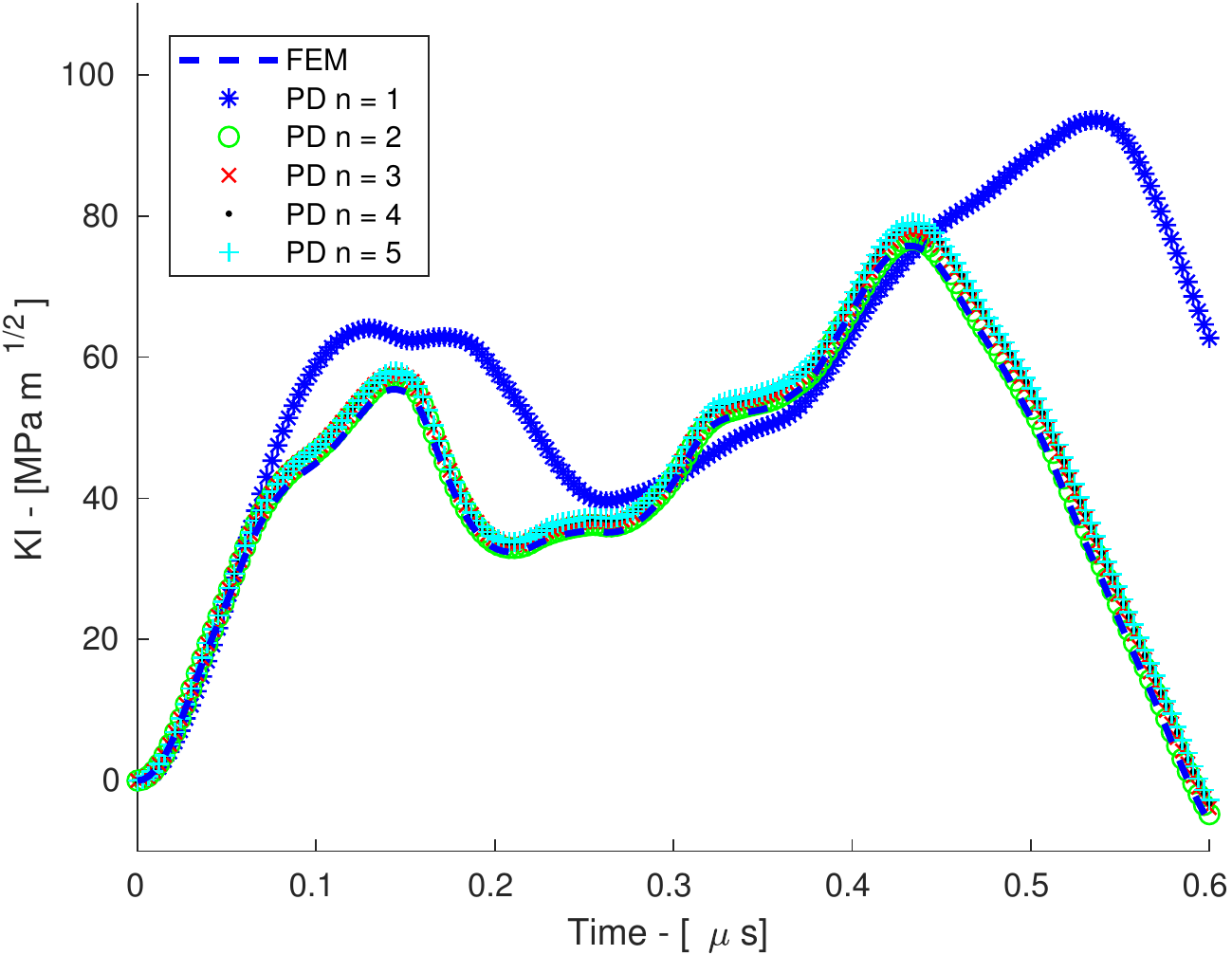} \label{fig:horizon_KI_401}}
\subfigure[$500 \times 500$ particles]{\includegraphics[scale=0.55]{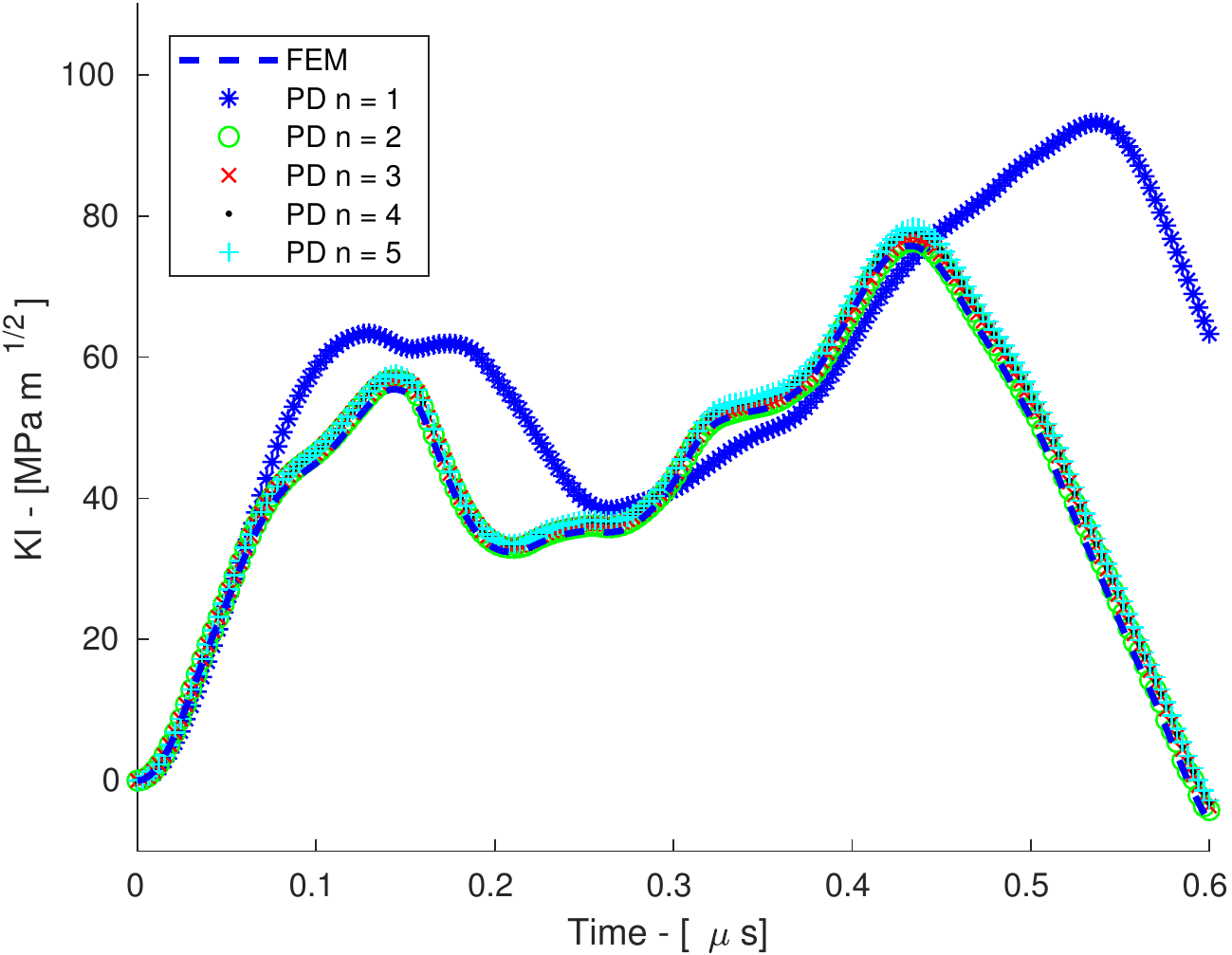} \label{fig:horizon_KI_501}}
\caption{Edge crack: comparison different grid spacing and horizon size - mode I.}
\label{fig:comparison_horizon_KI}
\end{figure}

\begin{figure}[!htb]
\centering
\subfigure[$200 \times 200$ particles]{\includegraphics[scale=0.551]{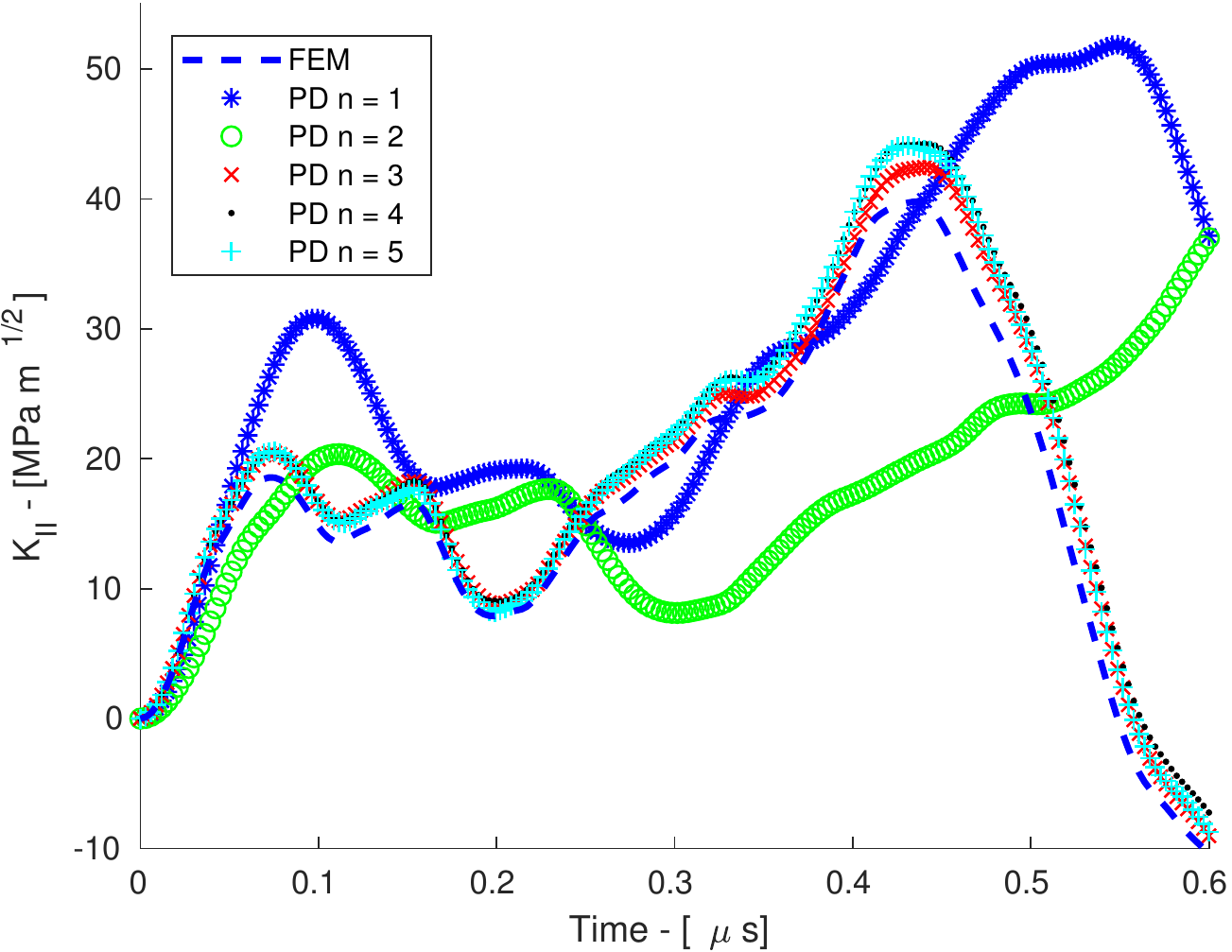} \label{fig:horizon_KII_201}}
\subfigure[$300 \times 300$ particles]{\includegraphics[scale=0.551]{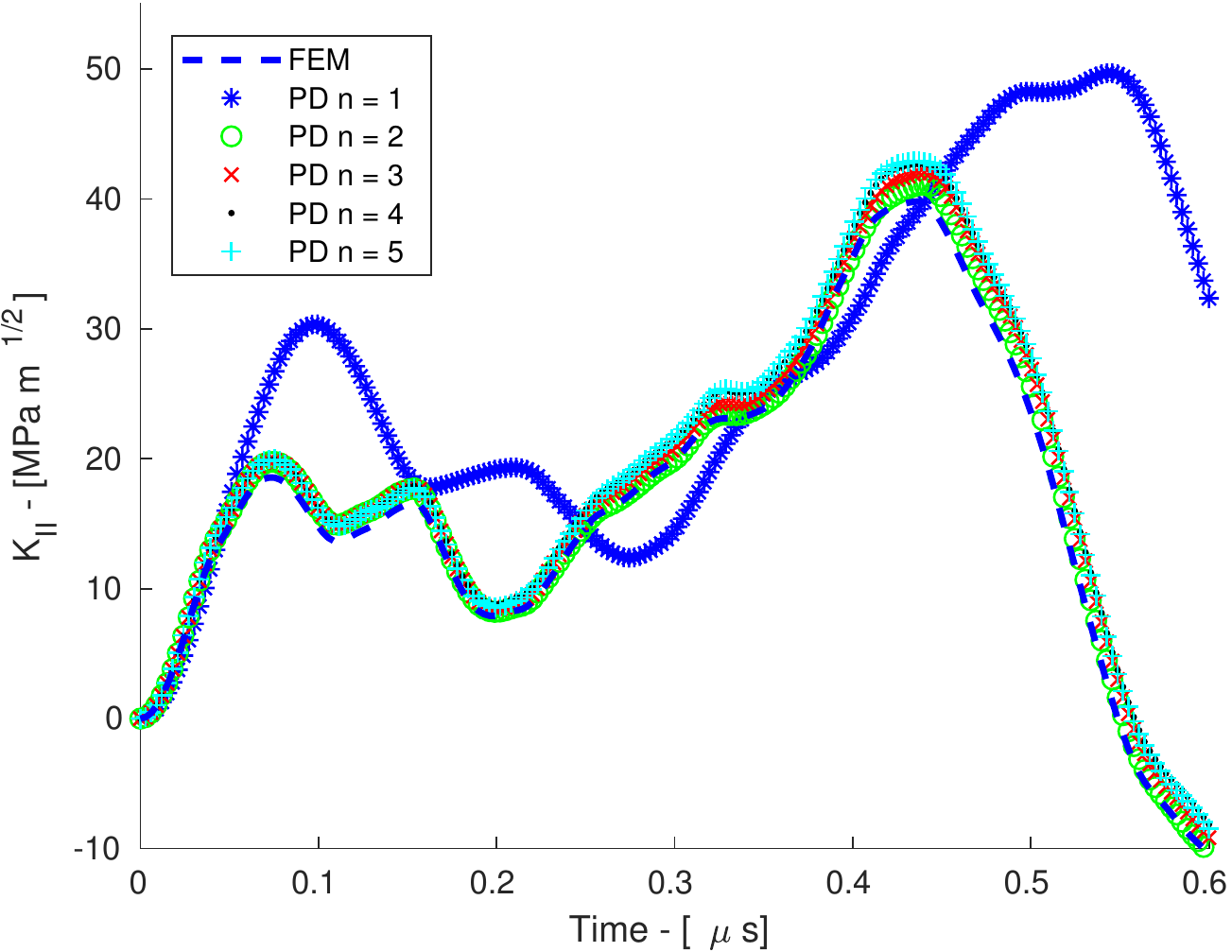} \label{fig:horizon_KII_301}}
\subfigure[$400 \times 400$ particles]{\includegraphics[scale=0.551]{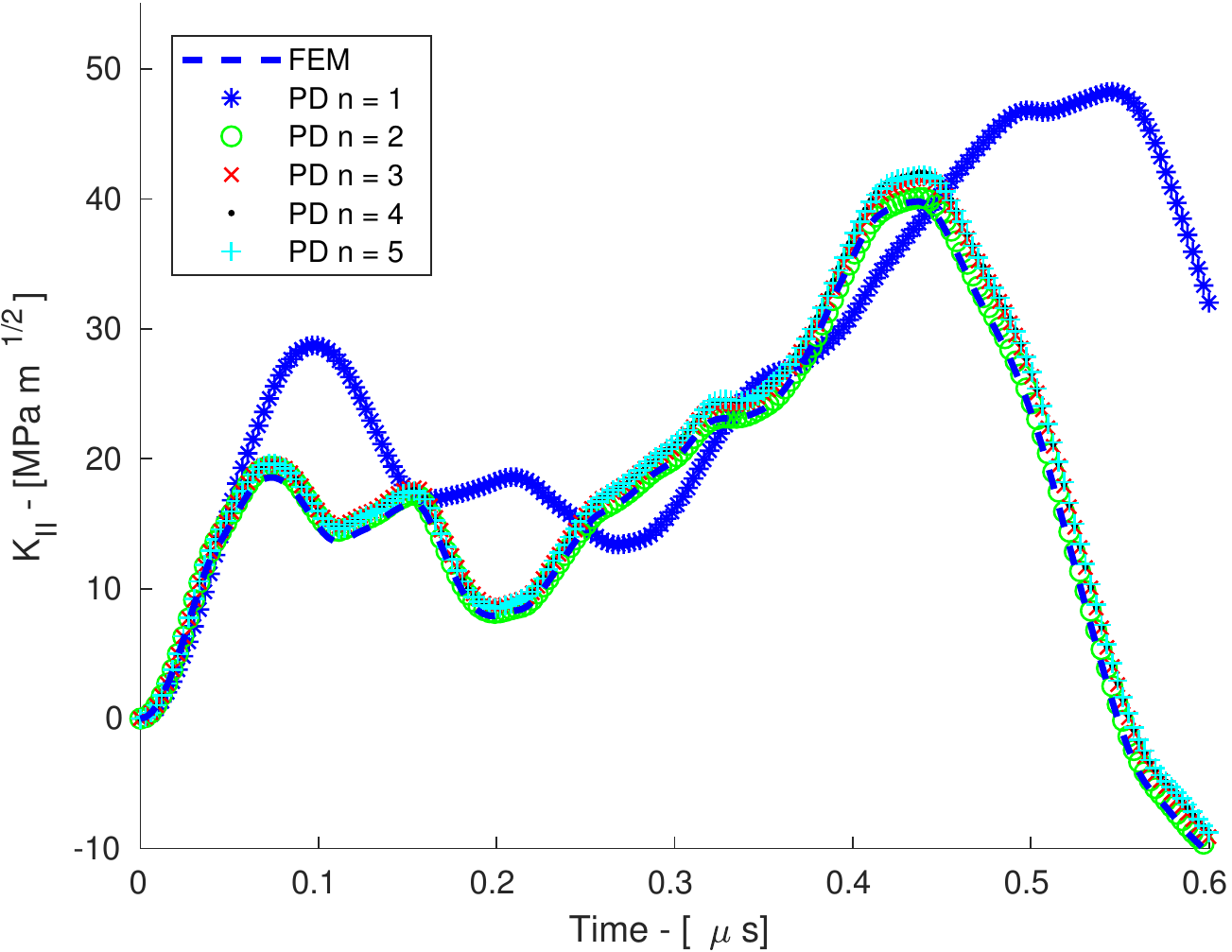} \label{fig:horizon_KII_401}}
\subfigure[$500 \times 500$ particles]{\includegraphics[scale=0.551]{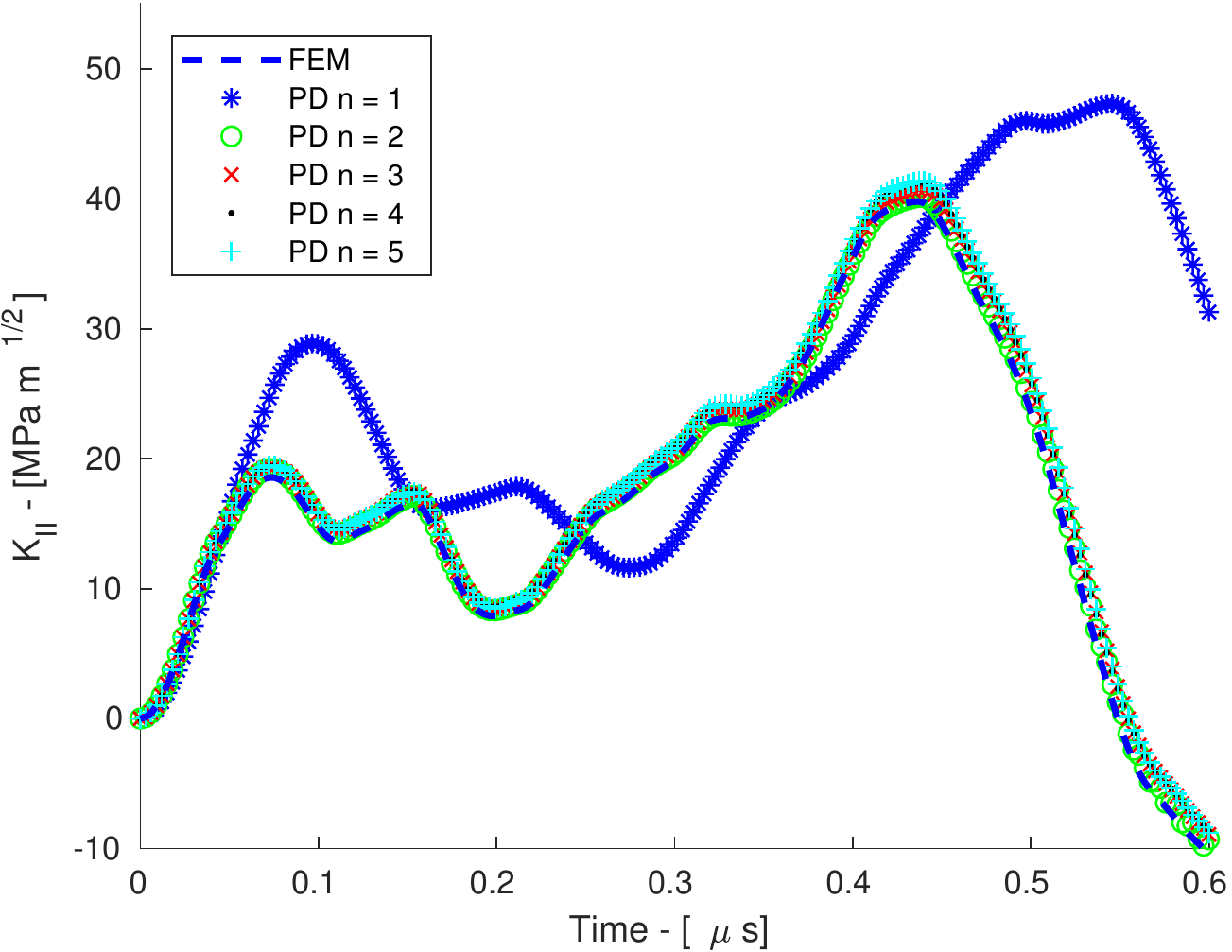} \label{fig:horizon_KII_501}}
\caption{Edge crack: comparison different grid spacing and horizon size - mode II.}
\label{fig:comparison_horizon_KII}
\end{figure}

Initially we investigate how the horizon size influences the DSIF for an anisotropic material. The material orientation is fixed at $\theta = 30^\circ$ and we calculate the DSIF for different horizon sizes. The dynamic mode I and mode II stress intensity factors are given in Figures \ref{fig:comparison_horizon_KI} and \ref{fig:comparison_horizon_KII} for four different particle discretisations. In all cases the DSIFs obtained with PD are compared with those obtained using a $500\times 500$ 4-node fully integrated quadrilateral finite element mesh. The horizon is defined as $\delta = n \Delta x$, where $n$ is a constant and $\Delta x$ is the grid spacing.
\psfrag{FEM}{\tiny FEM}
\psfrag{KI - [MPa m 0.5]}{\footnotesize $K_I$ - [MPa m$^{1/2}$]}
\psfrag{KII - [MPa m 0.5]}{\footnotesize $K_{II}$ - [MPa m$^{1/2}$]}
\psfrag{time - [s]}{\footnotesize Time - [$\mu $s]}
\psfrag{time - [micro s]}{\footnotesize Time - [$\mu $s]}
\psfrag{mesh 200 particles}{\tiny $200$ particles}
\psfrag{mesh 300 particles}{\tiny $300$ particles}
\psfrag{mesh 400 particles}{\tiny $400$ particles}
\psfrag{mesh 500 particles}{\tiny $500$ particles}
\psfrag{PD delta = 1}{\tiny $n = 1$}
\psfrag{PD delta = 2}{\tiny $n = 2$}
\psfrag{PD delta = 3}{\tiny $n = 3$}
\psfrag{PD delta = 4}{\tiny $n = 4$}
\psfrag{PD delta = 5}{\tiny $n = 5$}

Table \ref{tab:error_DSIF} shows the comparison between the PD and FEM solutions. The error in the $L_2$ norm is calculated as
\begin{equation}
\text{Error} = \frac{\sqrt{\sum_{i=1}^N (K_{\alpha_i}^{PD} - K_{\alpha_i}^{FEM})^2 }}{ \sqrt{\sum_{i=1}^N (K_{\alpha_i}^{FEM})^2} } 
\end{equation}
where $N$ is the total number of time steps, $K_\alpha^{PD}$ and $K_\alpha^{FEM}$ are the DSIFs for the PD and FEM formulations, respectively, and $\alpha = I, II$.

\begin{table}[!htb]
\centering
\caption{Relative error between PD and FEM for the dynamic mode I.}
\begin{tabular}{|c|c|c|c|c|c|}
\hline 
& \multicolumn{5}{|c|}{Horizon $\delta = n \Delta x$} \\ \hline 
Particles & $n = 1$ & $n = 2$ & $n = 3$ & $n = 4$ & $n = 5$ \\ \hline
$200 \times 200$  & $0.9898$ & $0.7387$ & $0.1137$ & $0.1643$ & $0.1374$\\ \hline
$300 \times 300$ & $0.9317$ & $0.0520$ & $0.0861$ & $0.1122$ & $0.1130$\\ \hline
$400 \times 400$ & $0.8990$ & $0.0273$ & $0.0751$ & $0.0844$ & $0.0847$\\ \hline
$500 \times 500$ & $0.8808$ & $0.0306$ & $0.0580$ & $0.0654$ & $0.0719$\\ \hline
\end{tabular}
\label{tab:error_DSIF}
\end{table}

In Figure \ref{fig:horizon_KI_201}, the DSIFs calculated for $n = 1$ and $n = 2$ are in total disagreement with the reference FEM solution and the other PD solutions as well. This is due to the fact that the dynamics are not modelled properly in this case; the horizon size is not adequate for this particular grid spacing. When the grid spacing decreases, the quality of the solution calculated with smaller horizon sizes also increases, agreeing with the reference solution, as can be seen in Figure \ref{fig:horizon_KI_301}, where the error solution for $n = 2$ decreases from $0.7387$ to $0.0520$, representing a better correlation to the reference solution, while $n = 1$ shows a small reduction in the error but still provides a poor approximation. In Figures \ref{fig:horizon_KI_401} and \ref{fig:horizon_KI_501}, there are negligible differences between the DSIFs obtained with PD for any horizon size. The same analysis is valid for the dynamic mode II results depicted in Figure \ref{fig:comparison_horizon_KII}.

\begin{figure}[!htb]
\centering
\includegraphics[scale=0.68]{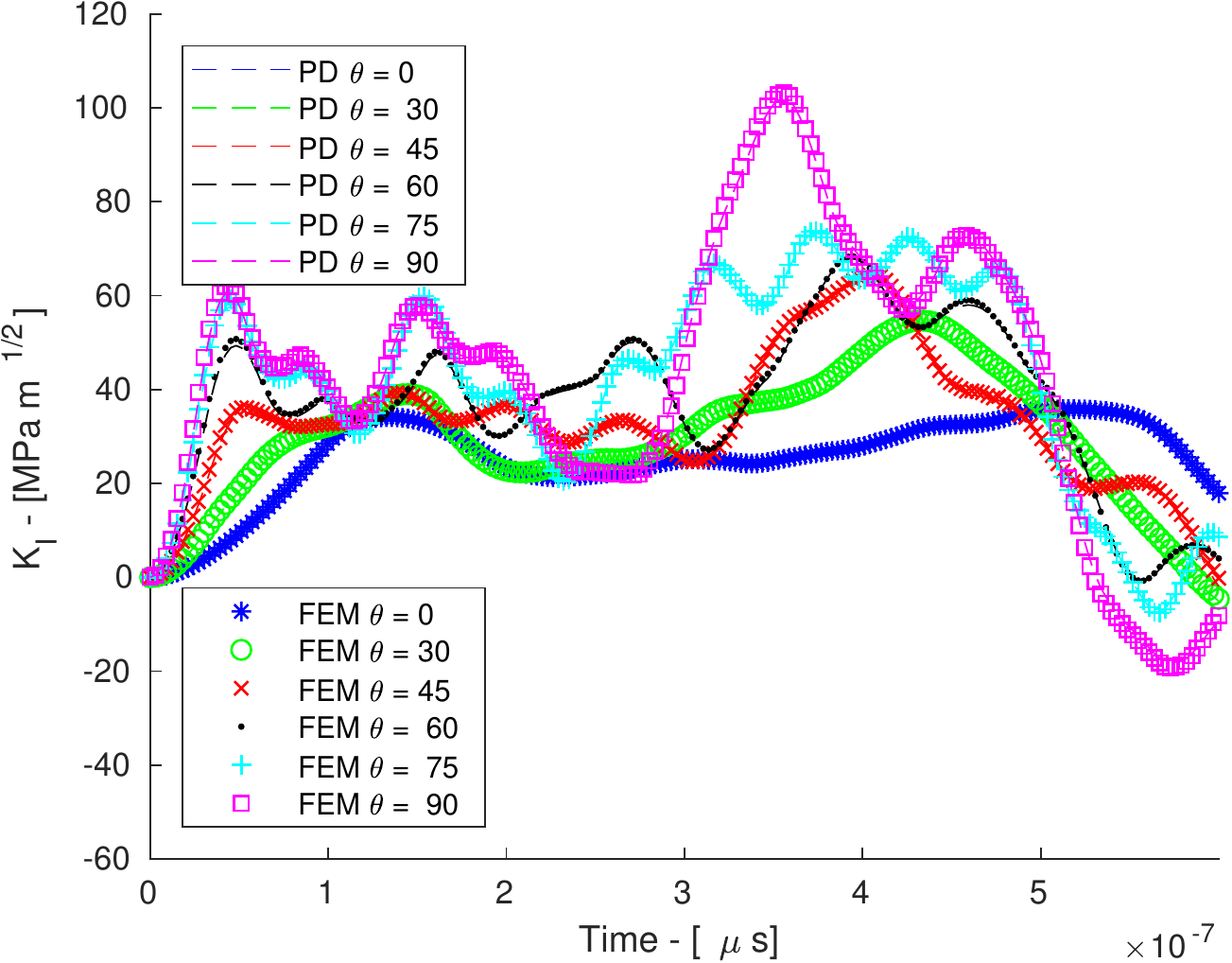}
\caption{Edge crack: DSIF for different values of $\theta$ - mode I.}
\label{fig:DSIF_thetaI}
\end{figure}

\begin{figure}[!htb]
\centering
\includegraphics[scale=0.68]{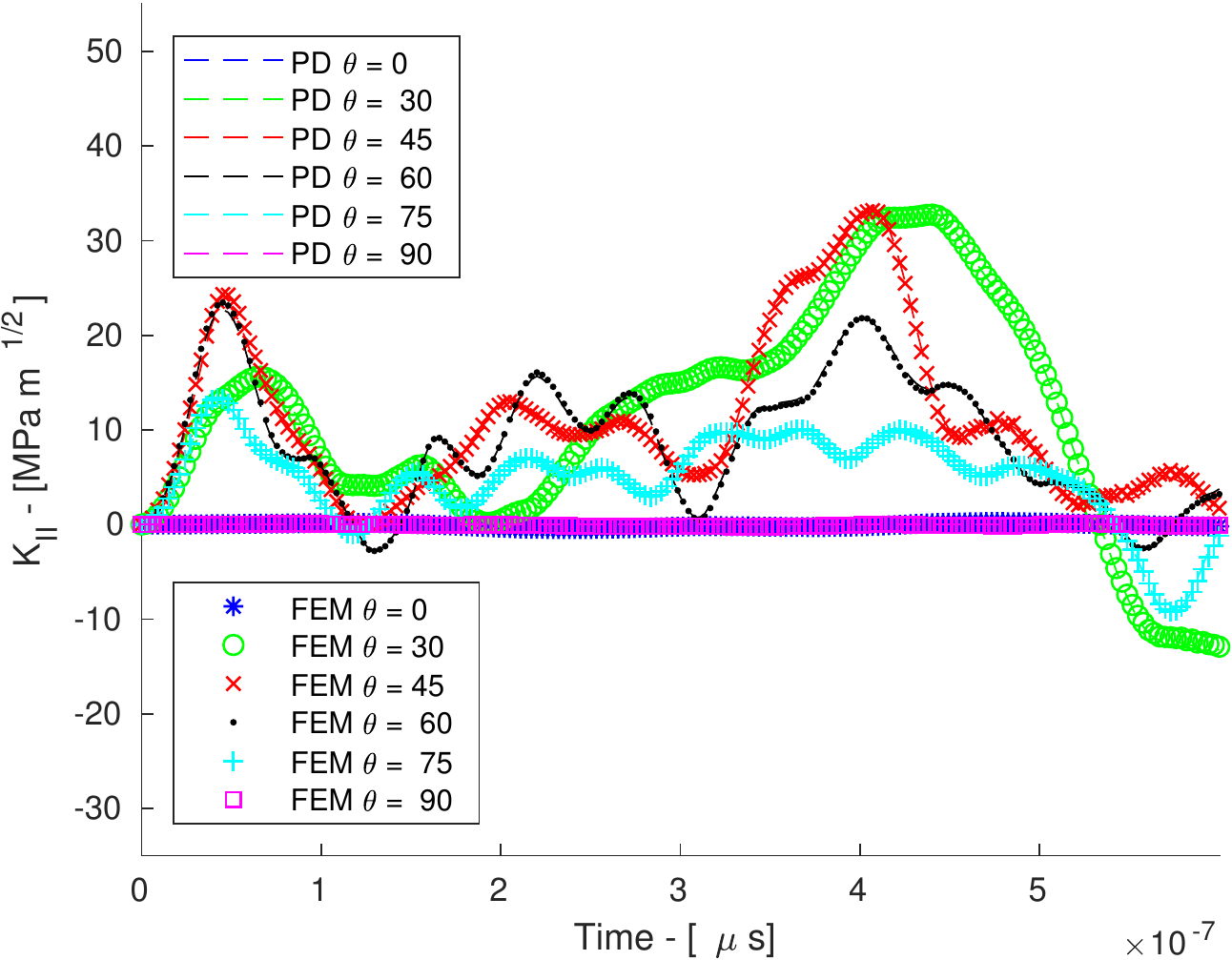}
\caption{Edge crack: DSIF for different values of $\theta$ - mode II.}
\label{fig:DSIF_thetaII}
\end{figure}

From the previous analysis and the results from Table \ref{tab:error_DSIF}, we adopt the following parameters: $400 \times 400$ particle discretisation and $n = 2$, as this has the lowest error between the FEM and PD solutions. Next we evaluate the DSIF for these parameters and varying the anisotropy angle $\theta$. The dynamic mode I and II stress intensity factors are shown in Figures \ref{fig:DSIF_thetaI} and \ref{fig:DSIF_thetaII}, respectively. An excellent agreement is obtained for the PD solution compared to a FEM approach.

\psfrag{Peridynamic theta 0}{\tiny PD $\theta = 0^\circ$}
\psfrag{Peridynamic theta 30}{\tiny PD $\theta = 30^\circ$}
\psfrag{Peridynamic theta 45}{\tiny PD $\theta = 45^\circ$}
\psfrag{Peridynamic theta 60}{\tiny PD $\theta = 60^\circ$}
\psfrag{Peridynamic theta 75}{\tiny PD $\theta = 75^\circ$}
\psfrag{Peridynamic theta 90}{\tiny PD $\theta = 90^\circ$}
\psfrag{Finite elem theta 0}{\tiny FEM $\theta = 0^\circ$}
\psfrag{Finite elem theta 30}{\tiny FEM $\theta = 30^\circ$}
\psfrag{Finite elem theta 45}{\tiny FEM $\theta = 45^\circ$}
\psfrag{Finite elem theta 60}{\tiny FEM $\theta = 60^\circ$}
\psfrag{Finite elem theta 75}{\tiny FEM $\theta = 75^\circ$}
\psfrag{Finite elem theta 90}{\tiny FEM $\theta = 90^\circ$}

\subsection{Crack propagation of centred crack in anisotropic 2D plate}

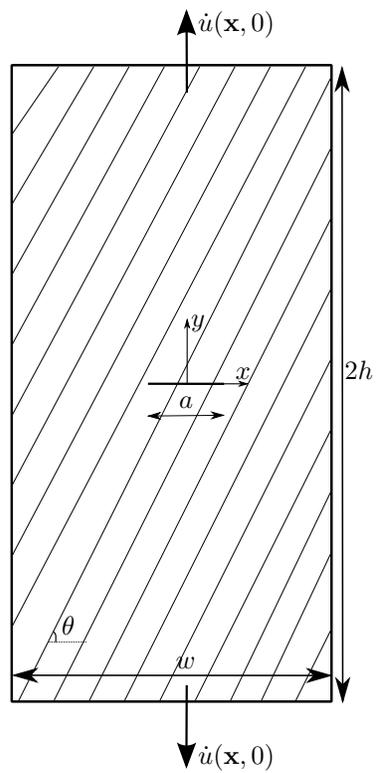
\begin{figure}[!htb]
\centering
\vspace{0.5cm}
\def\svgwidth{0.3\linewidth}
\begingroup%
  \makeatletter%
  \providecommand\color[2][]{%
    \errmessage{(Inkscape) Color is used for the text in Inkscape, but the package 'color.sty' is not loaded}%
    \renewcommand\color[2][]{}%
  }%
  \providecommand\transparent[1]{%
    \errmessage{(Inkscape) Transparency is used (non-zero) for the text in Inkscape, but the package 'transparent.sty' is not loaded}%
    \renewcommand\transparent[1]{}%
  }%
  \providecommand\rotatebox[2]{#2}%
  \ifx\svgwidth\undefined%
    \setlength{\unitlength}{166.3360806bp}%
    \ifx\svgscale\undefined%
      \relax%
    \else%
      \setlength{\unitlength}{\unitlength * \real{\svgscale}}%
    \fi%
  \else%
    \setlength{\unitlength}{\svgwidth}%
  \fi%
  \global\let\svgwidth\undefined%
  \global\let\svgscale\undefined%
  \makeatother%
  \begin{picture}(1,2.08673907)%
    %\put(0,0){\includegraphics[width=\unitlength]{centred_crack_anisot2.eps}}%
    \put(0,0){\includegraphics[width=\unitlength,page=1]{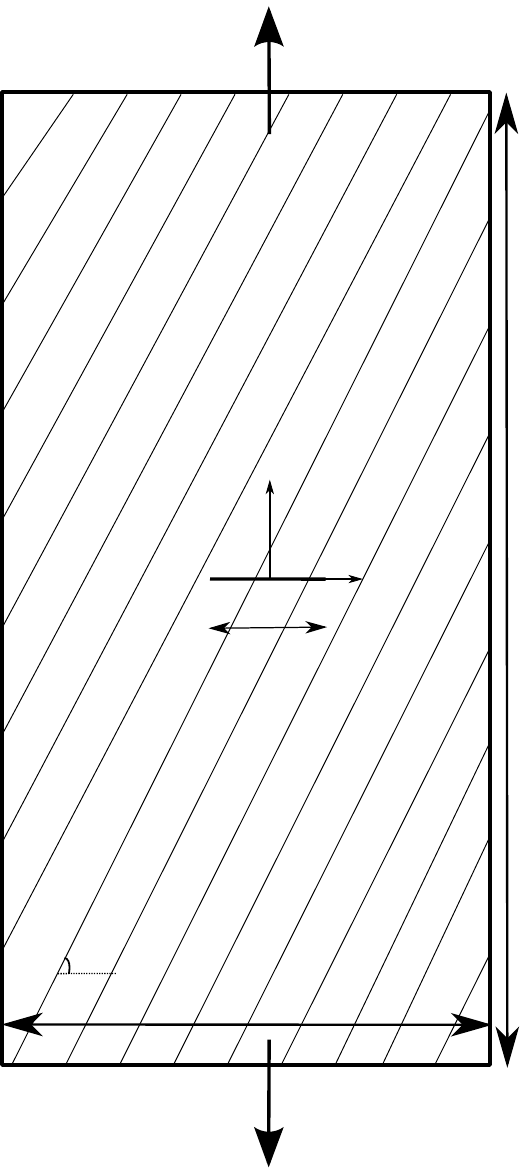}}%
    \put(0.485,0.30){\color[rgb]{0,0,0}\makebox(0,0)[lb]{$w$}}%
    \put(0.98,1.15){\color[rgb]{0,0,0}\makebox(0,0)[lb]{$2h$}}%
    \put(0.495,1.065){\color[rgb]{0,0,0}\makebox(0,0)[lb]{$a$}}%
    \put(0.55,0.001){\color[rgb]{0,0,0}\makebox(0,0)[lb]{$\dot u (\mathbf x,0)$}}%
    \put(0.55,2.15){\color[rgb]{0,0,0}\makebox(0,0)[lb]{$\dot u (\mathbf x,0)$}}%
    \put(0.15,0.40){\color[rgb]{0,0,0}\makebox(0,0)[lb]{$\theta$}}%
    \put(0.66,1.15){\color[rgb]{0,0,0}\makebox(0,0)[lb]{$x$}}%
    \put(0.53,1.29){\color[rgb]{0,0,0}\makebox(0,0)[lb]{$y$}}%
  \end{picture}%
\endgroup%
\caption{Anisotropic plate with centred crack.}
\label{fig:centred_crack_anisot}
\end{figure}

In this section we analyse the crack propagation in an anisotropic rectangular plate as depicted in Figure \ref{fig:centred_crack_anisot}. The plate has aspect ratio $h/w=2$ and contains a centred crack such that $a/w=0.2$. The plate has dimensions $w=h=125$ mm. The plate is made from a unidirectional HTA/6376 composite laminate and the material properties are given in Table \ref{tab:mat_HTA}, with $G_{12}=5.5$ GPa, $\sigma_{LTu}=70$ MPa and $\nu_{12} = 0.3$. The plate is subjected to an initial velocity gradient $\dot u(\mathbf x,0) = 50 \frac{y}{2h}$ m/s.

\begin{table}[!htb]
\caption{Material properties of HTA/6376 composite}
\begin{tabular}{|c|c|c|c|c|}
\hline
& HTA fibre & 6376 epoxy & laminate ($\theta = 0^\circ$) & laminate ($\theta = 90^\circ$) \\ \hline
Young's modulus [GPa] & $235$ & $3.6$ & $136$ & $8.75$ \\ \hline
Tensile strength [MPa] & $3920$ & $105$ & $1670$ & $60$ \\ \hline
Maximum elongation & $1.7\%$& $3.1\%$ & - & - \\ \hline
Density [kg/m$^3$] & $1770$ & $1310$ & $1586$ & $1586$ \\ \hline
\end{tabular}
\label{tab:mat_HTA}
\end{table}

\begin{figure}[!htb]
\centering
\subfigure[$n = 2$]{\includegraphics[scale=0.0835]{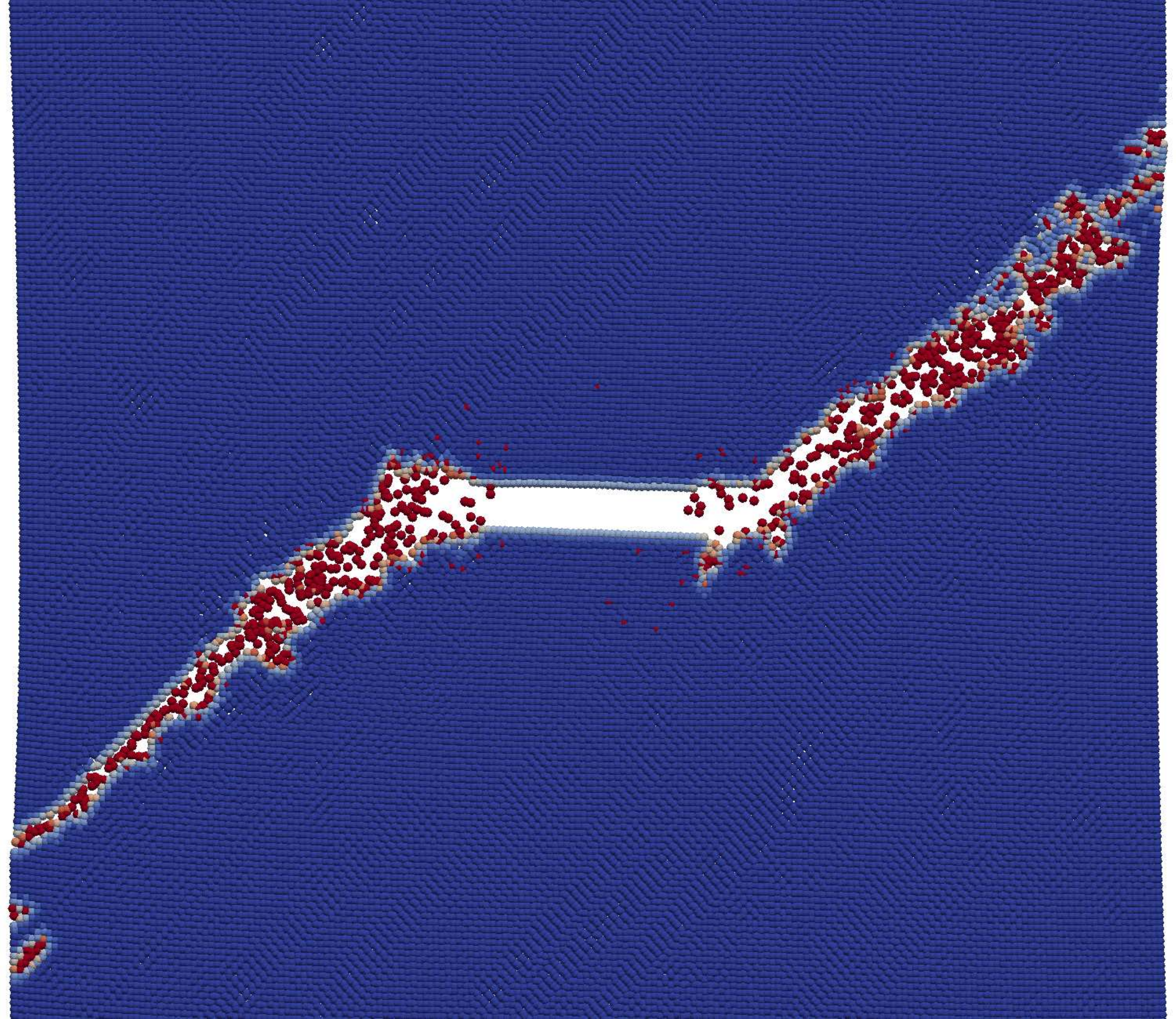} } 
\subfigure[$n = 3$]{\includegraphics[scale=0.0835]{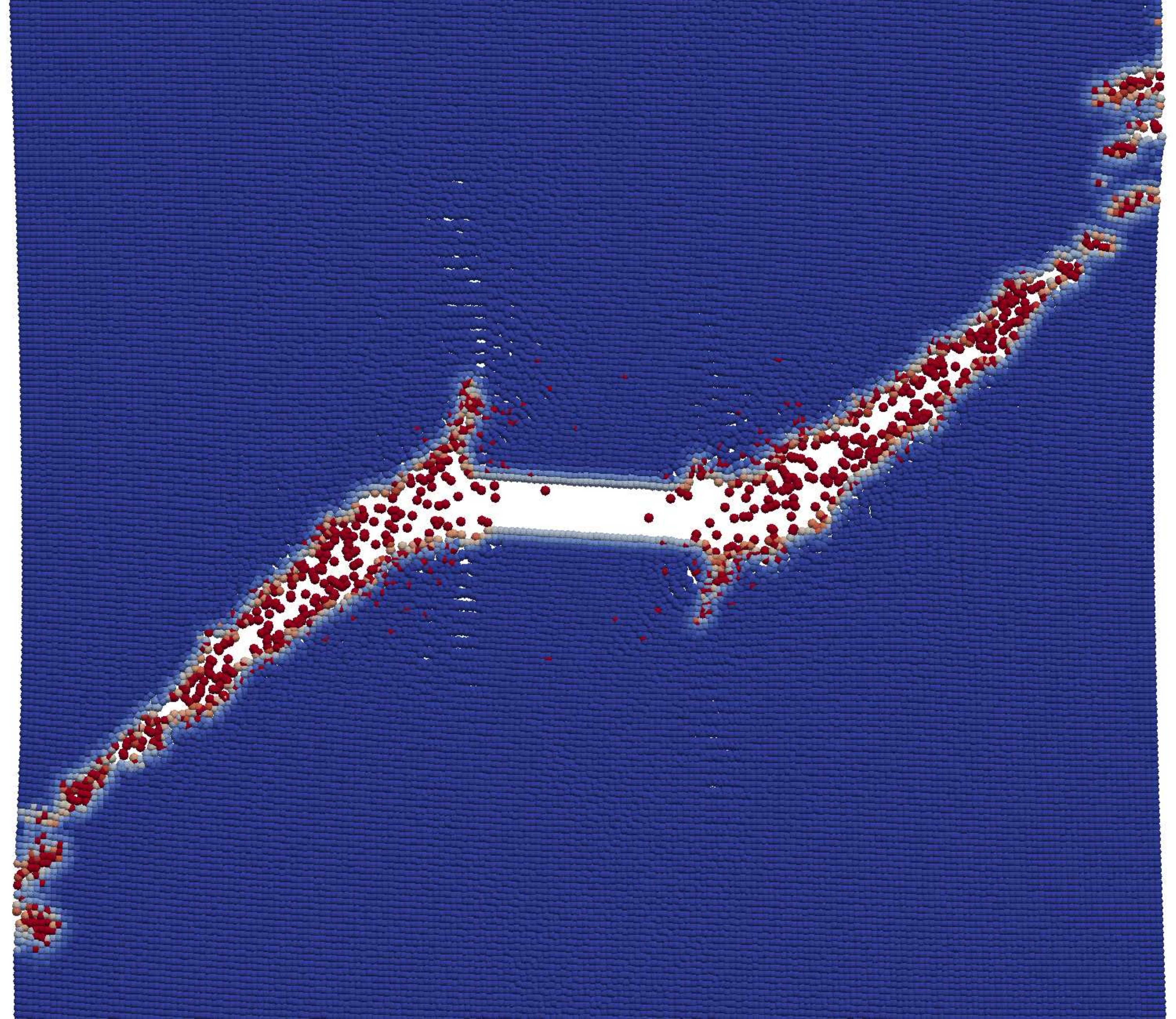} } 
\subfigure[$n = 4$]{\includegraphics[scale=0.0835]{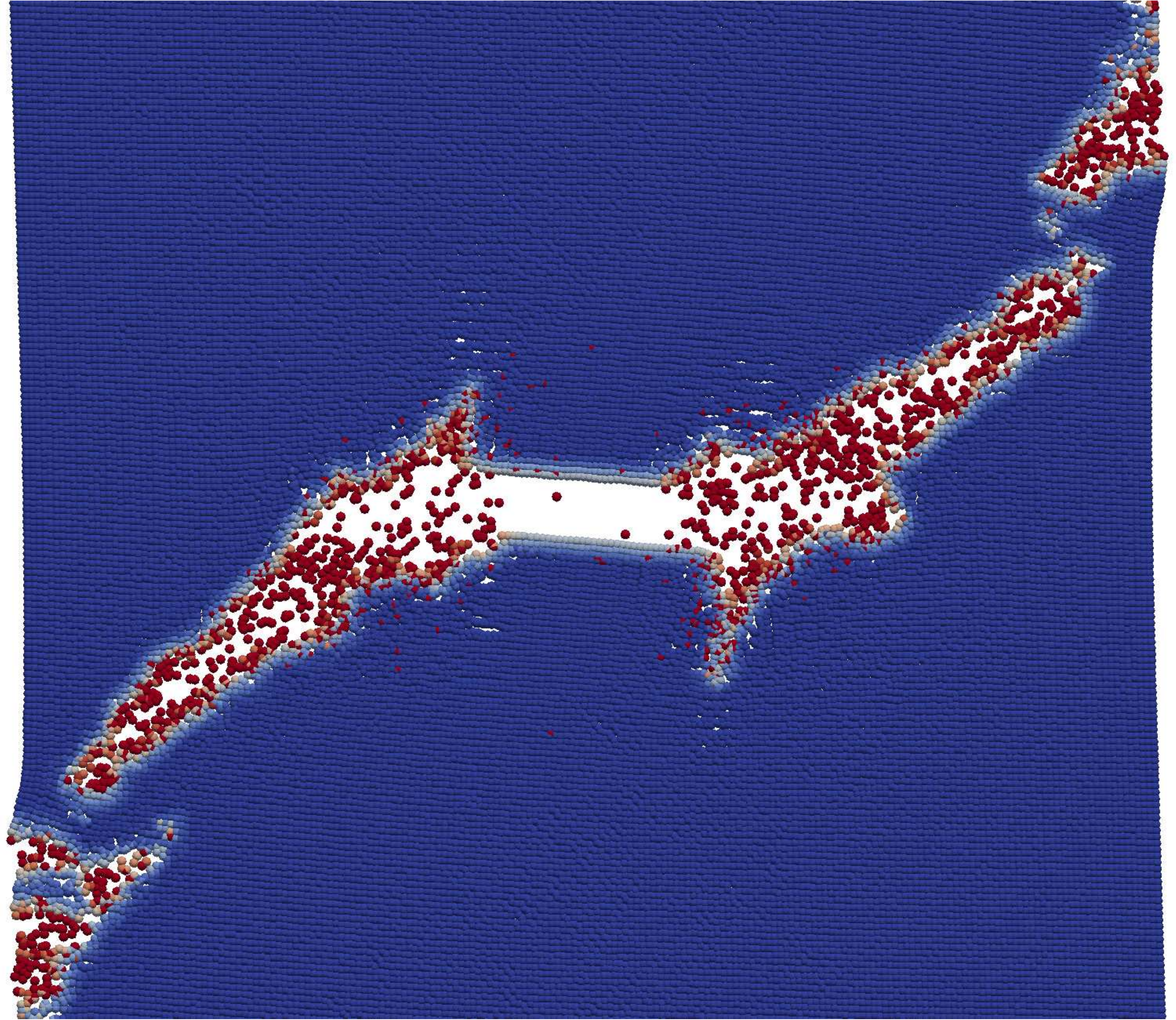} \label{fig:201_theta45_delta4}} 
\subfigure[$n = 5$]{\includegraphics[scale=0.0835]{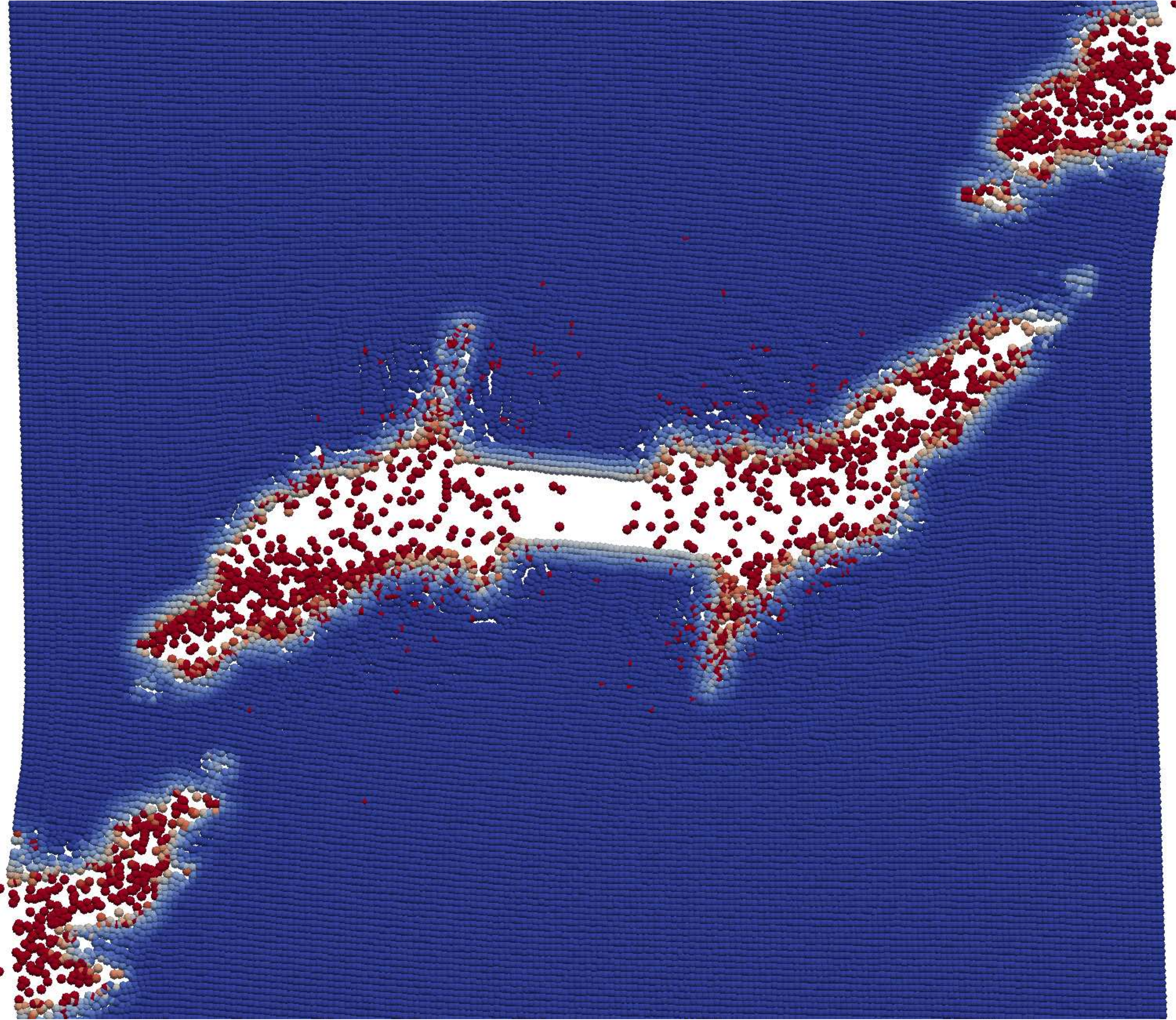} \label{fig:201_theta45_delta5}} 
\caption{Crack propagation of centred crack for different horizon - $\theta = 45^\circ$ - $200\times 400$ particles.}
\label{fig:theta45_201_401mesh}
\end{figure}

\begin{figure}[!htb]
\centering
\subfigure[$n = 2$]{\includegraphics[scale=0.084]{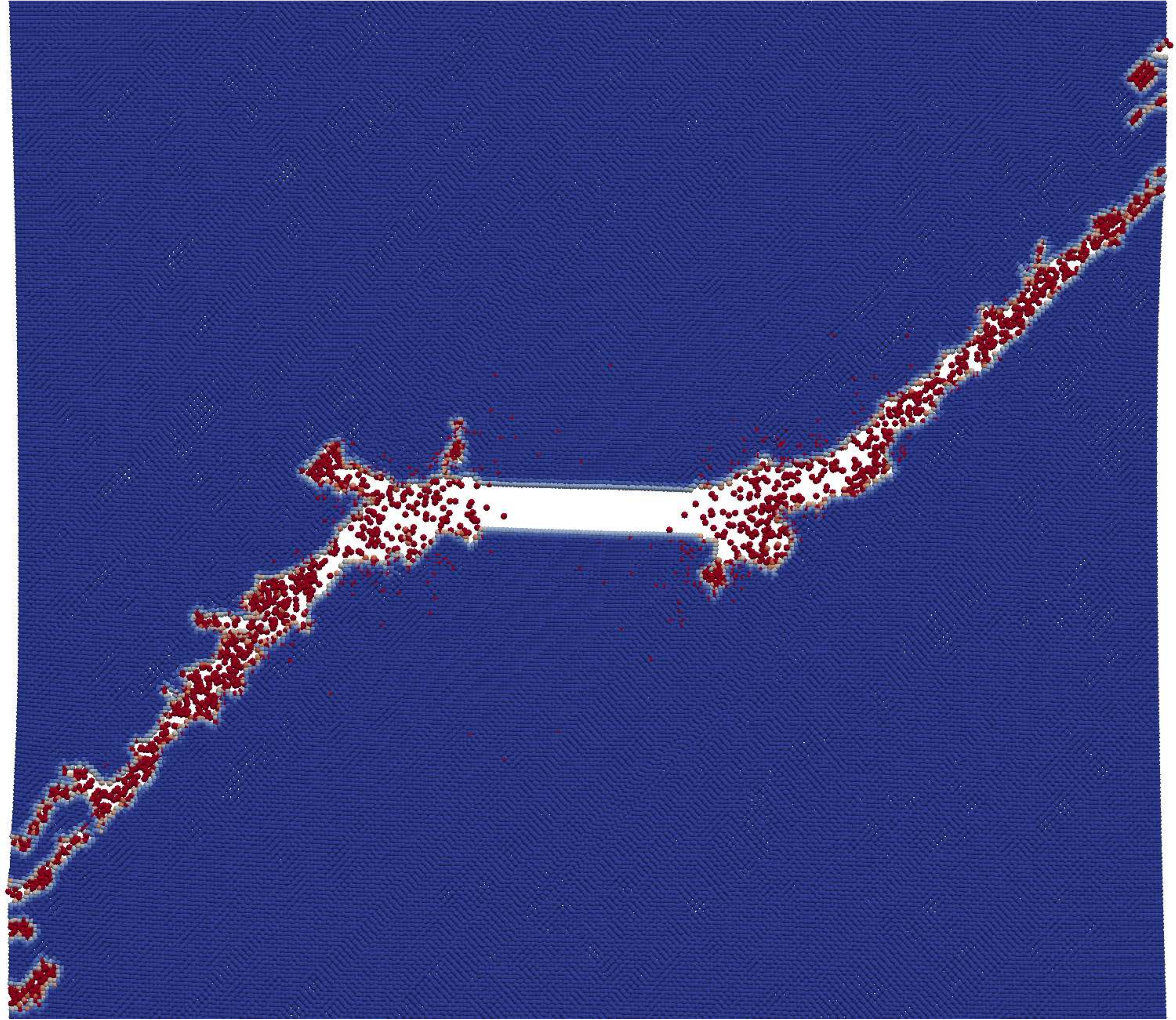} } 
\subfigure[$n = 3$]{\includegraphics[scale=0.084]{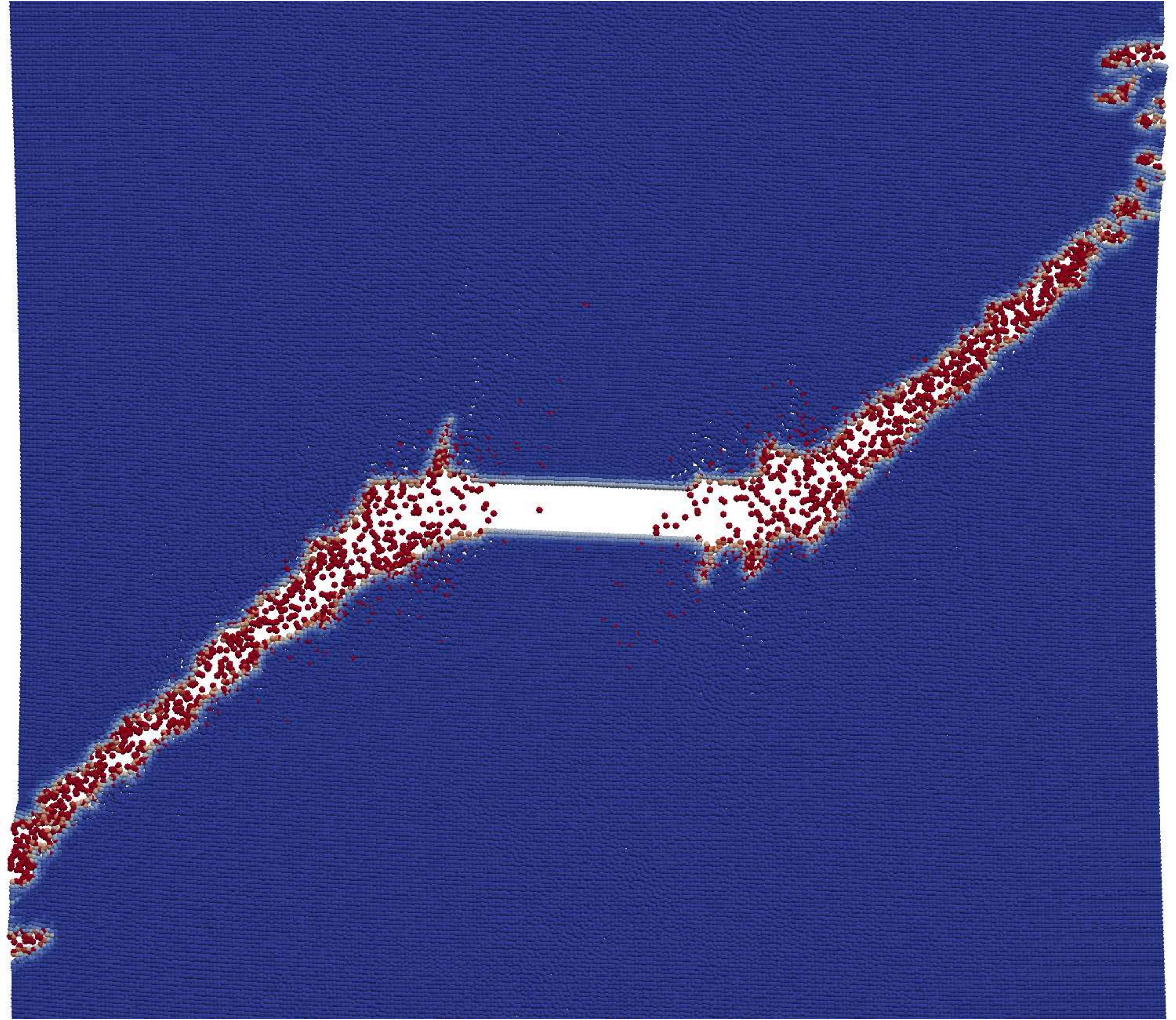} \label{fig:301_601_delta3}} 
\subfigure[$n = 4$]{\includegraphics[scale=0.084]{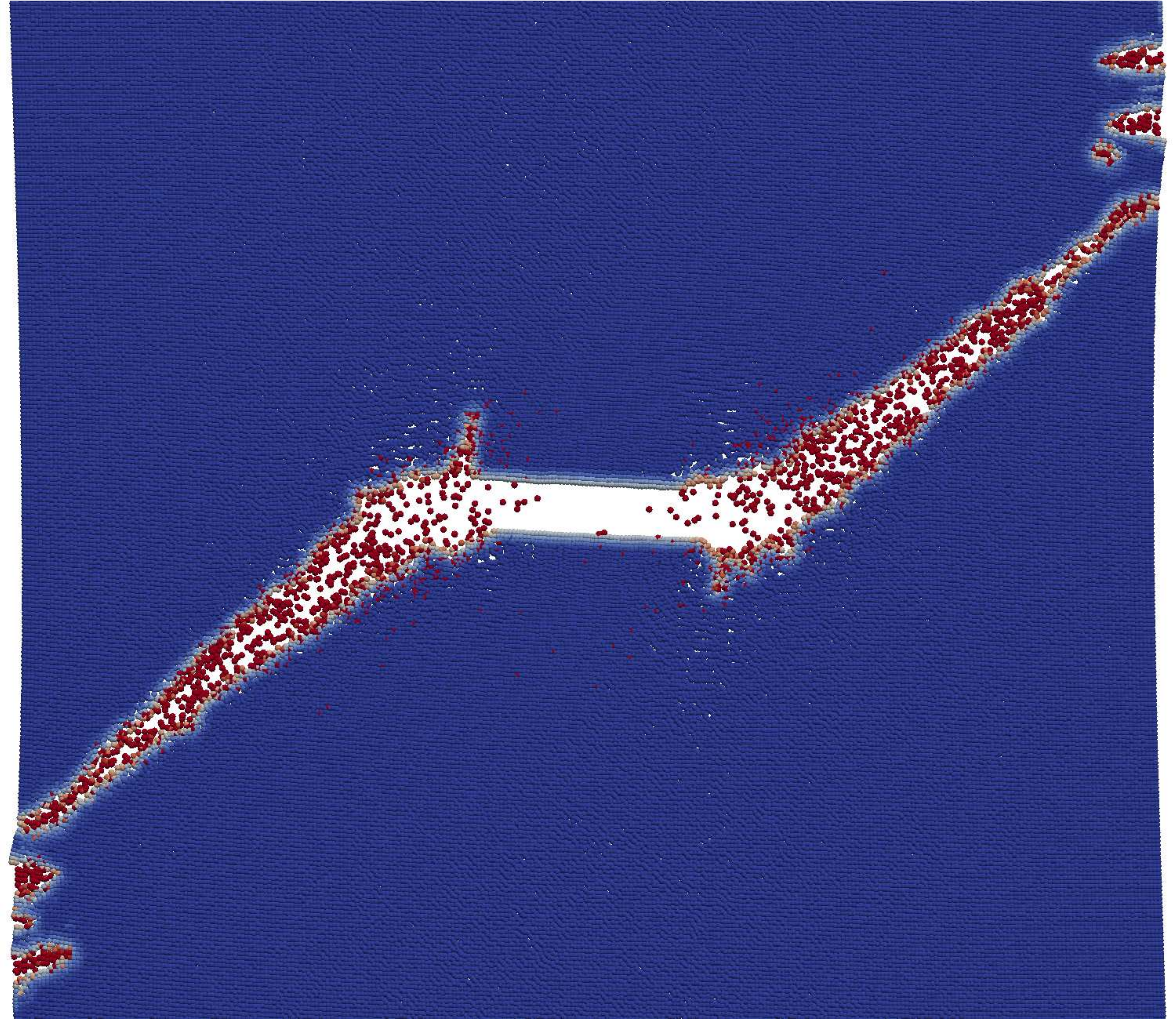} \label{fig:301_601_delta4}} 
\subfigure[$n = 5$]{\includegraphics[scale=0.084]{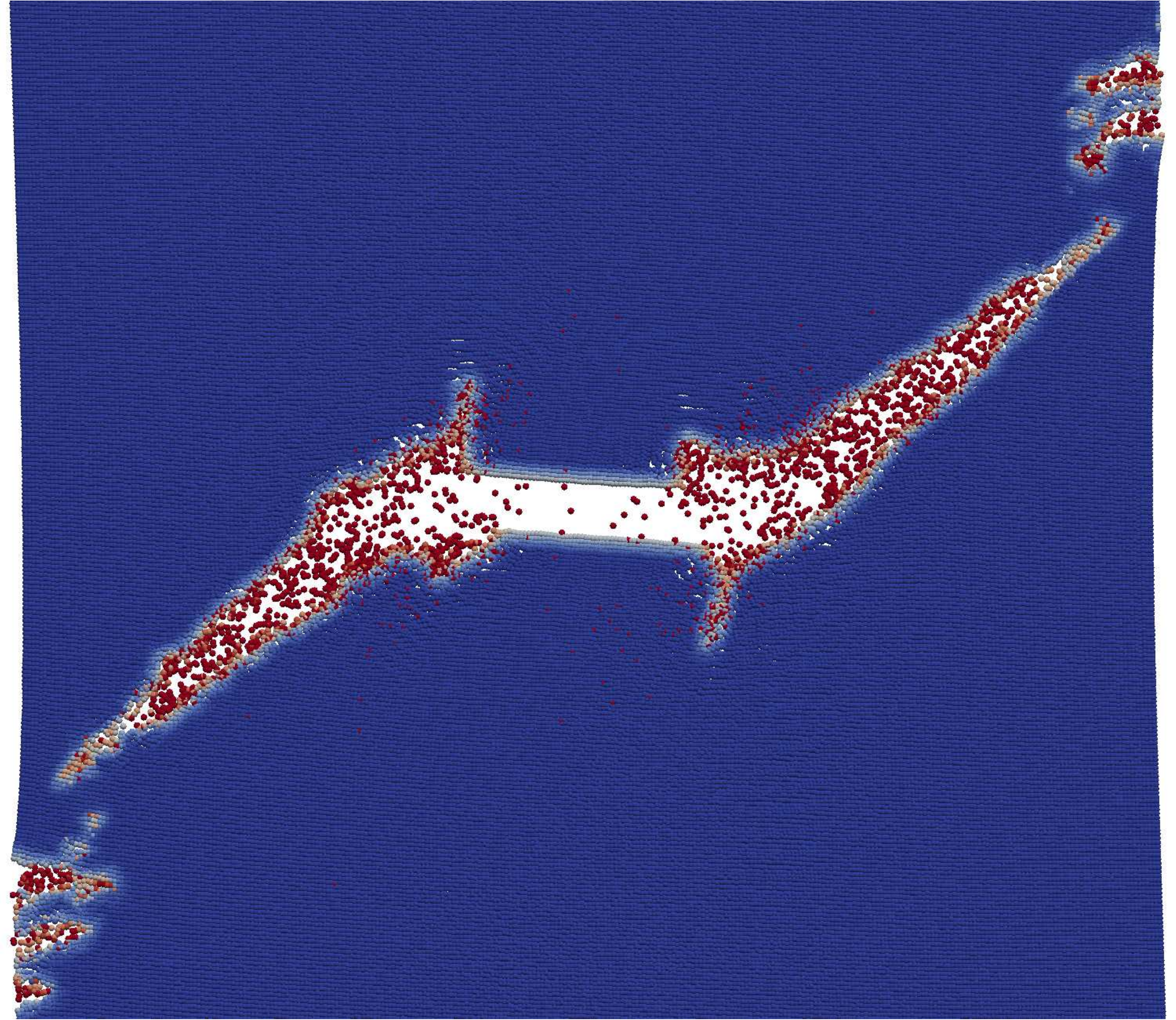} \label{fig:301_601_delta5} } 
\caption{Crack propagation of centred crack for different horizon - $\theta = 45^\circ$ - $300\times 600$ particles.}
\label{fig:theta45_301_601mesh}
\end{figure}

\begin{figure}[!htb]
\centering
\subfigure[$n = 2$]{\includegraphics[scale=0.085]{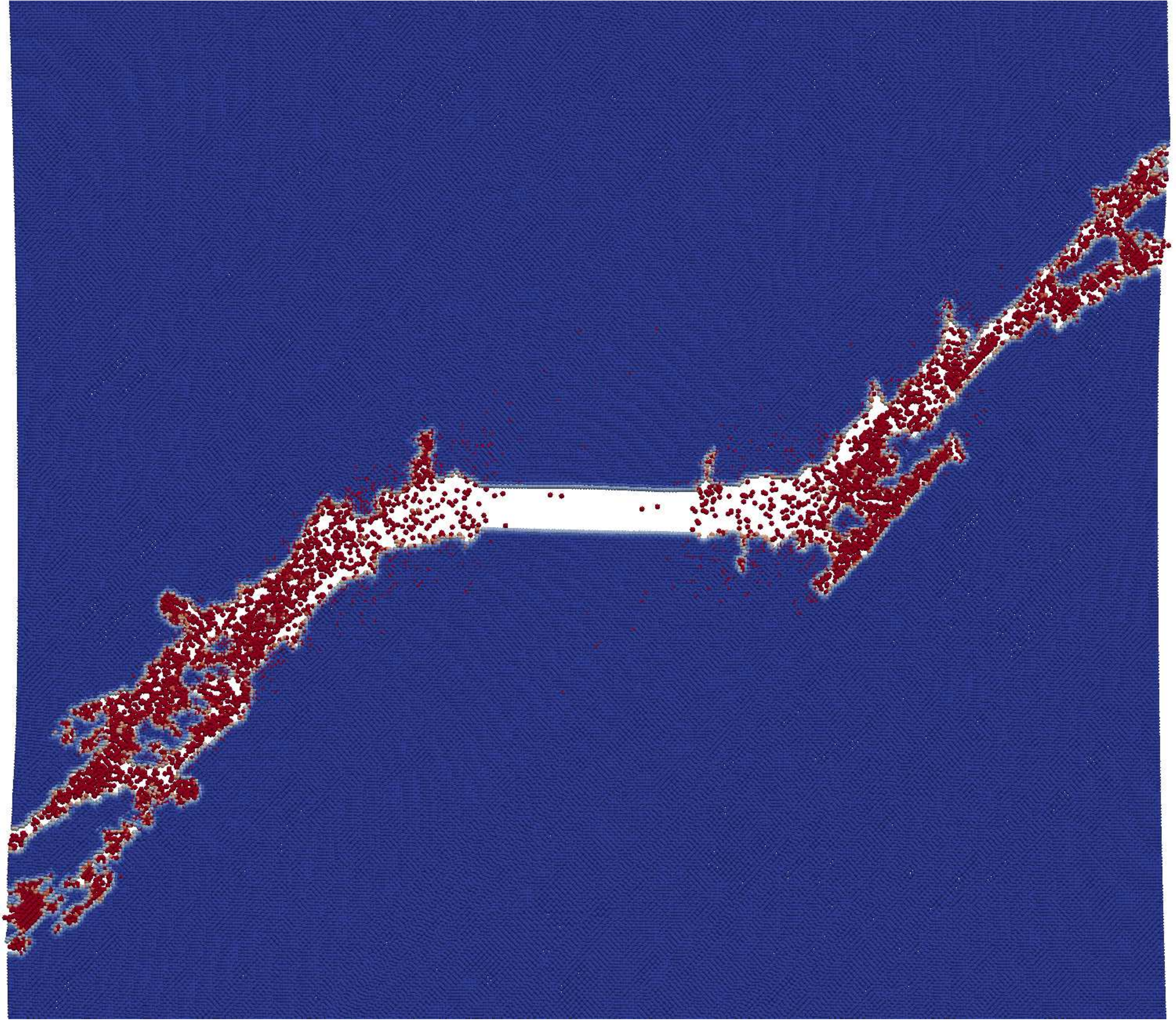} } 
\subfigure[$n = 3$]{\includegraphics[scale=0.085]{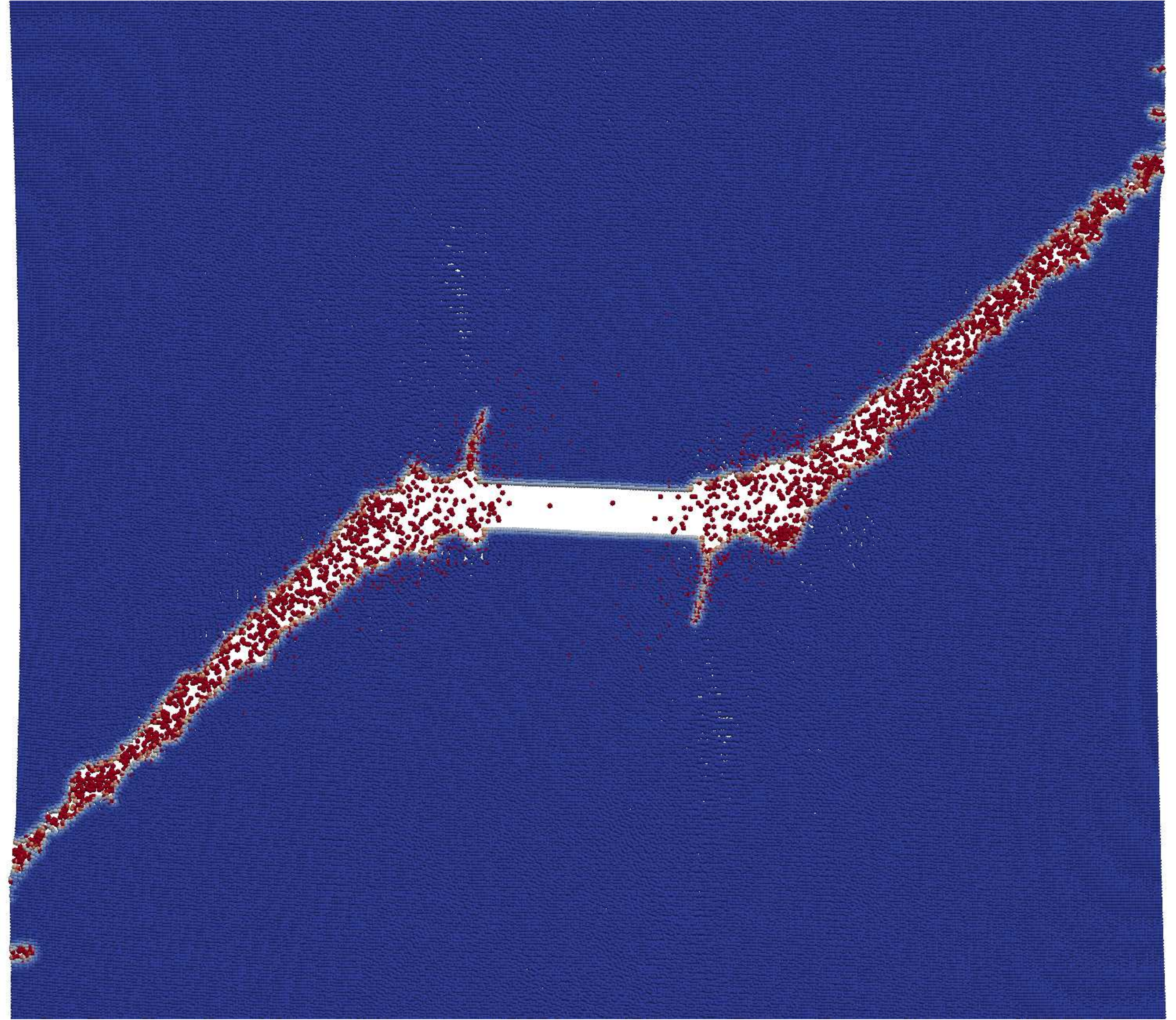} \label{fig:401_801_delta3}} 
\subfigure[$n = 4$]{\includegraphics[scale=0.085]{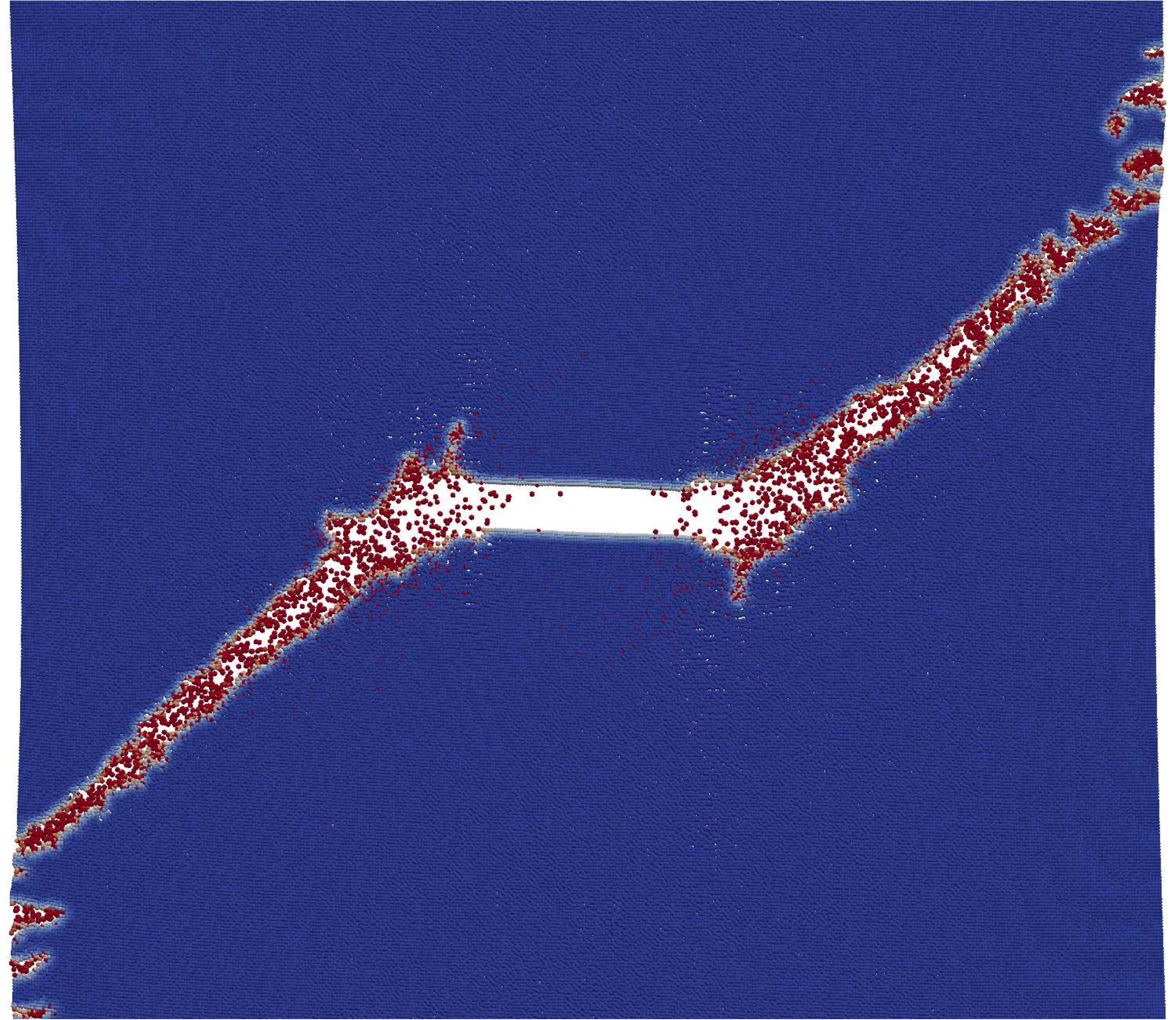} \label{fig:401_801_delta4}} 
\subfigure[$n = 5$]{\includegraphics[scale=0.085]{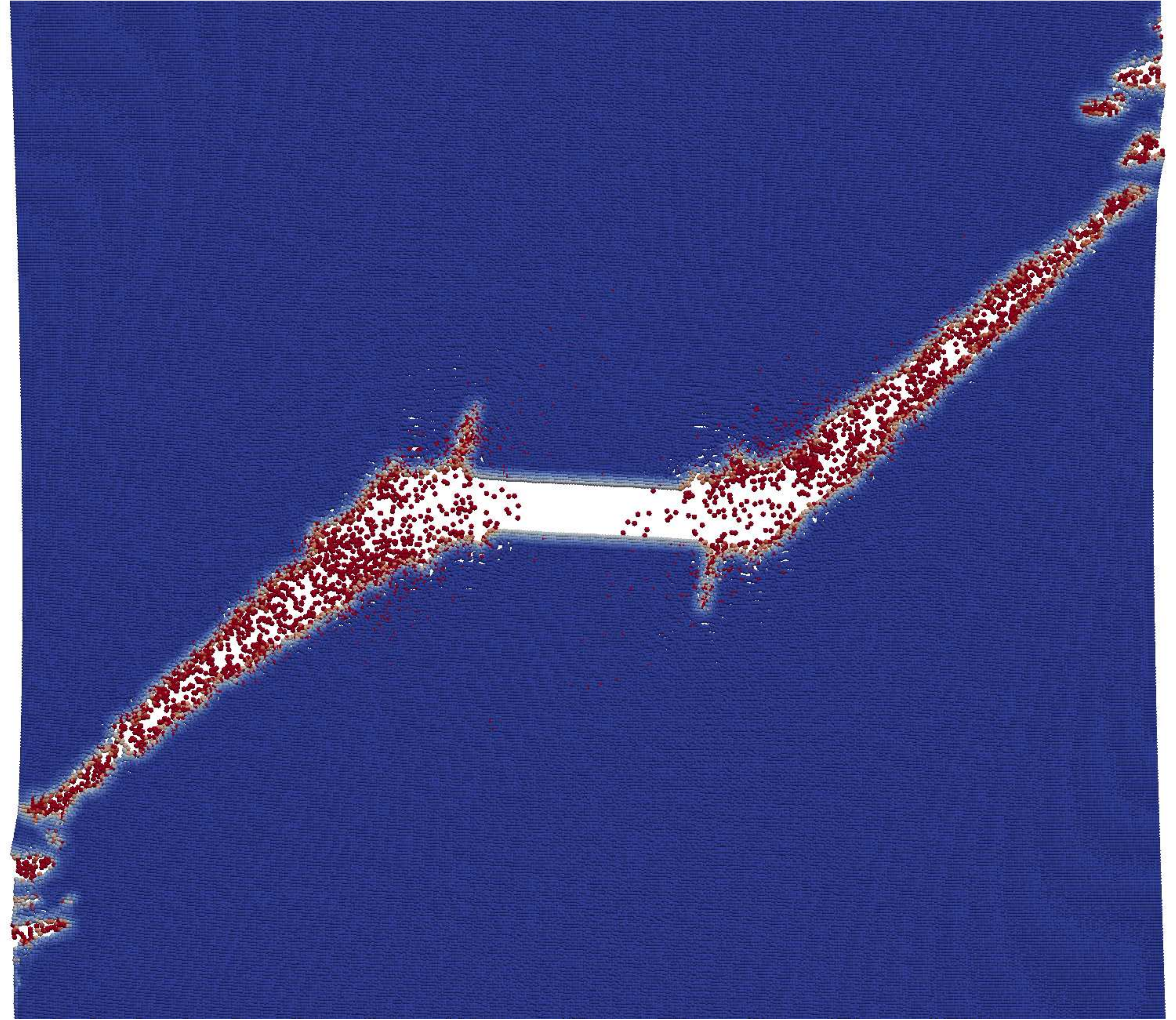}} 
\caption{Crack propagation of centred crack for different horizon - $\theta = 45^\circ$ - $400\times 800$ particles.}
\label{fig:theta45_401_801mesh}
\end{figure}

We study the effect of the particle discretisation and the horizon size for $\theta = 45^\circ$. Cahill et al. \cite{cahill2014experimental} give experimentally found paths of crack propagation in unidirectional composite materials. They have shown that the crack propagation path grows parallel to the fibre direction, indicating that the damage originates only through matrix failure. Figures  \ref{fig:theta45_201_401mesh}, \ref{fig:theta45_301_601mesh} and \ref{fig:theta45_401_801mesh} illustrate the crack propagation for different grid spacing and horizon sizes. Only the region around the crack is represented. The Figures represent the damage index $\phi$ for the deformed configuration, where blue stands for $\phi = 0$ while the red colour stands for $\phi = 1$. The displacements are scaled by a factor of 20.

It is clear that for $n = 2$, the crack path is irregular and some branching can occur at the crack tips. One reason for this behaviour is that the energy of a broken bond is redistributed to the remaining active bonds in that particle, which also lead these bonds to break. Hence, a larger horizon will stabilise the crack propagation, since there are more particles to re-balance the energy from broken bonds. However, larger horizon in coarse particle discretisation can lead to problematic results. Some oscillations and crack nucleation sites are visible at the edges of the plate, and these are attributed to the dynamics of the problem as the crack approaches the edge of the plate. However, as illustrated in Figures \ref{fig:201_theta45_delta4}, \ref{fig:201_theta45_delta5} and \ref{fig:301_601_delta5}, this effect takes place too early for the coarse discretisations and is presenting an unrealistic result. Moreover, $n = 3$ and $n = 4$ seem to provide stable results for the $300\times 600$ discretisation, shown in Figures \ref{fig:301_601_delta3} and \ref{fig:301_601_delta4}, and $400\times 800$ discretisation illustrated in Figures \ref{fig:401_801_delta3} and \ref{fig:401_801_delta4}. 

% MAYBE RELEVANT WHEN SHOWING CRACK PATHS FOR DIFFERENT THETA
%However, a horizon too large may change the physical behaviour, as seen by BOBARU ??? \cite{XXX} when studying crack branching. Figure \ref{fig:201_theta45_delta5_4} represents this issue, since smaller cracks arise close to the crack propagation path. This issue is not seen for the same horizon and a smaller grid spacing, indicating that 

\begin{figure}[!htb]
\centering
\subfigure[$t = 14.08 \mu s$]{\includegraphics[scale=0.113]{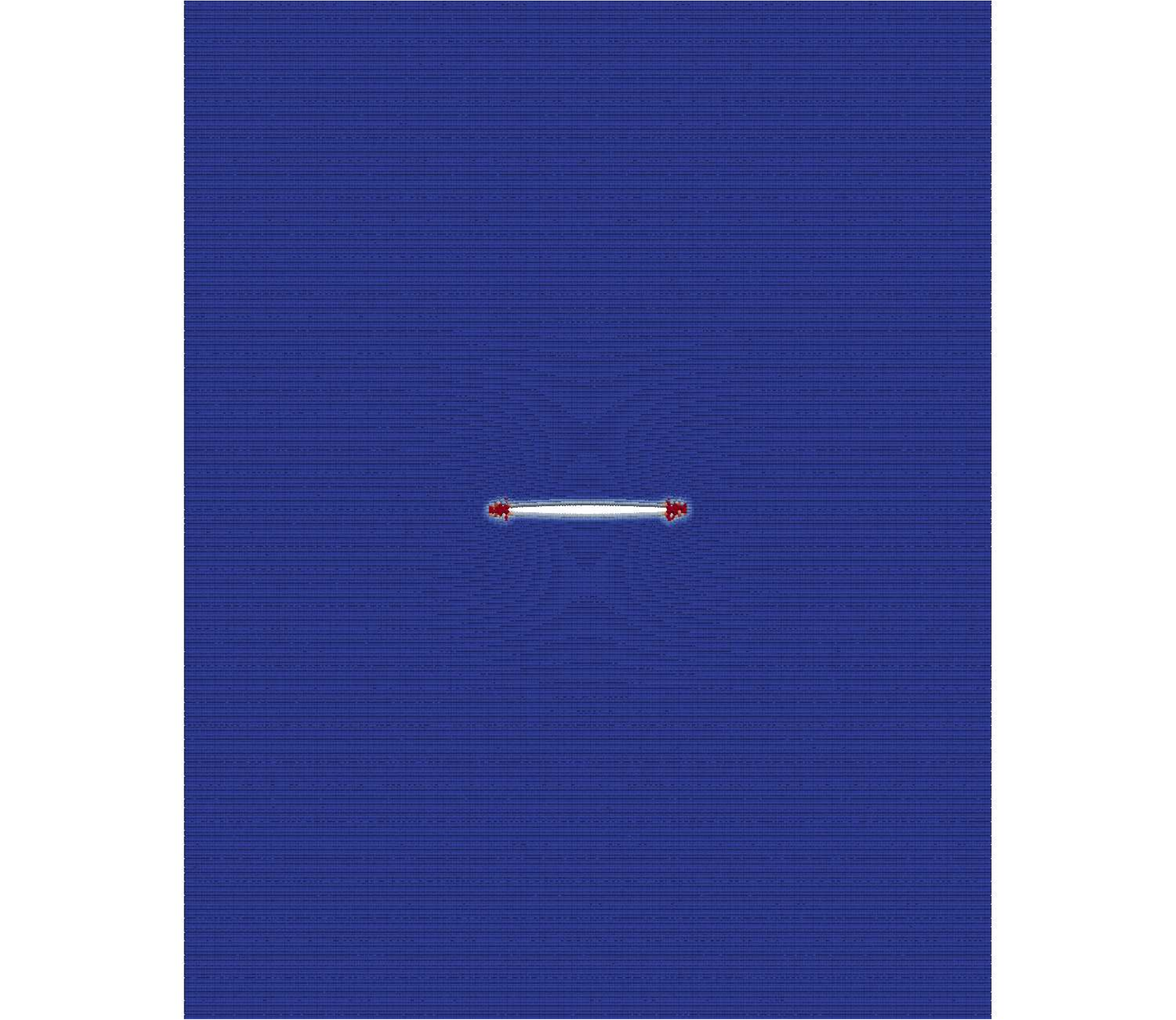} } 
\subfigure[$t = 23.46 \mu s$]{\includegraphics[scale=0.113]{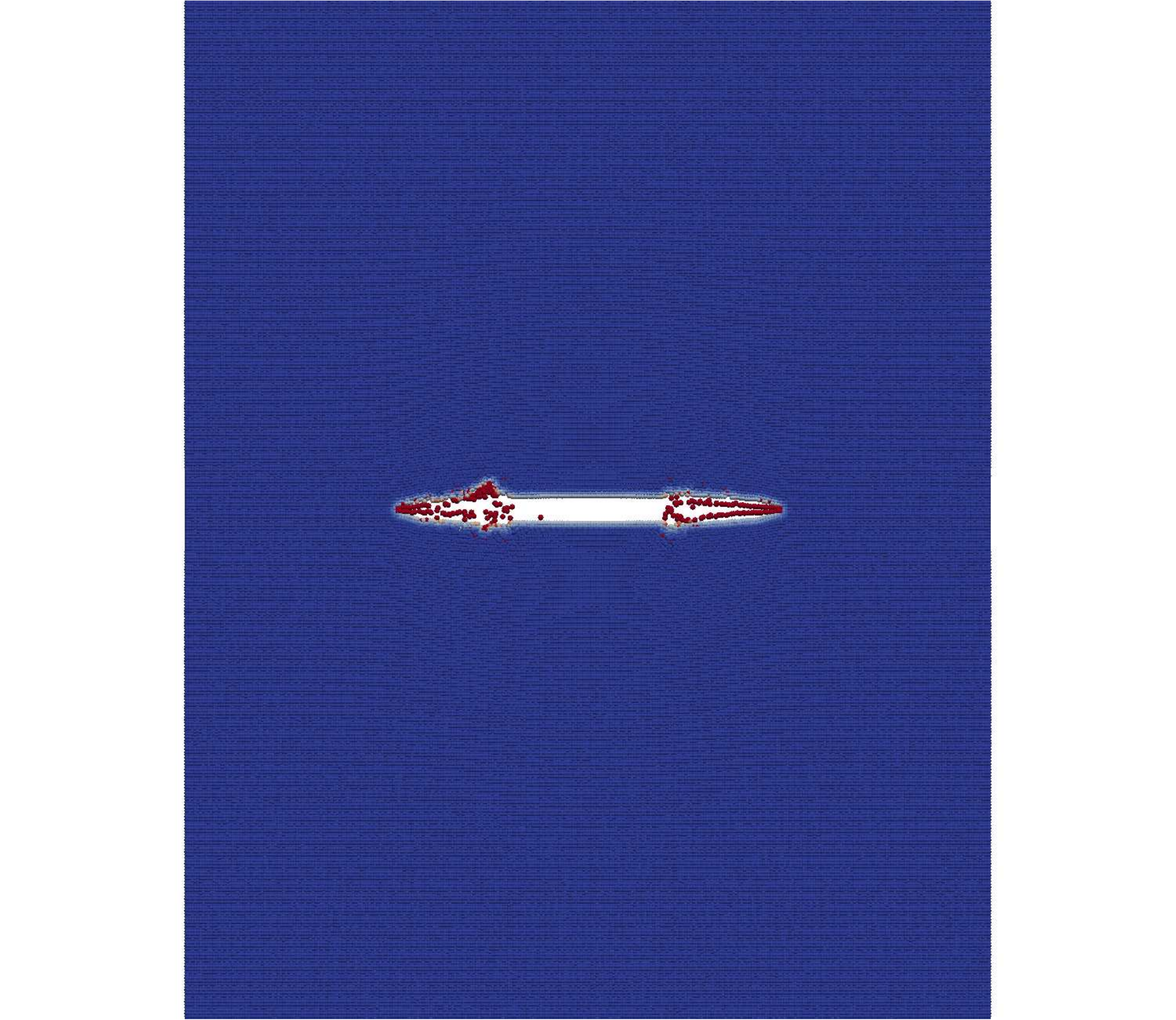} } 
\subfigure[$t = 32.85 \mu s$]{\includegraphics[scale=0.113]{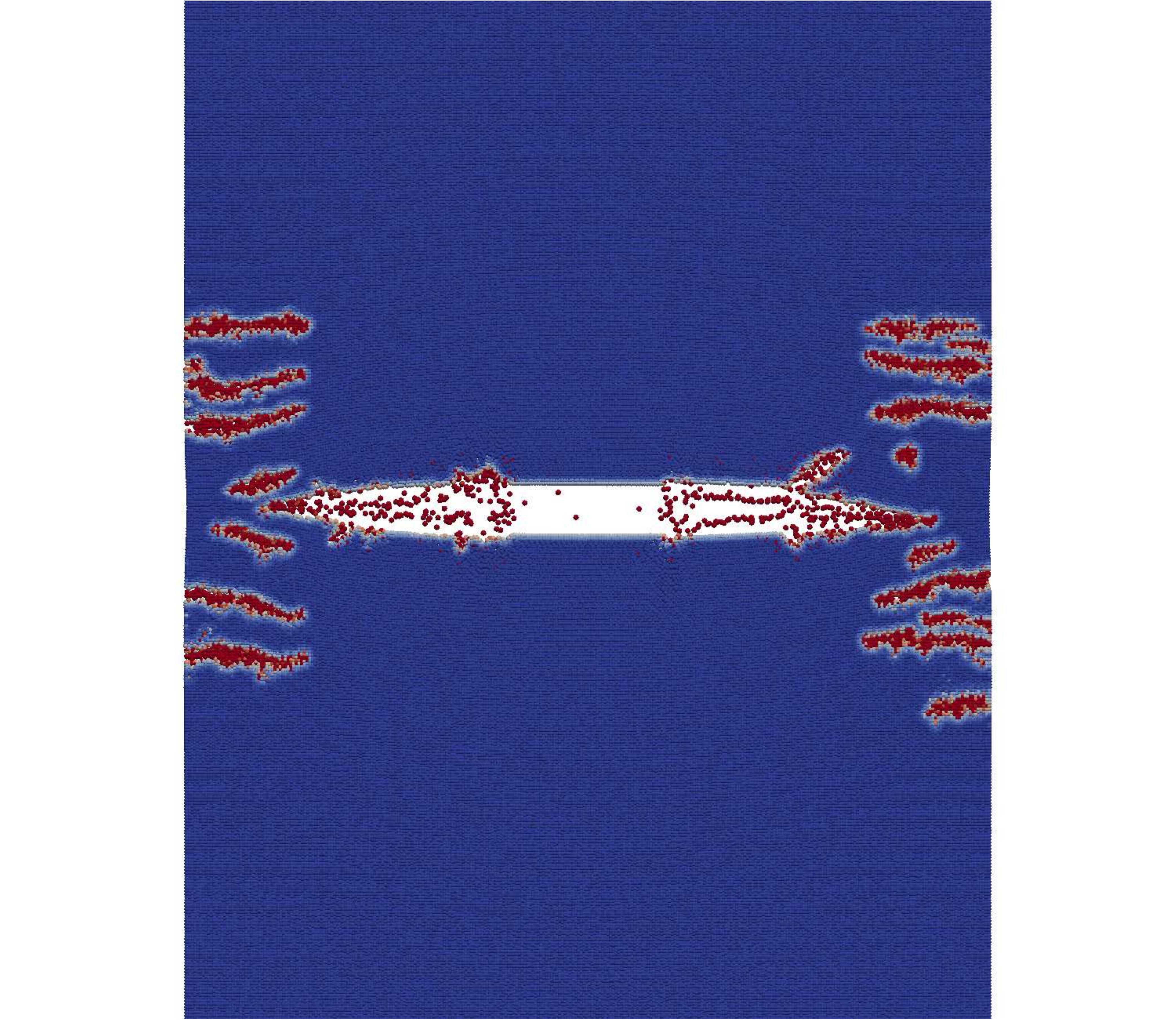} \label{fig:theta0_fibres}} 
\subfigure[Experiment from \cite{cahill2014experimental}]{\includegraphics[scale=1.5]{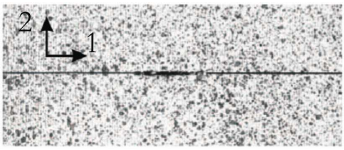} \label{fig:theta0_cahill}} 
\caption{Crack propagation of centred crack for $\theta = 0^\circ$ - $n = 3$ - $300\times 600$ particles.}
\label{fig:theta0}
\end{figure}

\begin{figure}[!htb]
\centering
\subfigure[$t = 16.38 \mu s$]{\includegraphics[scale=0.113]{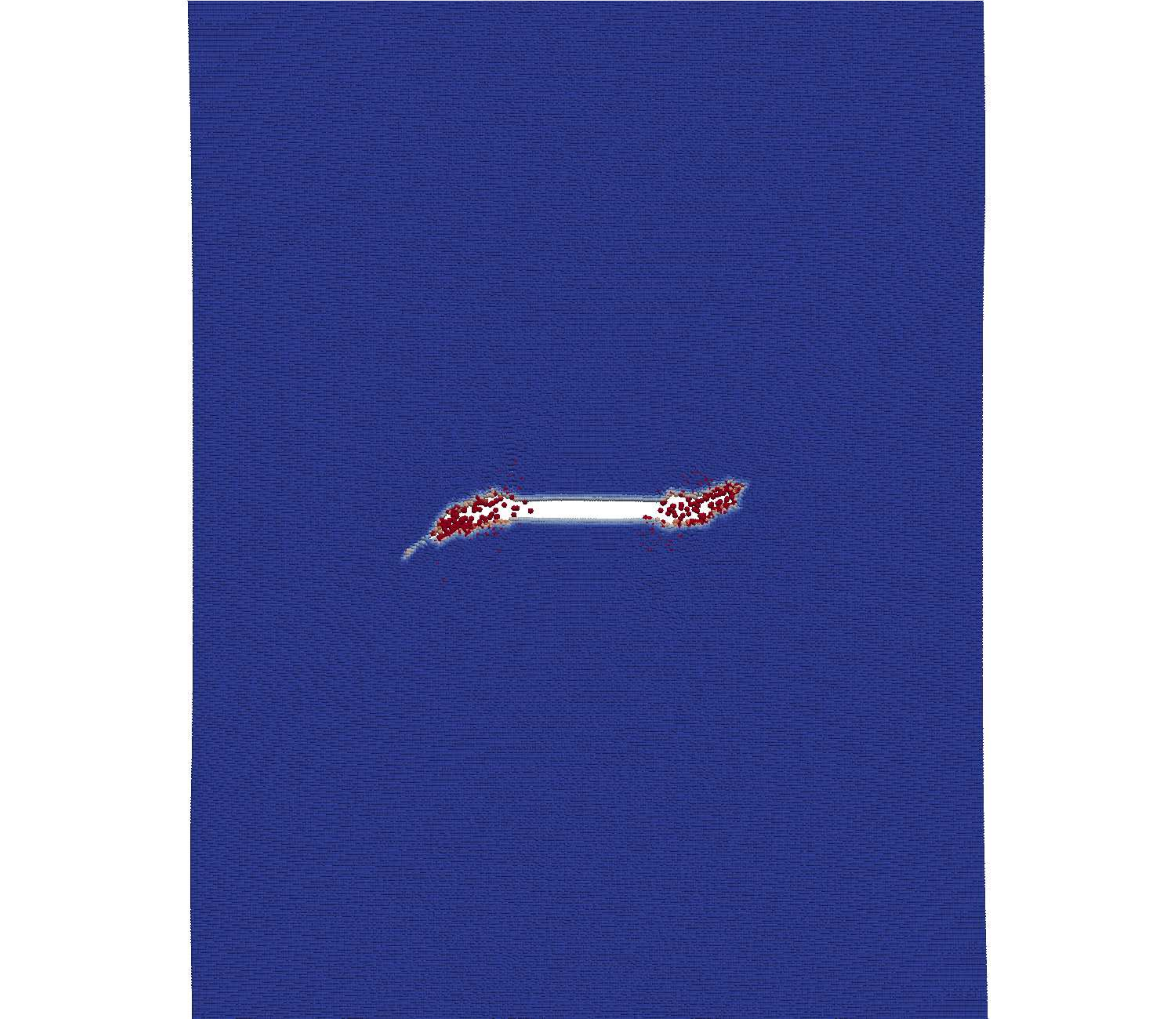} } 
\subfigure[$t = 20.13 \mu s$]{\includegraphics[scale=0.113]{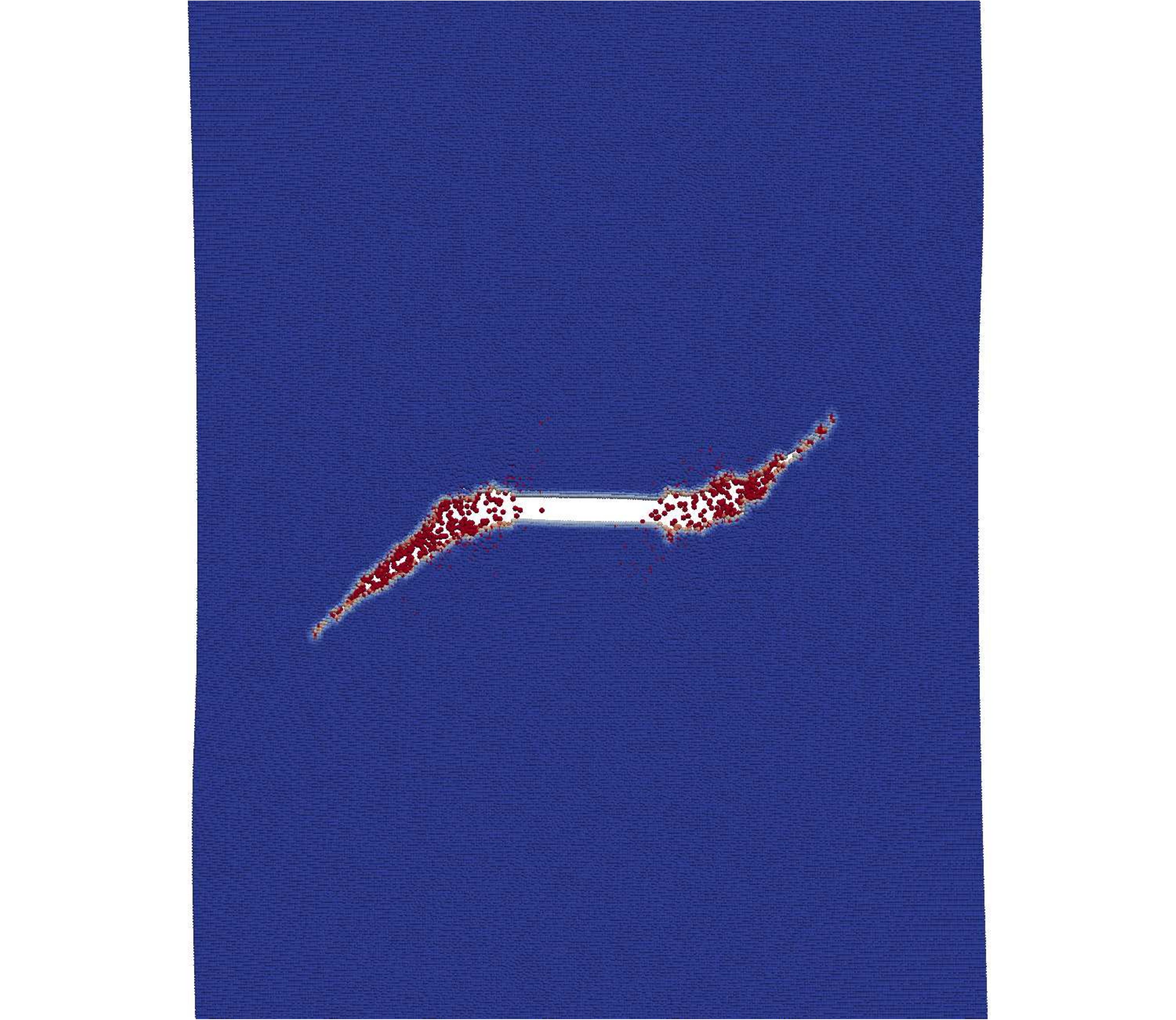} } 
\subfigure[$t = 23.87 \mu s$]{\includegraphics[scale=0.113]{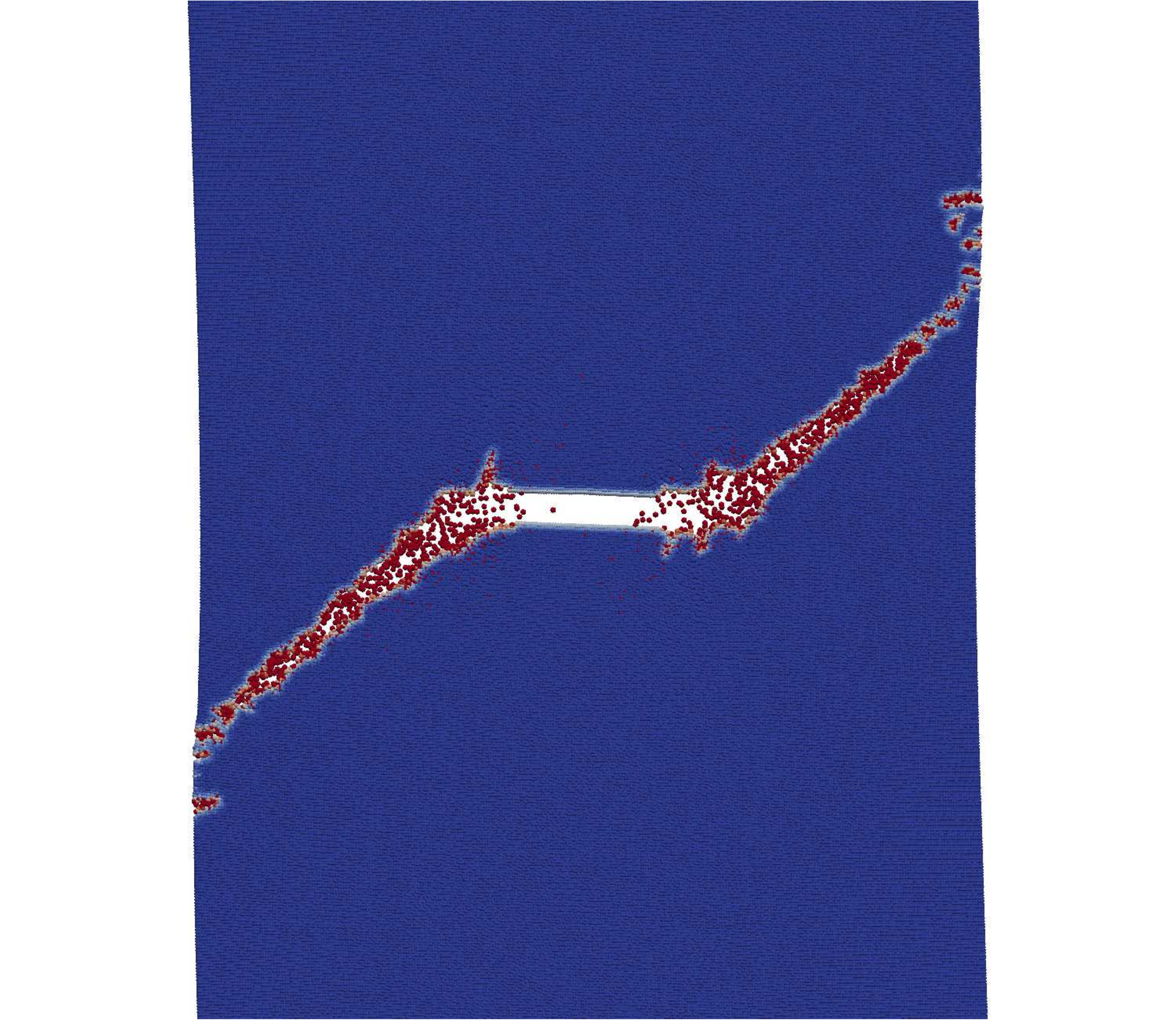} \label{fig:theta45_fibres}} 
\subfigure[Experiment from \cite{cahill2014experimental}]{\includegraphics[scale=1.5]{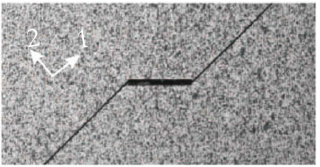} \label{fig:theta45_cahill}} 
\caption{Crack propagation of centred crack for $\theta = 45^\circ$ - $n = 3$ - $300\times 600$ particles.}
\label{fig:theta45}
\end{figure}

\begin{figure}[!htb]
\centering
\subfigure[$t = 12.88 \mu s$]{\includegraphics[scale=0.113]{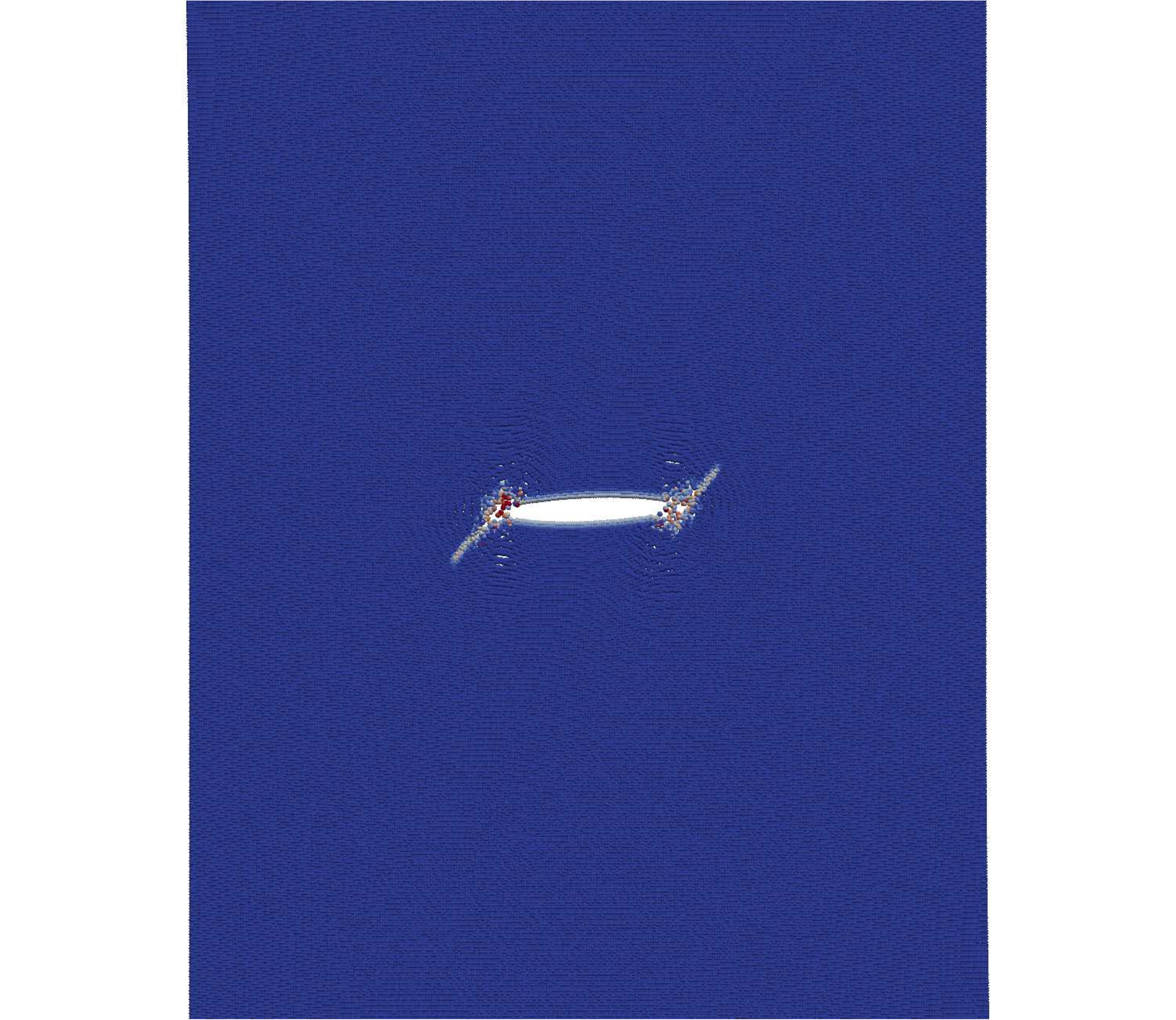} } 
\subfigure[$t = 17.56 \mu s$]{\includegraphics[scale=0.113]{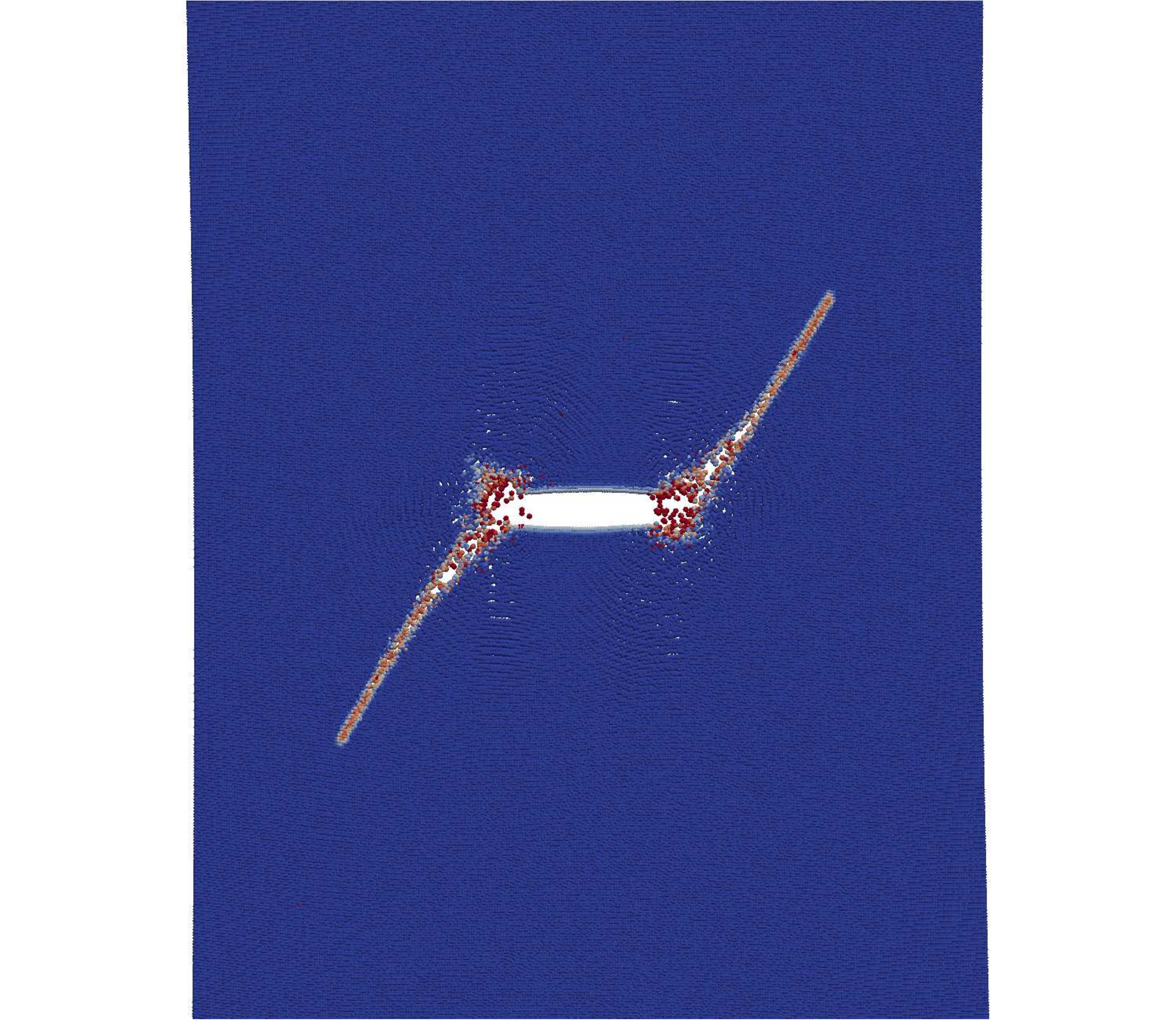} } 
\subfigure[$t = 22.24 \mu s$]{\includegraphics[scale=0.113]{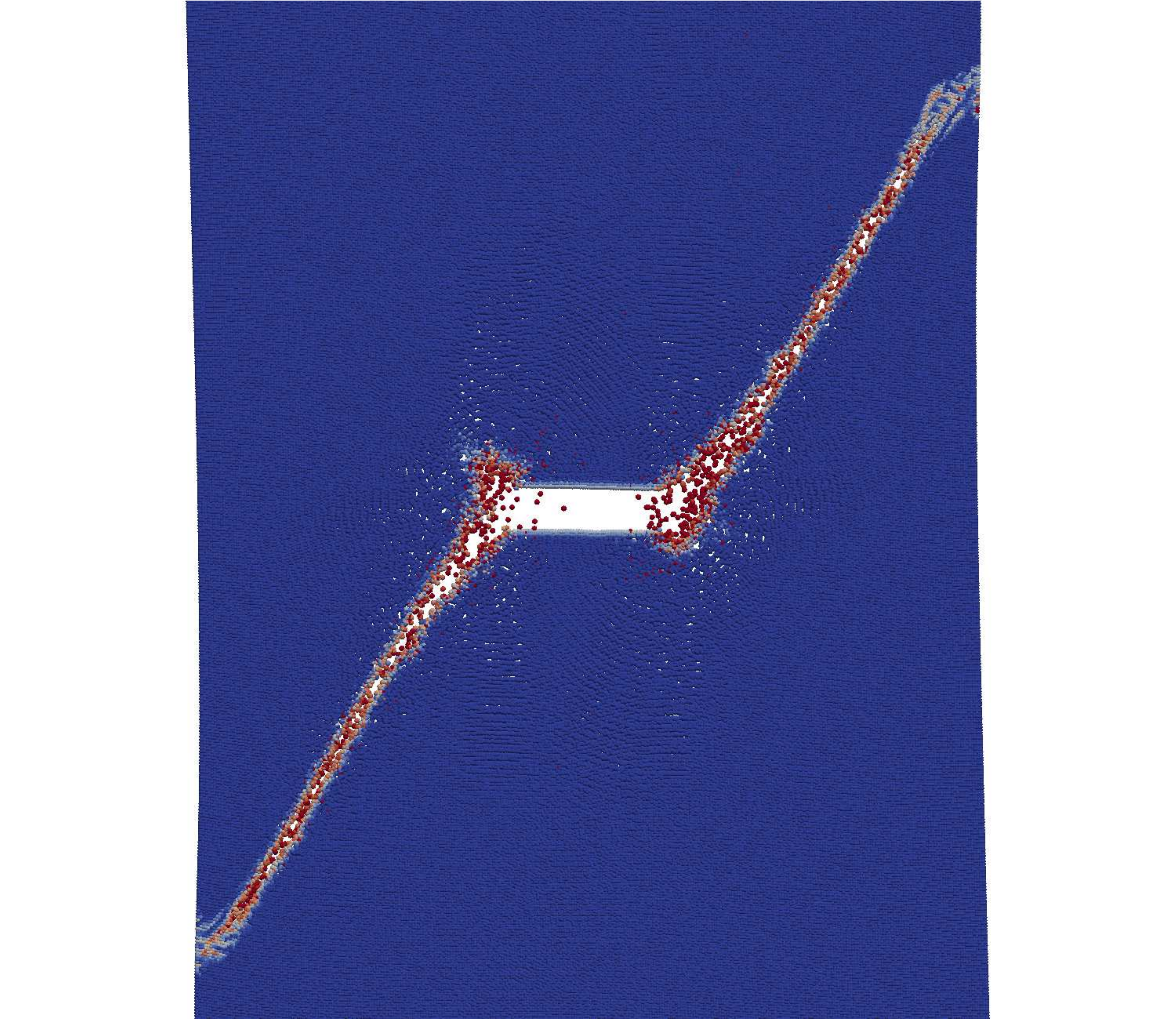} \label{fig:theta60_fibres}} 
\subfigure[Experiment from \cite{cahill2014experimental}]{\includegraphics[scale=1.5]{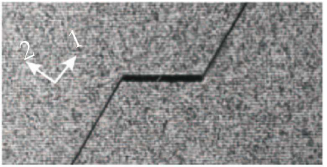} \label{fig:theta60_cahill}} 
\caption{Crack propagation of centred crack for $\theta = 60^\circ$ - $n = 3$ - $300\times 600$ particles.}
\label{fig:theta60}
\end{figure}

\begin{figure}[!htb]
\centering
\subfigure[$t = 10.53 \mu s$]{\includegraphics[scale=0.113]{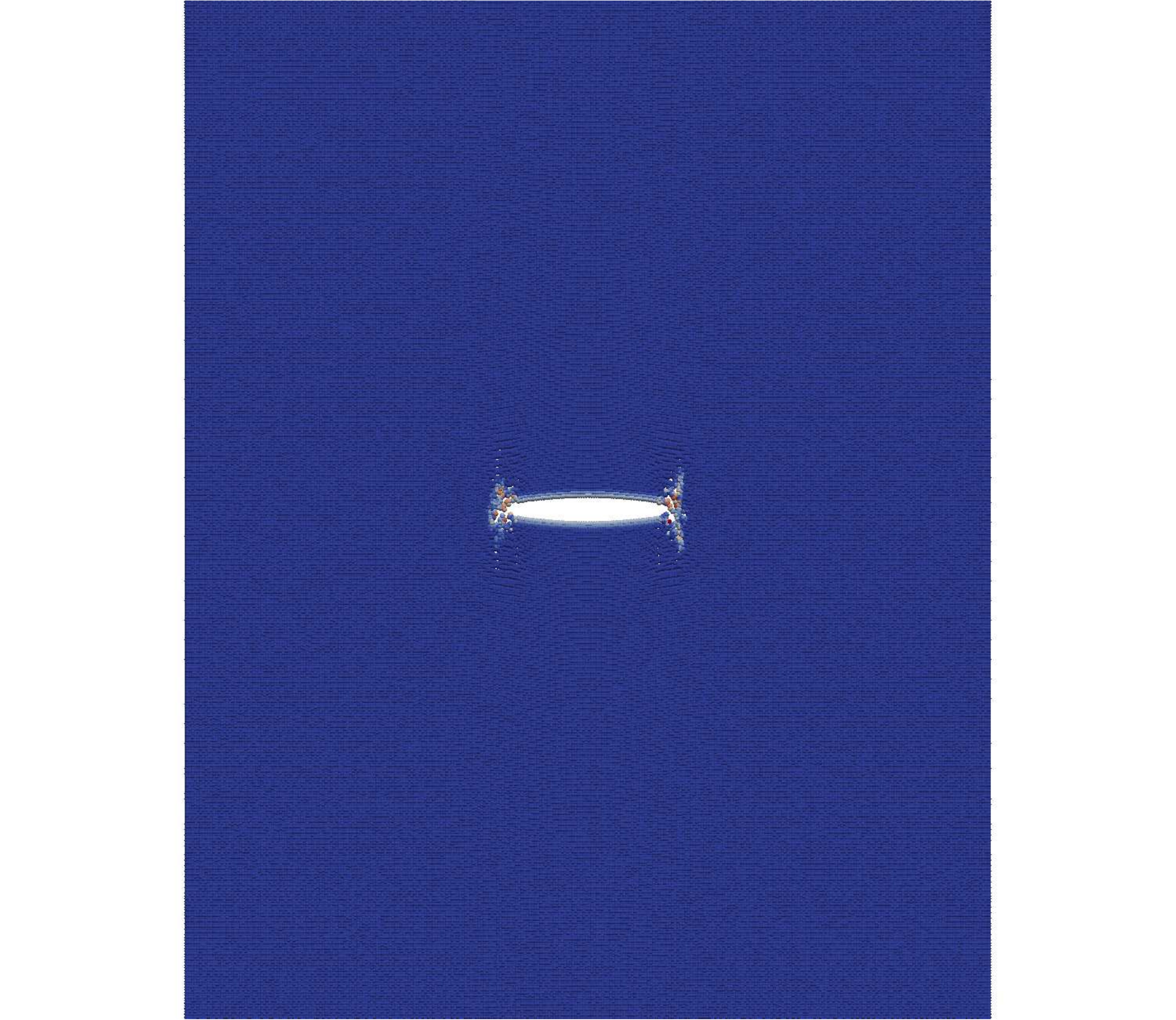} } 
\subfigure[$t = 17.78 \mu s$]{\includegraphics[scale=0.113]{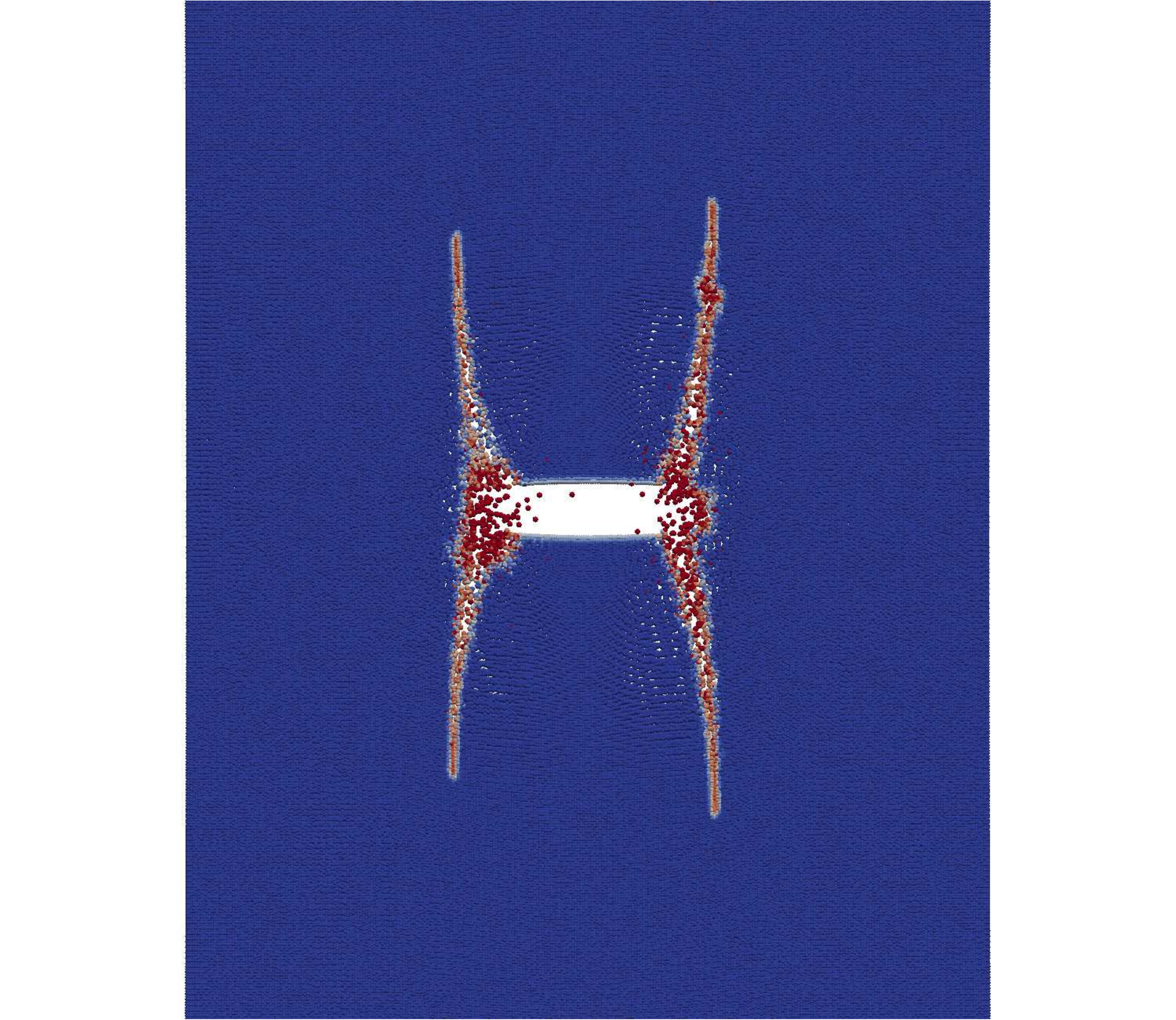} } 
\subfigure[$t = 23.17 \mu s$]{\includegraphics[scale=0.113]{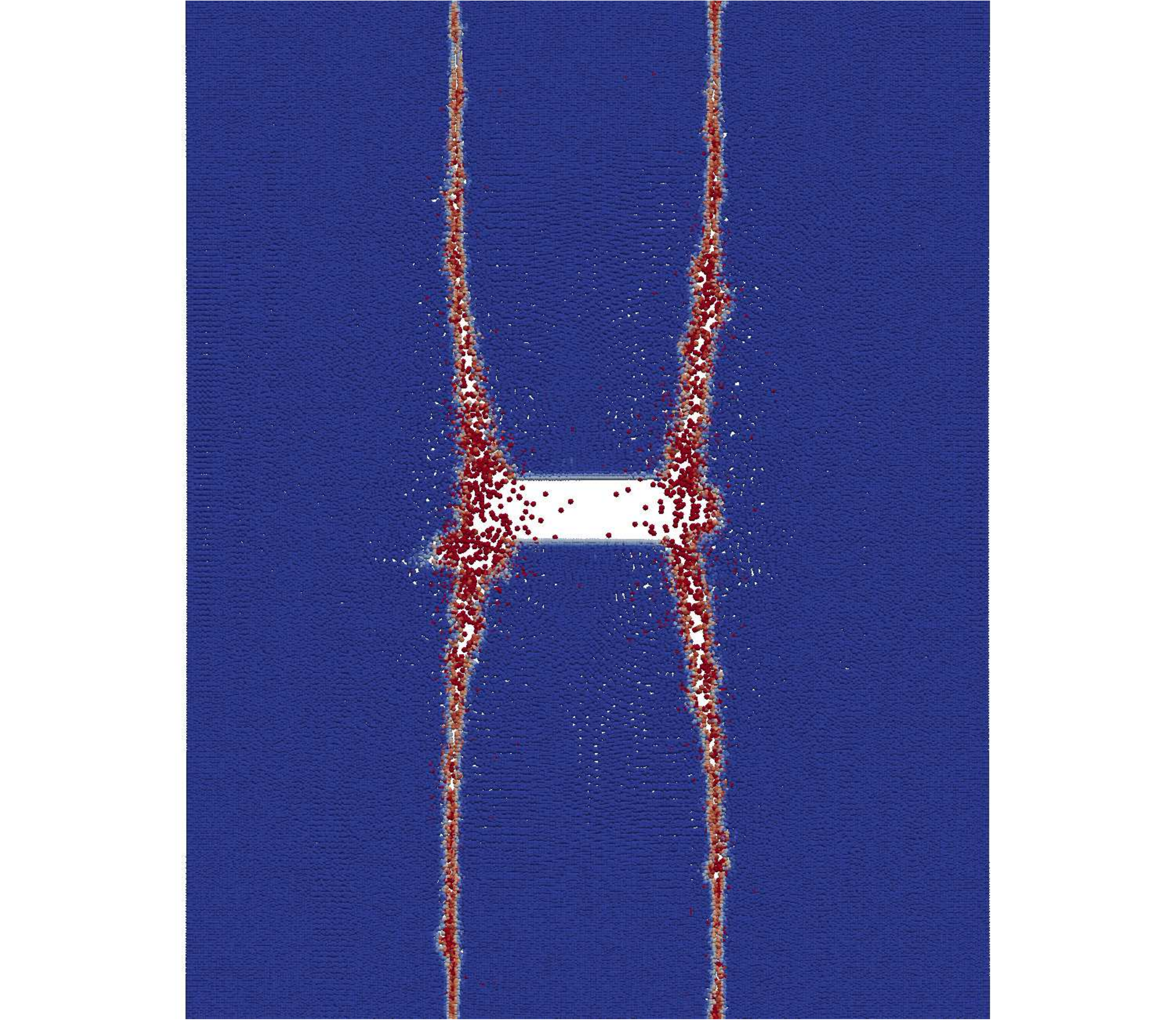} } 
\subfigure[Experiment from \cite{cahill2014experimental}]{\includegraphics[scale=1.5]{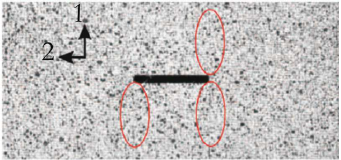} \label{fig:theta90_cahill}} 
\caption{Crack propagation of centred crack for $\theta = 90^\circ$ - $n = 3$ - $300\times 600$ particles.}
\label{fig:theta90}
\end{figure}

Figures \ref{fig:theta0}, \ref{fig:theta45}, \ref{fig:theta60} and \ref{fig:theta90} illustrate different fibre orientations for $n = 3$ and $300\times 600$ particle discretisation. The crack propagation paths are compared with experimental ones from \cite{cahill2014experimental} for uni-directional HTA/ 6376 composite.

From Figures \ref{fig:theta0} to \ref{fig:theta90}, the crack propagation paths in PD match those obtained  experimentally. However, since the plates are subjected to a dynamic load, some differences arise during the analysis. For instance, in Figure \ref{fig:theta0_fibres}, some parallel cracks appear as the centred crack propagates towards the edges of the plate. As the crack propagates, the newly formed crack surfaces increase the wave reflection inside the plate, which can lead to the formation of new cracks at the edge of the plate. Similar effects can also be seen in Figures \ref{fig:theta45_fibres} and \ref{fig:theta60_fibres}, where a small level of branching appears at the crack tip as it nears the edge of the plate. 

Cahill et al. \cite{cahill2014experimental} have mentioned that for the $\theta = 90^\circ$ case, the crack would propagate either up or down, and occasionally it would branch. In the PD framework, we have seen that the crack always branches, propagating in both directions.

The evolution of the crack propagation for the $\theta = 60^\circ$ fibre orientation can be visualised in the Supplemental Data available online.

\subsection{Edge crack in an anisotropic plate with inclusion and hole}

\begin{figure}[!htb]
\centering
\def\svgwidth{0.3\linewidth}
\begingroup%
  \makeatletter%
  \providecommand\color[2][]{%
    \errmessage{(Inkscape) Color is used for the text in Inkscape, but the package 'color.sty' is not loaded}%
    \renewcommand\color[2][]{}%
  }%
  \providecommand\transparent[1]{%
    \errmessage{(Inkscape) Transparency is used (non-zero) for the text in Inkscape, but the package 'transparent.sty' is not loaded}%
    \renewcommand\transparent[1]{}%
  }%
  \providecommand\rotatebox[2]{#2}%
  \ifx\svgwidth\undefined%
    \setlength{\unitlength}{166.33607927bp}%
    \ifx\svgscale\undefined%
      \relax%
    \else%
      \setlength{\unitlength}{\unitlength * \real{\svgscale}}%
    \fi%
  \else%
    \setlength{\unitlength}{\svgwidth}%
  \fi%
  \global\let\svgwidth\undefined%
  \global\let\svgscale\undefined%
  \makeatother%
  \begin{picture}(1,2.06990565)%
    %\put(0,0){\includegraphics[width=\unitlength]{edge_crack_inclusion_void.eps}}%
    \put(0,0){\includegraphics[width=\unitlength,page=1]{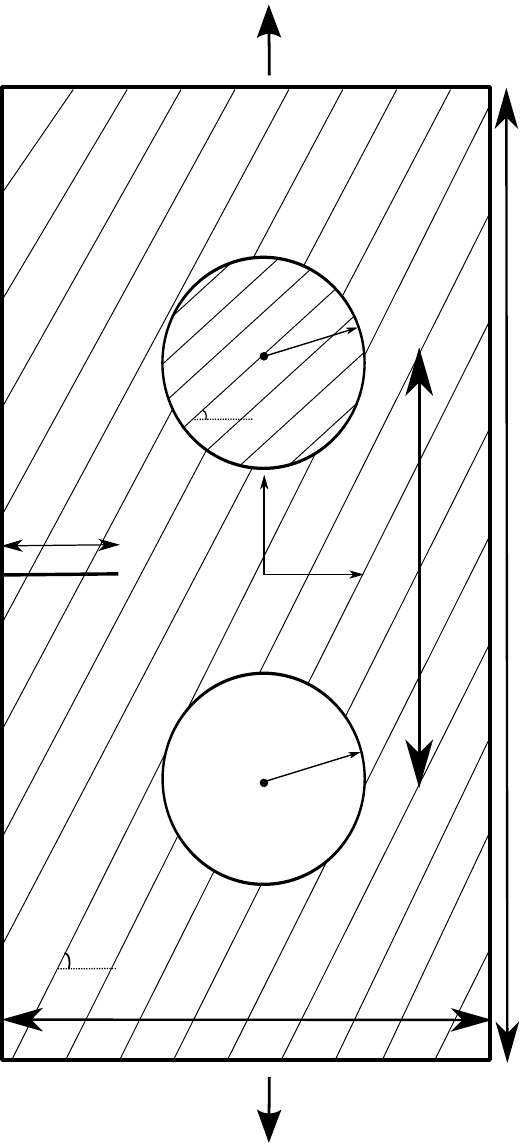}}%
    \put(0.505,0.27){\color[rgb]{0,0,0}\makebox(0,0)[lb]{$w$}}%
    \put(0.98,1.11){\color[rgb]{0,0,0}\makebox(0,0)[lb]{$2h$}}%
    \put(0.82,1.11){\color[rgb]{0,0,0}\makebox(0,0)[lb]{$2b$}}%
    \put(0.11,1.18){\color[rgb]{0,0,0}\makebox(0,0)[lb]{$a$}}%
    \put(0.53,-0.01){\color[rgb]{0,0,0}\makebox(0,0)[lb]{$v$}}%
    \put(0.53,2.17){\color[rgb]{0,0,0}\makebox(0,0)[lb]{$v$}}%
    \put(0.16,0.35){\color[rgb]{0,0,0}\makebox(0,0)[lb]{\footnotesize $\theta_1$}}%
    \put(0.43,1.41){\color[rgb]{0,0,0}\makebox(0,0)[lb]{\footnotesize $\theta_2$}}%
    \put(0.595,0.75){\color[rgb]{0,0,0}\makebox(0,0)[lb]{$r$}}%
    \put(0.58,1.56){\color[rgb]{0,0,0}\makebox(0,0)[lb]{$r$}}%
    \put(0.67,1.12){\color[rgb]{0,0,0}\makebox(0,0)[lb]{$x$}}%
    \put(0.52,1.235){\color[rgb]{0,0,0}\makebox(0,0)[lb]{$y$}}%
  \end{picture}%
\endgroup%
\caption{Anisotropic plate with inclusion and void.}
\label{fig:plate_inclusion}
\end{figure}
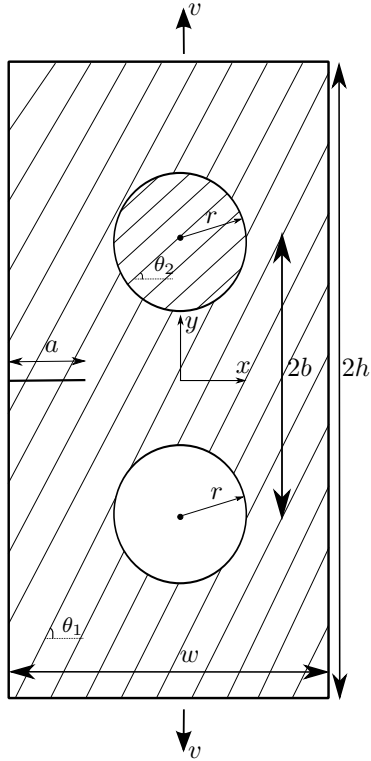

In this example we study a rectangular ($w=h=20$ mm) anisotropic plate with an edge crack of length $a=4$ mm. The plate has an inclusion of radius $r=4.5$ mm, shifted $b=8$ mm from the centre of the plate, and a hole of same radius shifted downwards from the centre of the plate, as illustrated in Figure \ref{fig:plate_inclusion}. The plate is subjected to an initial velocity defined across the plate, and given by $v = 50\frac{y}{2h}$ m/s.

The material properties of the plate and the inclusion are given in Voigt notation by
\begin{equation}
\mathbf C_{IJ}^{plate} = \left(
\begin{array}{ccc}
155.43 & 3.72 & 0 \\
3.72 & 16.34 & 0 \\
0 & 0 & 7.48 
\end{array} \right) \text{GPa}
\end{equation}
\begin{equation}
\mathbf C_{IJ}^{inc} = \left(
\begin{array}{ccc}
235 & 3.69 & 0 \\
3.69 & 2 & 0 \\
0 & 0 & 28.2
\end{array} \right) \text{GPa}
\end{equation}

The density of the plate is $\rho^{plate} = 1600$ kg/m$^3$ and the density of the inclusion is $\rho^{inc} = 5670$ kg/m$^3$. The material properties of the plate are rotated by an angle $\theta_1$ with respect to the horizontal axis, while the inclusion represents an orthotropic material ($\theta_2 = 0^\circ$).

Initially we study the behaviour of the problem with no crack propagation. Table \ref{tab:error_DSIF_inc} shows the relative error of the DSIF obtained with a finite element mesh and the PD formulation for $\theta_1=45^\circ$. The finite element mesh has 165728 3-node triangular elements, and has been defined using the MESH2D algorithm (for details see reference \cite{mesh2d2014}). For this configuration, the discretisation with $400\times 800$ particles and $n = 2$ presents the lowest relative error. A possible explanation for this fact is that the horizon size is also dependent on the material properties. 

In most works, it has been shown that the horizon size is chosen according to the analysed problem, however Bobaru et al. \cite{bobaru2012meaning} have shown that the horizon size affects the dynamics of crack branching, where a horizon too large causes the elastic wave to propagate too fast, leading to differences with respect to experimental results. The influence of the horizon size in the analysis would lead to larger errors in anisotropic materials, implying that there is an optimum horizon size for a given material. For the material in the present study, Table \ref{tab:error_DSIF_inc} suggests that the optimum horizon size lies between $\Delta x$ and $2\Delta x$.
\begin{table}[!htb]
\centering
\caption{Relative error between PD and FEM for DSIF I.}
\begin{tabular}{|c|c|c|c|c|c|}
\hline 
& \multicolumn{5}{|c|}{Horizon $\delta = n \Delta x$} \\ \hline 
Particles & $n = 1$ & $n = 2$ & $n = 3$ & $n = 4$ & $n = 5$ \\ \hline
$200\times 200$ & $0.8124$ & $0.0937$ & $0.1168$ & $0.1227$ & $0.1391$ \\ \hline
$300\times 300$ & $0.8115$ & $0.0574$ & $0.0751$ & $0.0789$ & $0.0913$ \\ \hline
$400\times 400$ & $0.8087$ & $0.0487$ & $0.0580$ & $0.0634$ & $0.0706$ \\ \hline
\end{tabular}
\label{tab:error_DSIF_inc}
\end{table}

\begin{figure}[!htb]
\centering
\subfigure[dynamic mode I]{\includegraphics[scale=0.55]{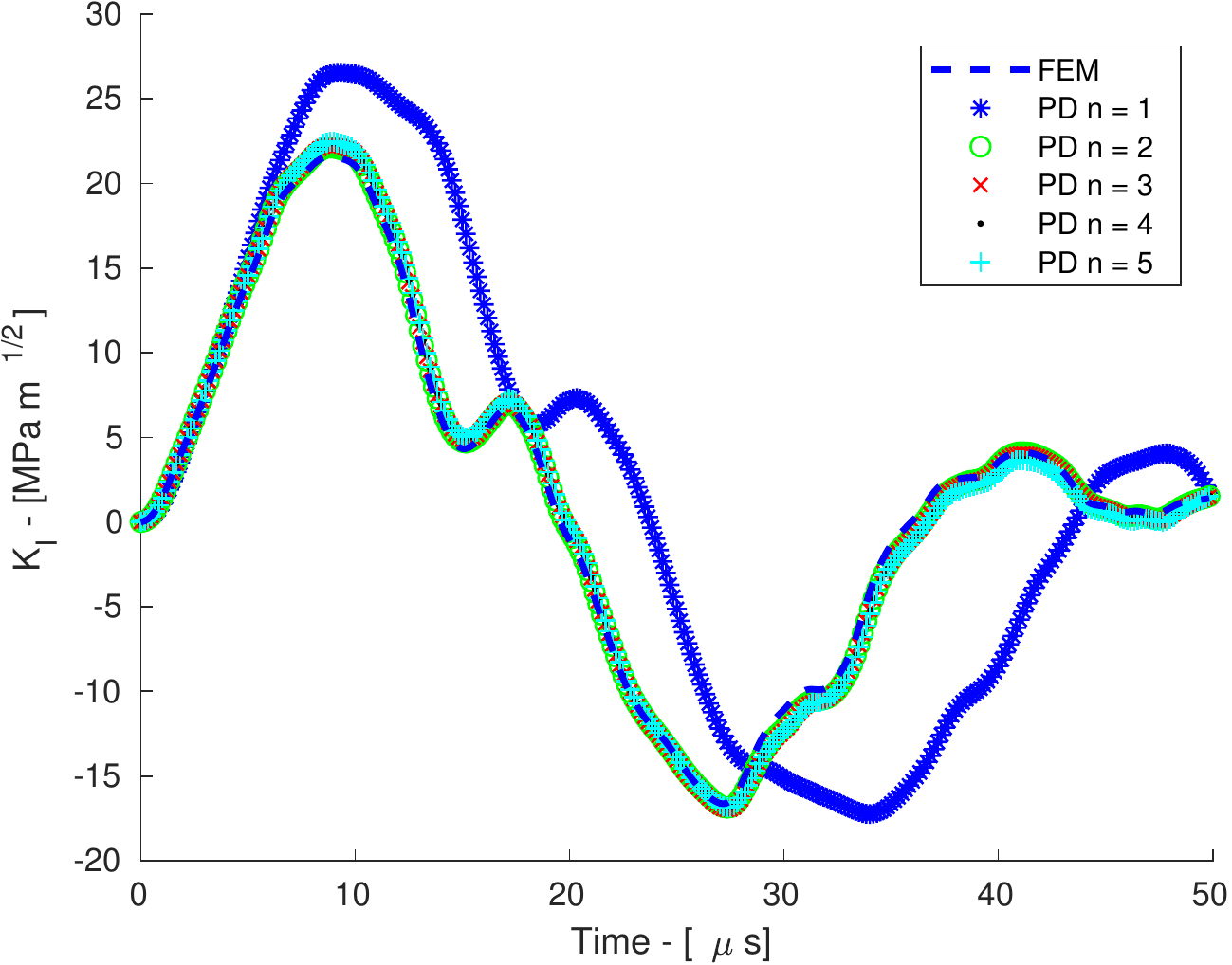} \label{fig:comparison_horizon_inc_KI}}
\subfigure[dynamic mode II]{\includegraphics[scale=0.55]{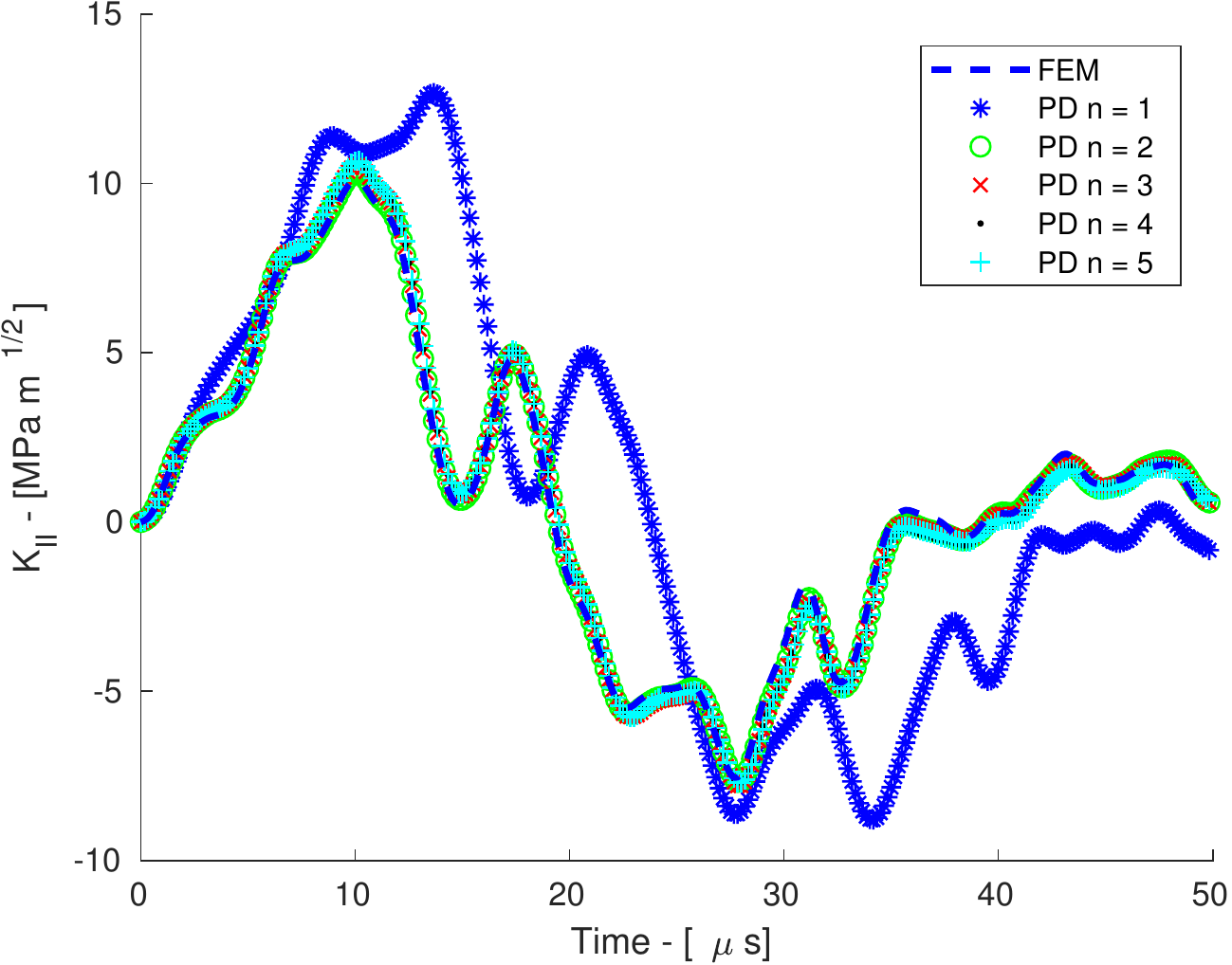} \label{fig:comparison_horizon_inc_KII}}
\caption{Edge crack with inclusion and void: comparison between different horizon size - $400\times 800$ particle discretisation - $\theta_1 = 45^\circ$.}
\label{fig:comparison_horizon_inc}
\end{figure}

Figures \ref{fig:comparison_horizon_inc_KI} and \ref{fig:comparison_horizon_inc_KII} depict the mode I and mode II DSIFs for this particular PD configuration, respectively. One can observe that the DSIFs calculated for different horizon sizes provide similar values of the DSIF for $n \geq 2$.
\psfrag{time - [micro s]}{\footnotesize Time - [$\mu $s]}

\begin{figure}[!htb]
\centering
\includegraphics[scale=0.68]{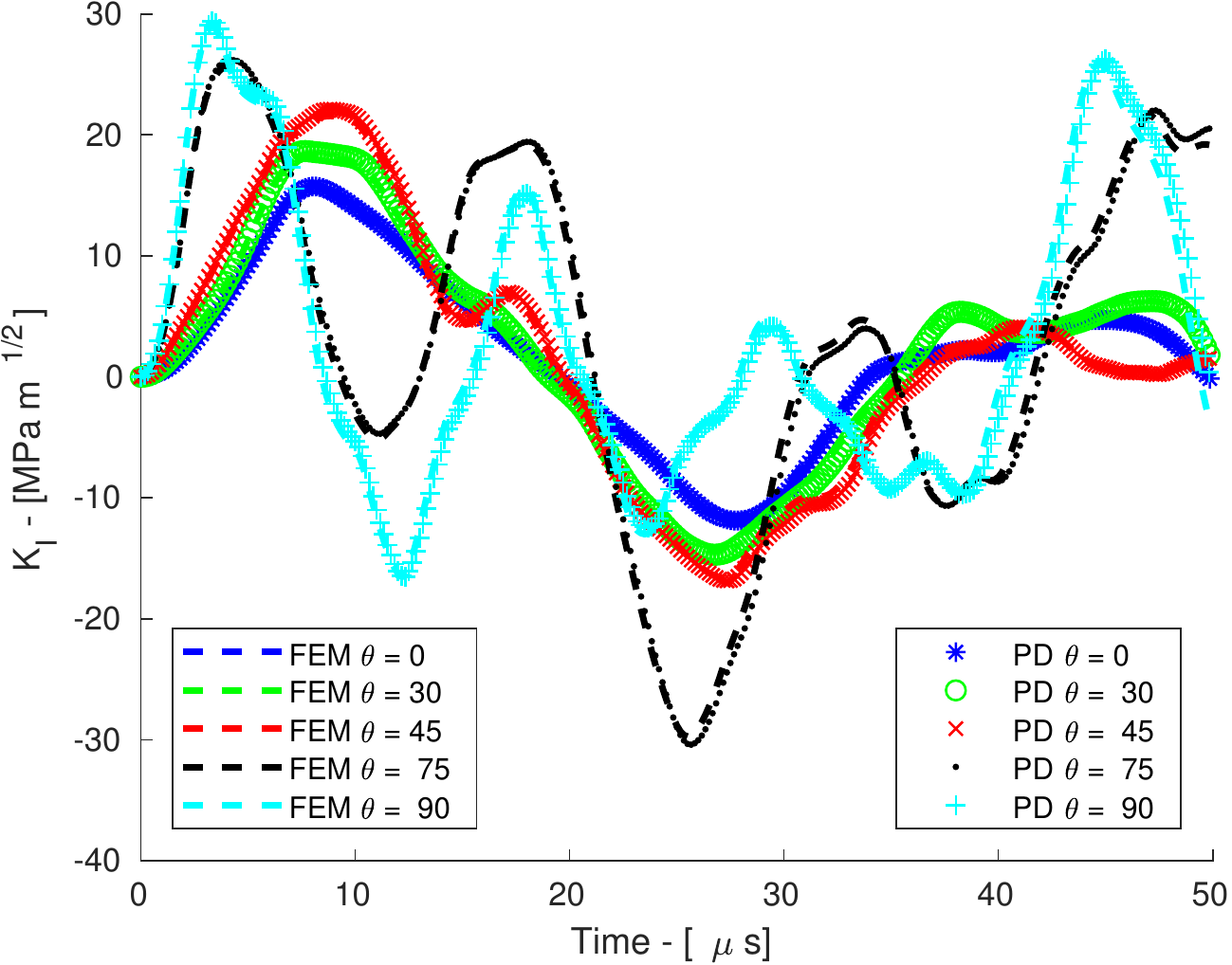} 
\caption{Edge crack with inclusion and void: DSIF for different values of $\theta$ - $400\times 400$ particles - $n = 2$ - mode I.}
\label{fig:inc_theta_KI}
\end{figure}

\begin{figure}[!htb]
\centering
\includegraphics[scale=0.68]{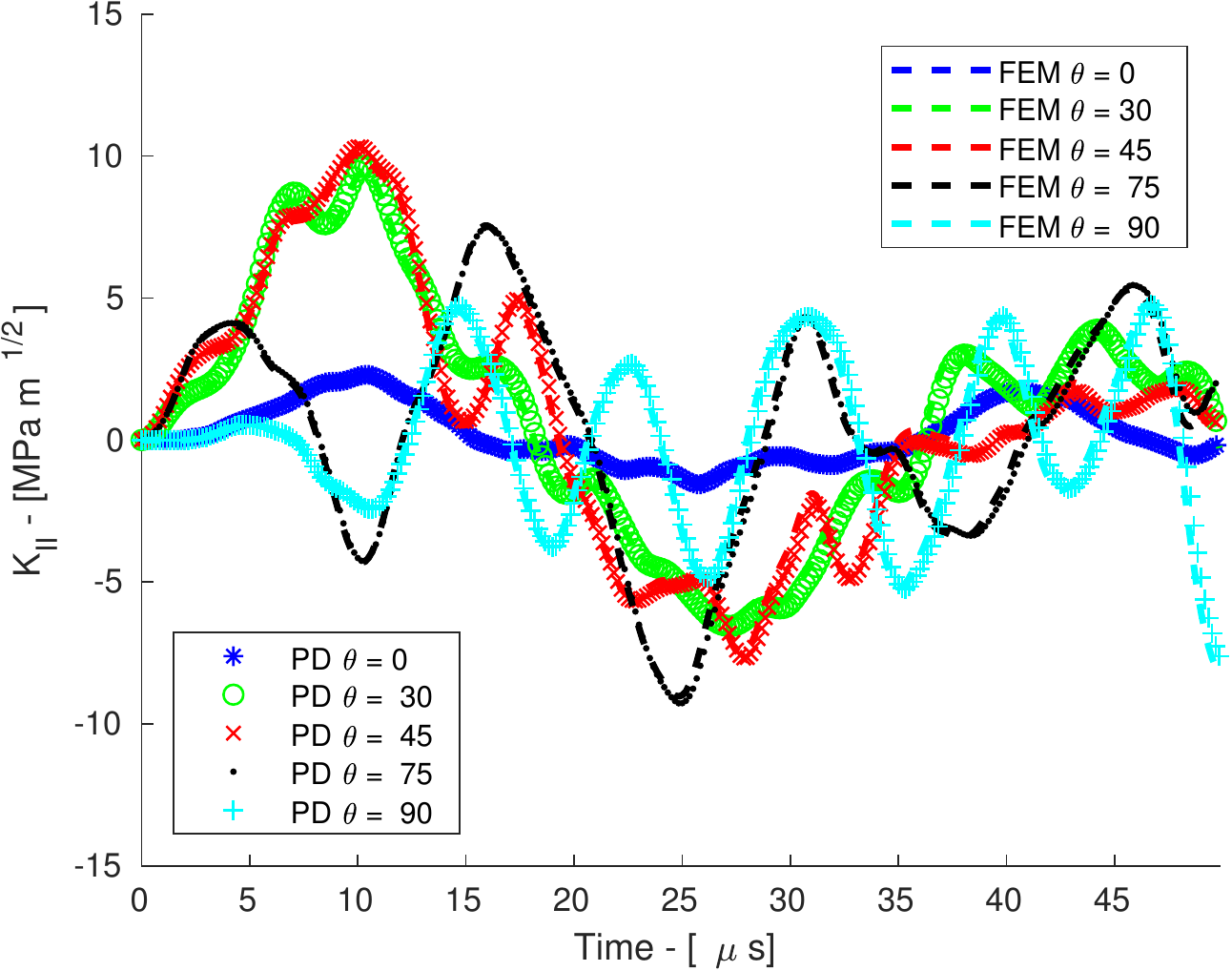} 
\caption{Edge crack with inclusion and void: DSIF for different values of $\theta$ - $400\times 400$ particles - $n = 2$ - mode II.}
\label{fig:inc_theta_KII}
\end{figure}

Figures \ref{fig:inc_theta_KI} and \ref{fig:inc_theta_KII} depict the DSIF for different values of $\theta$. Very good agreement between the FEM and PD solutions is achieved. One can remark that the oscillation behaviour increases as $\theta$ increases, since the elastic P-wave speed in the $y$-direction increases with increasing $\theta$.

From Figure \ref{fig:inc_theta_KII}, we observe that $K_{II}$ is not zero when $\theta = 0^\circ$ and $\theta = 90^\circ$. Since the problem is no longer symmetric due to the presence of the inclusion and the hole, there is an acting mode II behaviour. In case where there would be a double inclusion (or double void), $K_{II}$ will be zero for these values of $\theta$.

Next we evaluate the crack propagation patterns in this problem. We assume the same material properties given in Table \ref{tab:mat_HTA} for the plate, while the material properties of the inclusion remain the same. The tensile strength in the fibre, matrix and shear direction of the inclusion are given by: $\sigma_{Lu}=2100$ MPa, $\sigma_{Tu}=120$ MPa and $\tau_{LTu}=135$ MPa, respectively. For the interface between the plate and the inclusion, we consider the tensile strength parameters of the plate, since they assume lower values than the corresponding parameters for the inclusion. We analyse the crack propagation for two different initial velocities.

\begin{figure}[!htb]
\centering
\subfigure[$t = 6.00 \mu s$]{\includegraphics[scale=0.175]{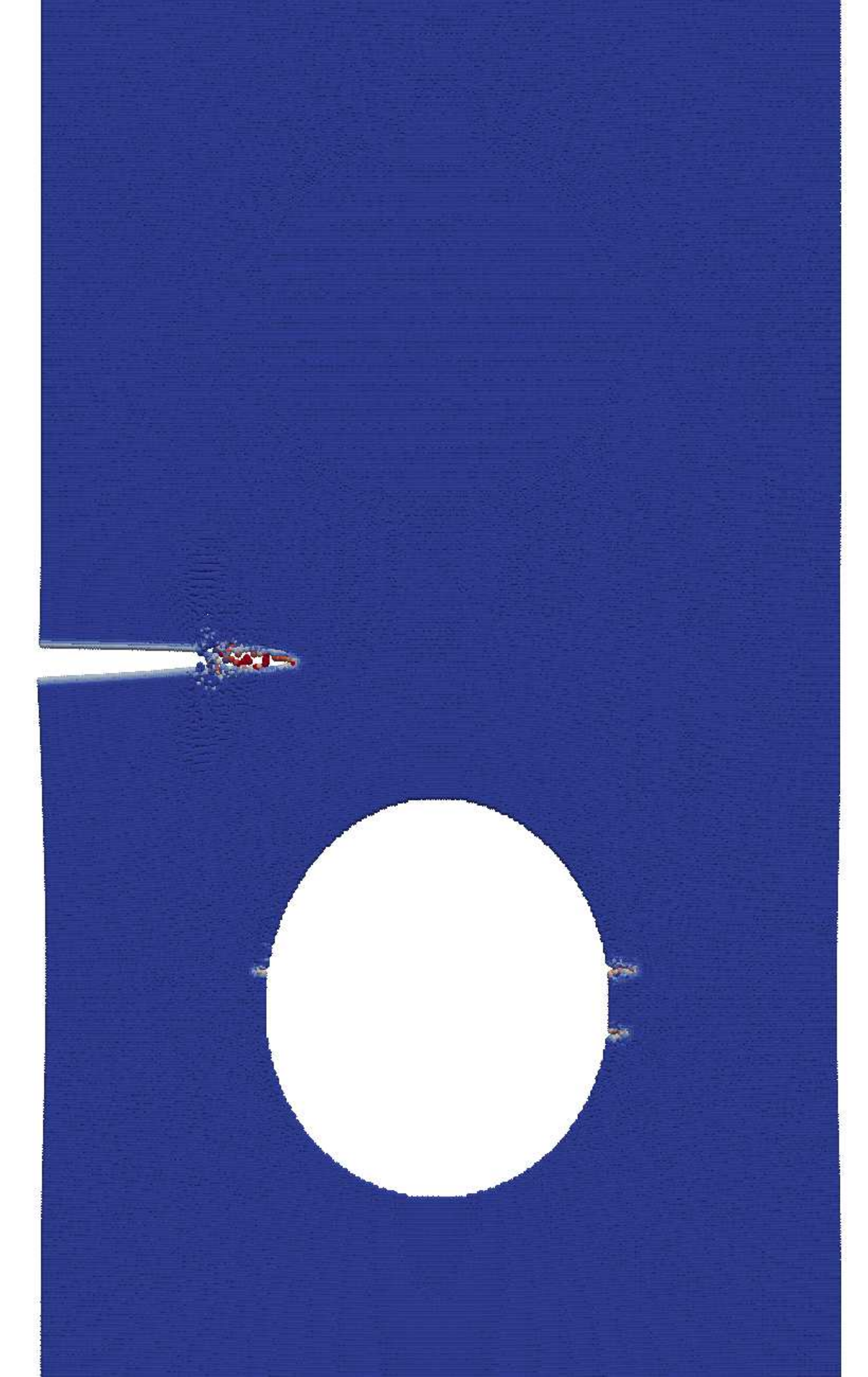} } 
\subfigure[$t = 8.00 \mu s$]{\includegraphics[scale=0.175]{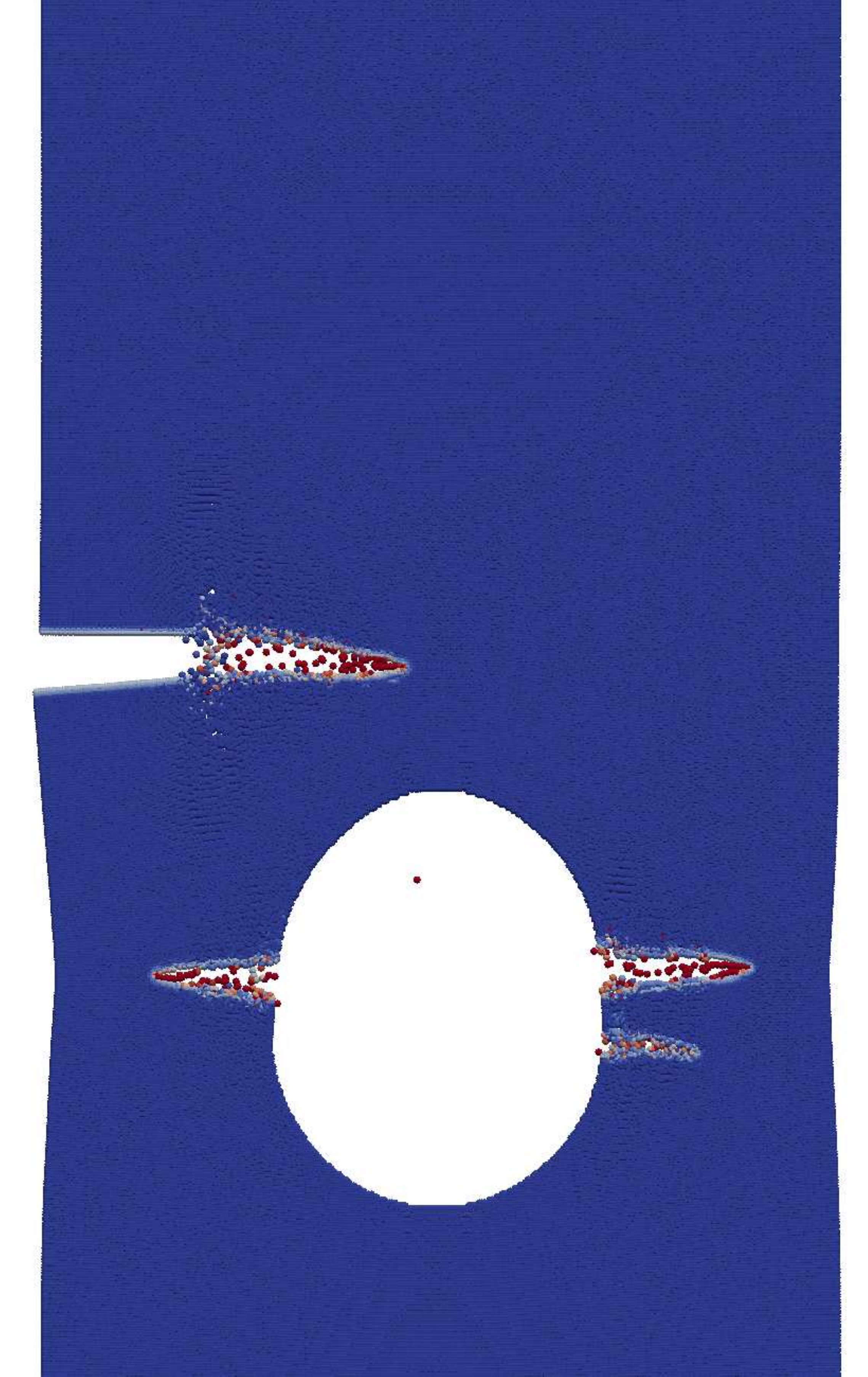} } 
\subfigure[$t = 10.00 \mu s$]{\includegraphics[scale=0.175]{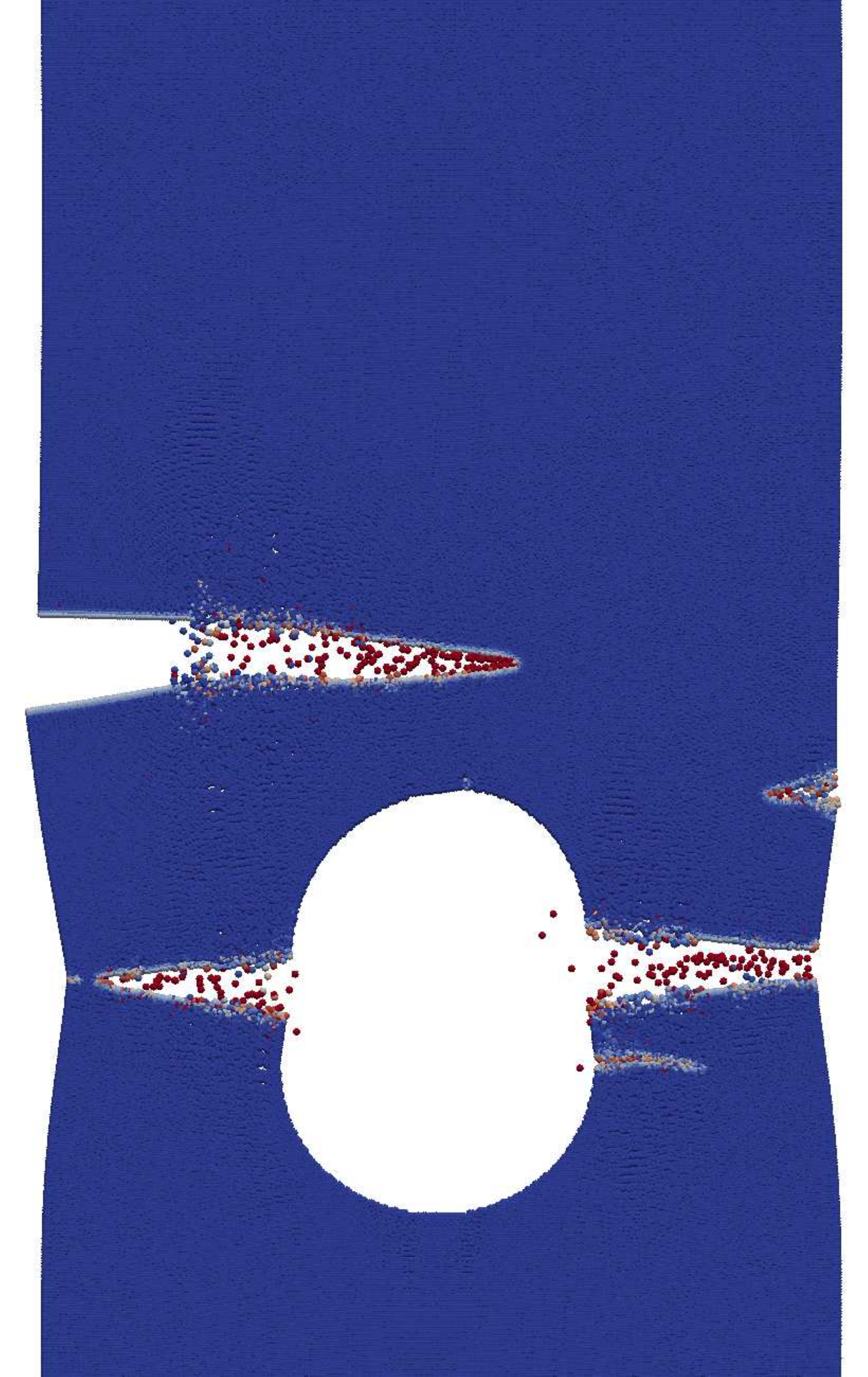} } 
\caption{Crack propagation for $\theta = 0^\circ$ - $v=25$ m/s - $n = 3$ - $300\times 600$ particles.}
\label{fig:theta0_inclusion_vel25}
\end{figure}

\begin{figure}[!htb]
\centering
\subfigure[$t = 4.00 \mu s$]{\includegraphics[scale=0.175]{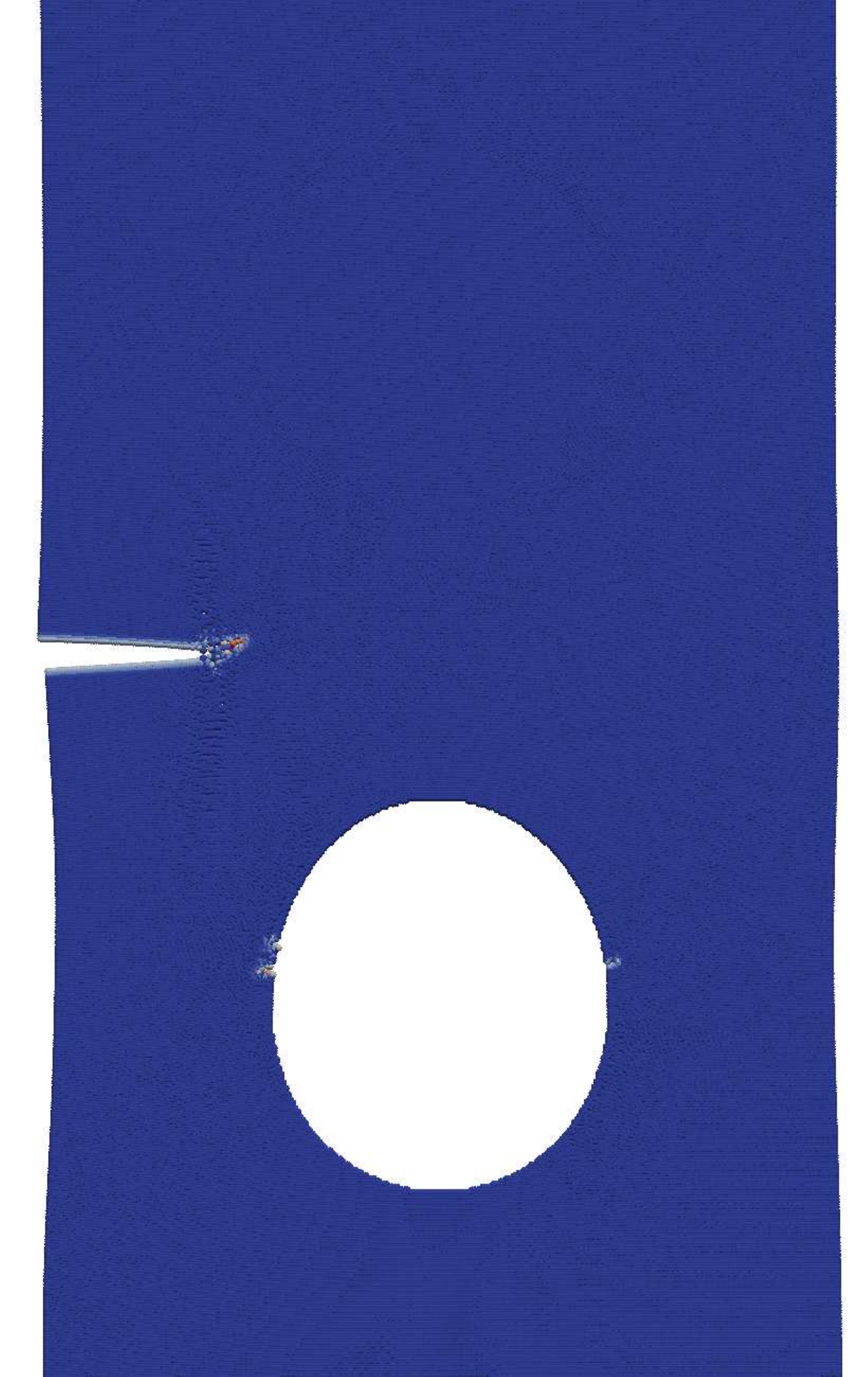} } 
\subfigure[$t = 9.50 \mu s$]{\includegraphics[scale=0.175]{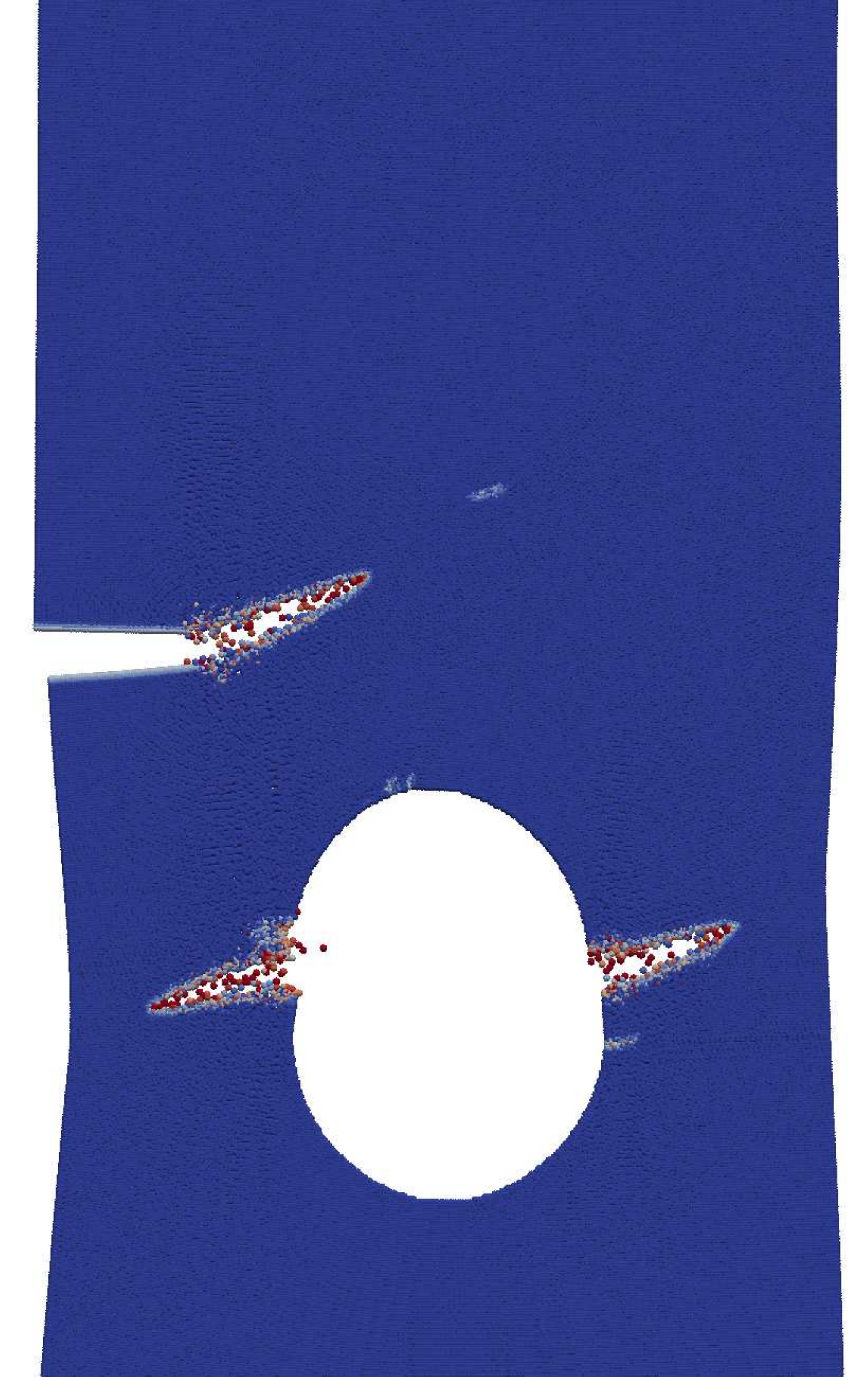} } 
\subfigure[$t = 13.50 \mu s$]{\includegraphics[scale=0.175]{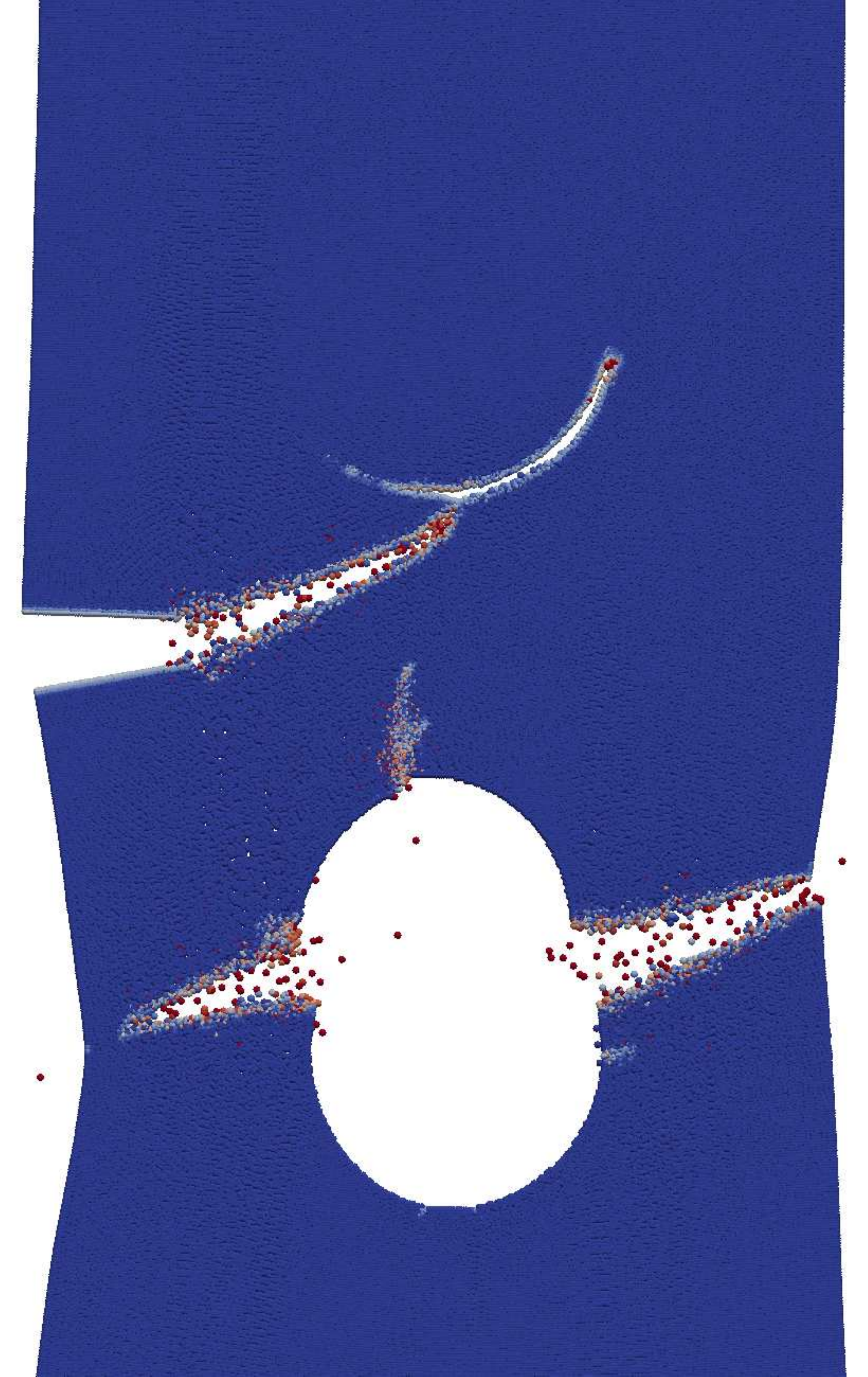} } 
\caption{Crack propagation for $\theta = 45^\circ$ - $v=25$ m/s - $n = 3$ - $300\times 600$ particles.}
\label{fig:theta45_inclusion_vel25}
\end{figure}

\begin{figure}[!htb]
\centering
\subfigure[$t = 4.00 \mu s$]{\includegraphics[scale=0.175]{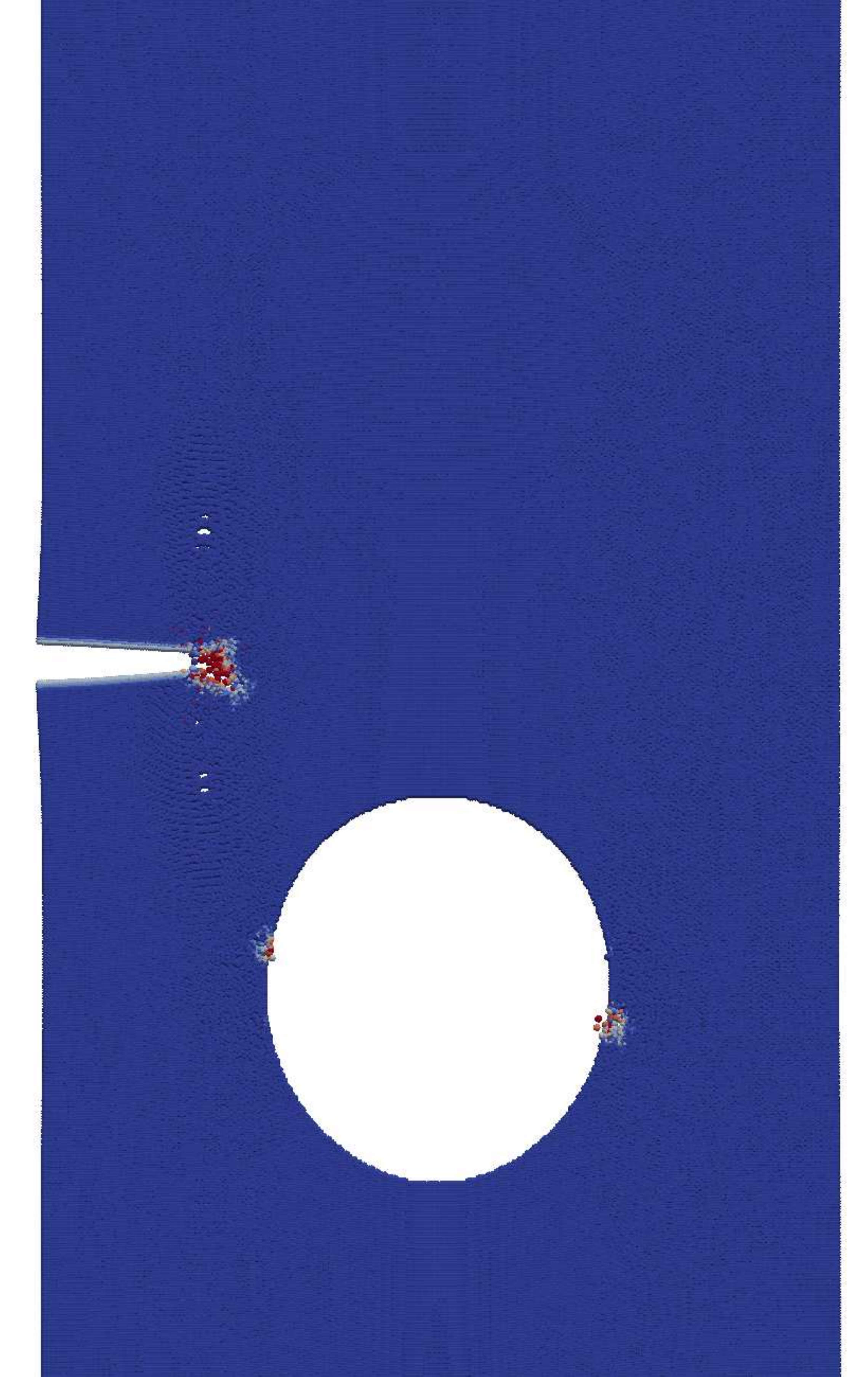} } 
\subfigure[$t = 9.50 \mu s$]{\includegraphics[scale=0.175]{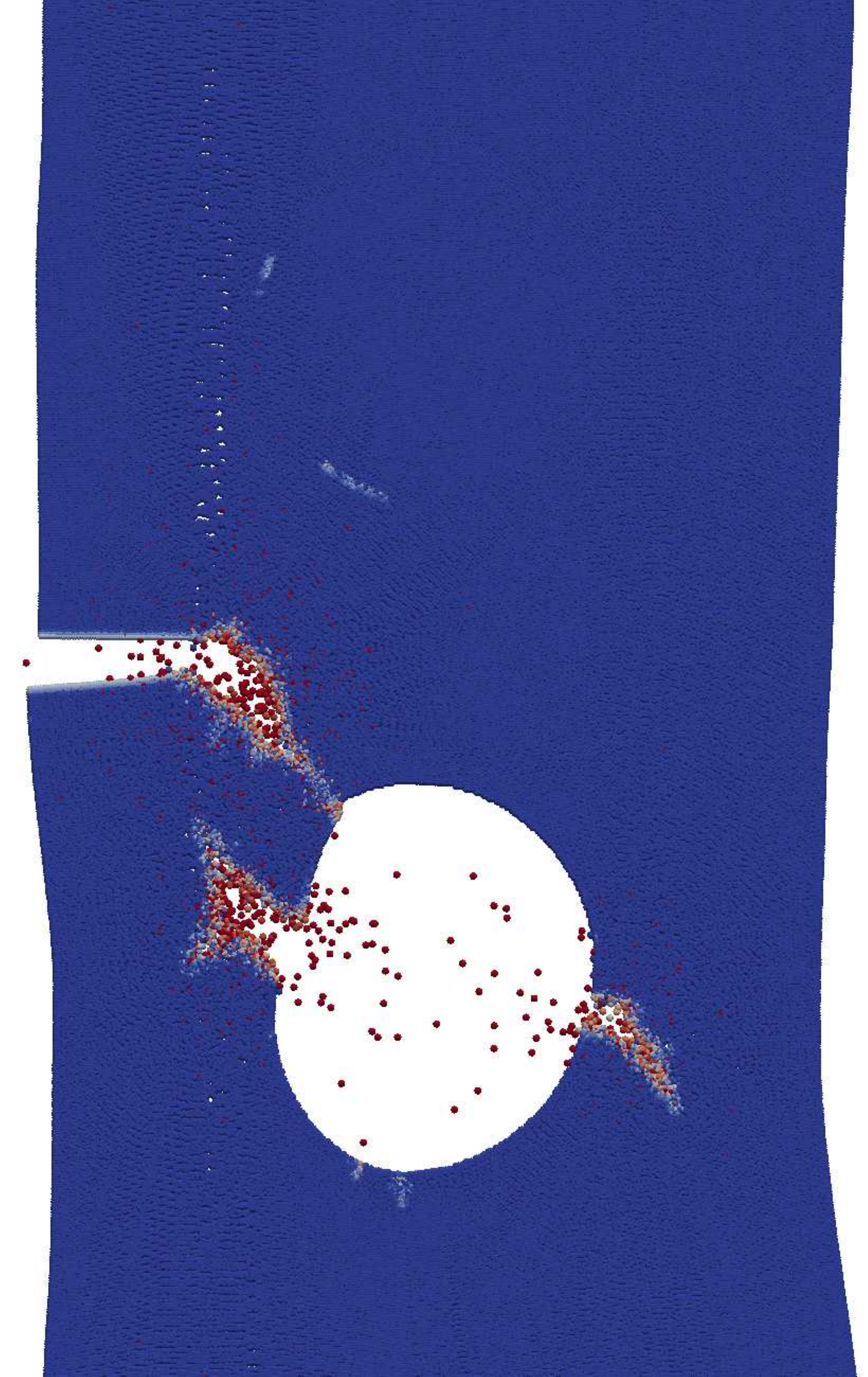} } 
\subfigure[$t = 13.50 \mu s$]{\includegraphics[scale=0.175]{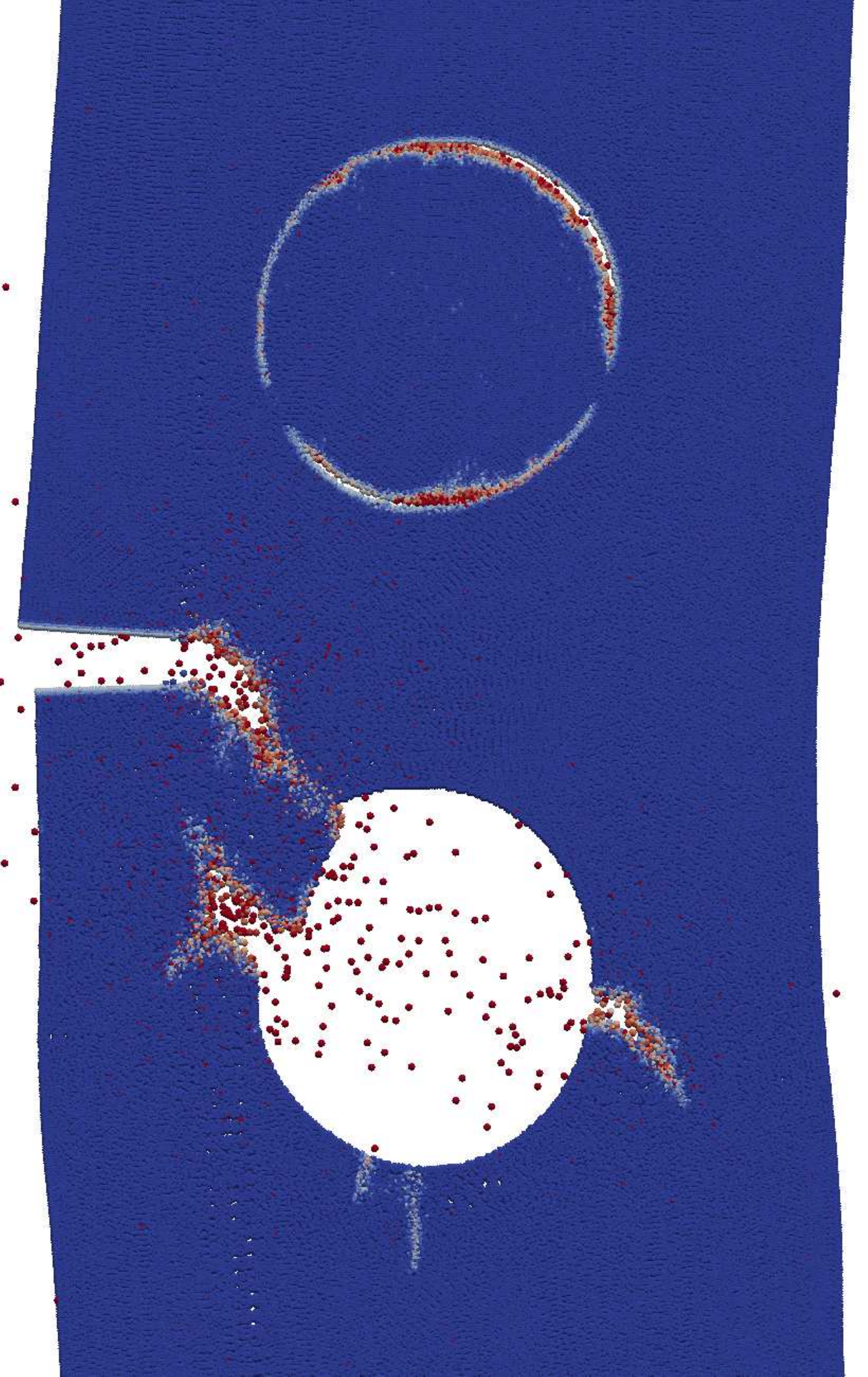} } 
\caption{Crack propagation for $\theta = 90^\circ$ - $v=25$ m/s - $n = 3$ - $300\times 600$ particles.}
\label{fig:theta90_inclusion_vel25}
\end{figure}

Figures \ref{fig:theta0_inclusion_vel25}, \ref{fig:theta45_inclusion_vel25} and \ref{fig:theta90_inclusion_vel25} depict the crack propagation for $\theta=0^\circ$, $\theta=45^\circ$ and $\theta=90^\circ$, respectively, under an initial velocity of $v=25 \frac{y}{2h}$ m/s. The different orientation of the material properties provide different crack propagation paths. In Figure \ref{fig:theta0_inclusion_vel25}, the inclusion is not damaged, but cracks appear on the hole. From Figure \ref{fig:theta45_inclusion_vel25}, there is some damage arising at the interface of the plate and the inclusion, as well as crack propagation from the edge crack and the hole. Figure \ref{fig:theta90_inclusion_vel25} shows a different crack propagation pattern, compared to those shown in Figure \ref{fig:theta90}. Additionally, the interface between the plate and the inclusion is almost fully damaged.

\begin{figure}[!htb]
\centering
\subfigure[$t = 3.57 \mu s$]{\includegraphics[scale=0.175]{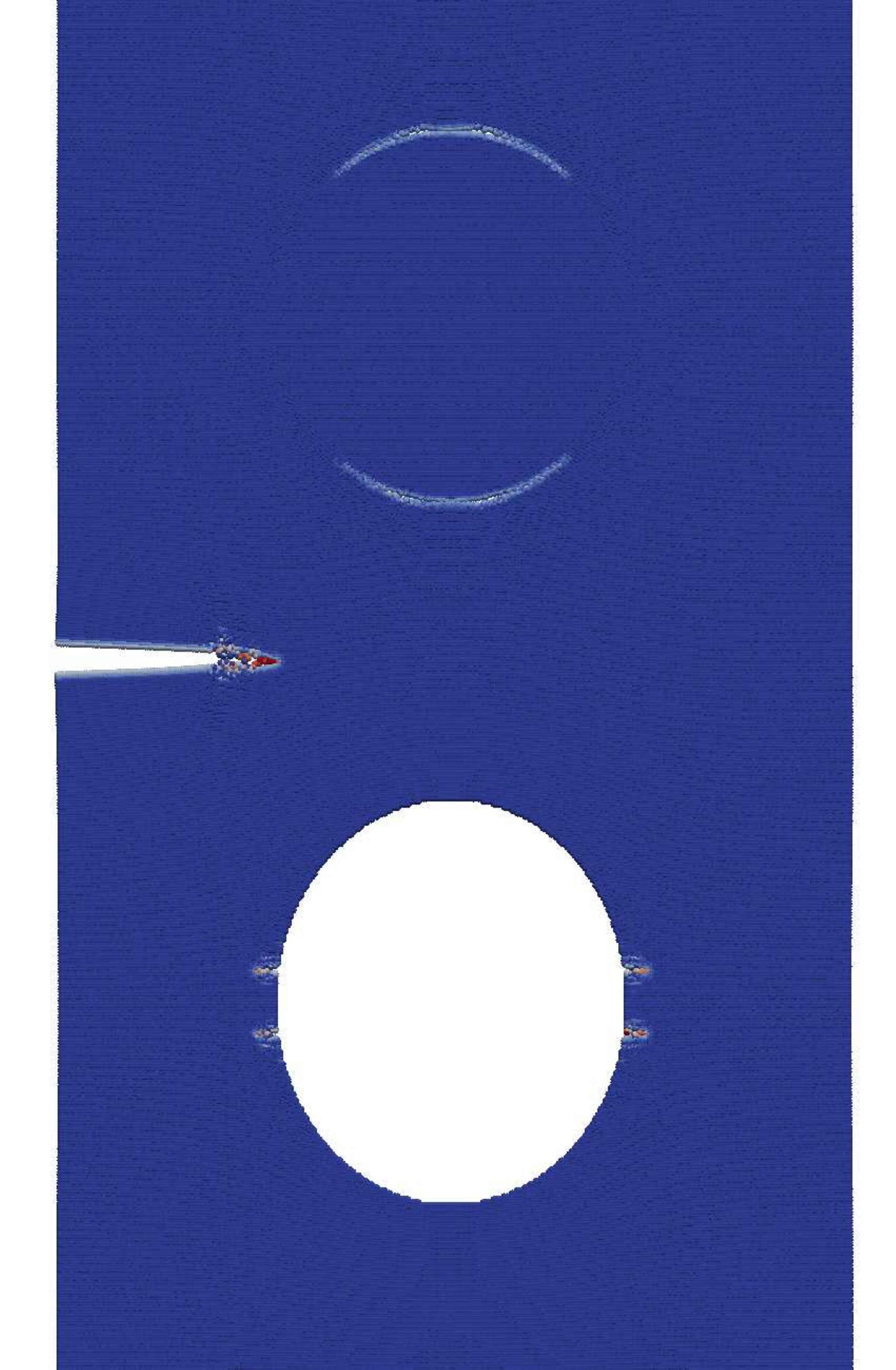} } 
\subfigure[$t = 4.94 \mu s$]{\includegraphics[scale=0.175]{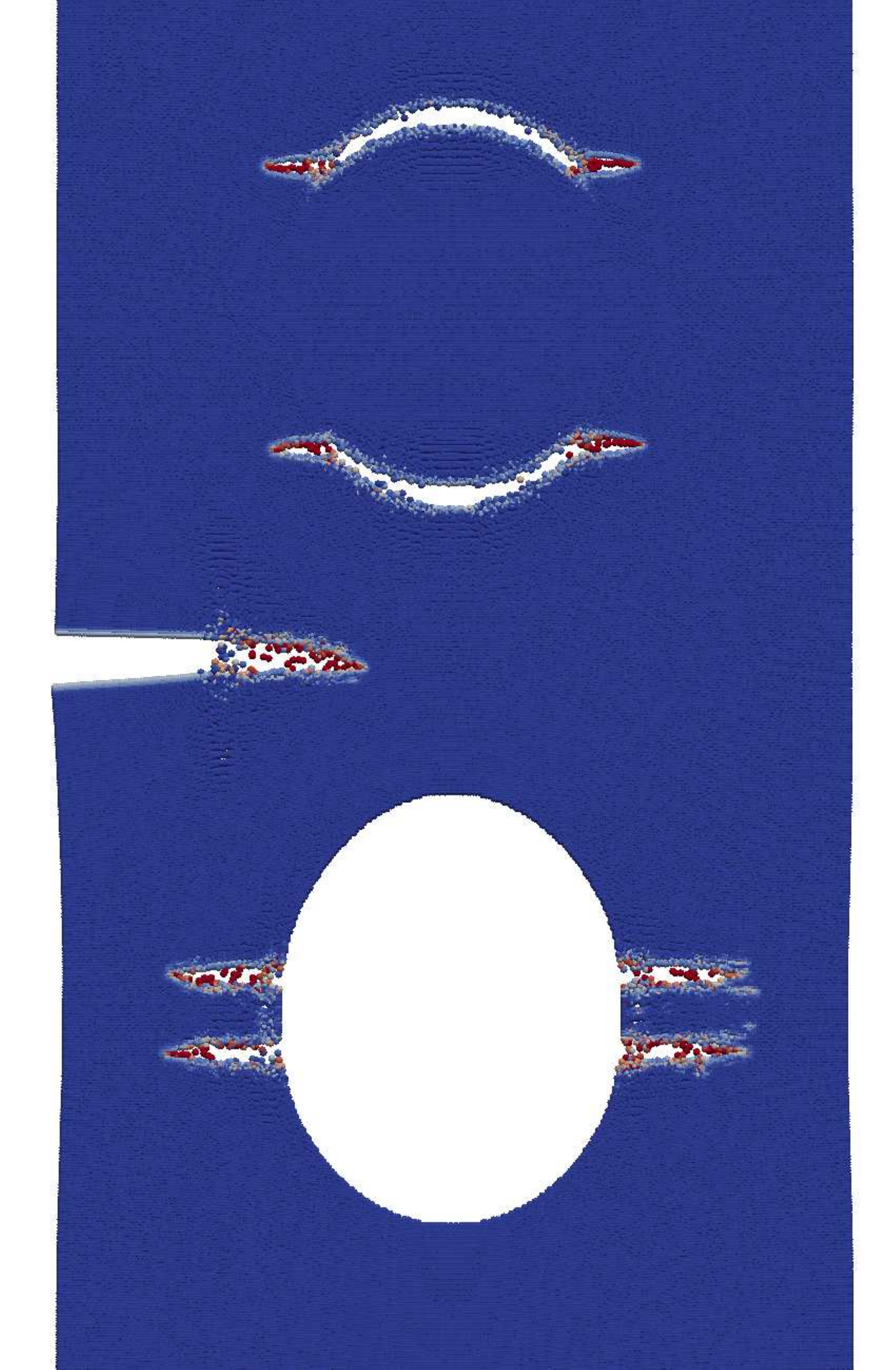} } 
\subfigure[$t = 6.98 \mu s$]{\includegraphics[scale=0.175]{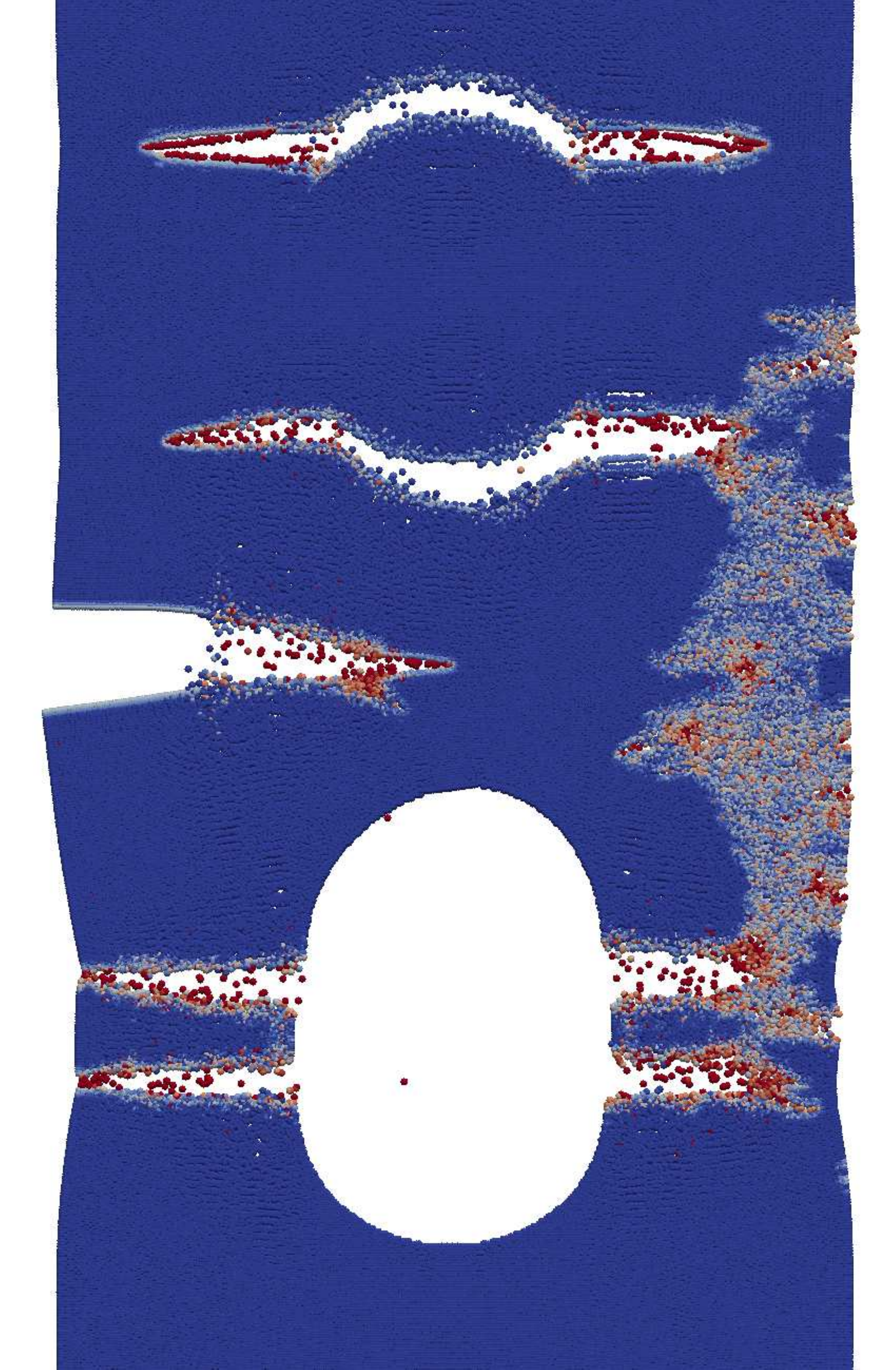} } 
\caption{Crack propagation for $\theta = 0^\circ$ - $v=50$ m/s- $n = 3$ - $300\times 600$ particles.}
\label{fig:theta0_inclusion_vel50}
\end{figure}

\begin{figure}[!htb]
\centering
\subfigure[$t = 3.06 \mu s$]{\includegraphics[scale=0.175]{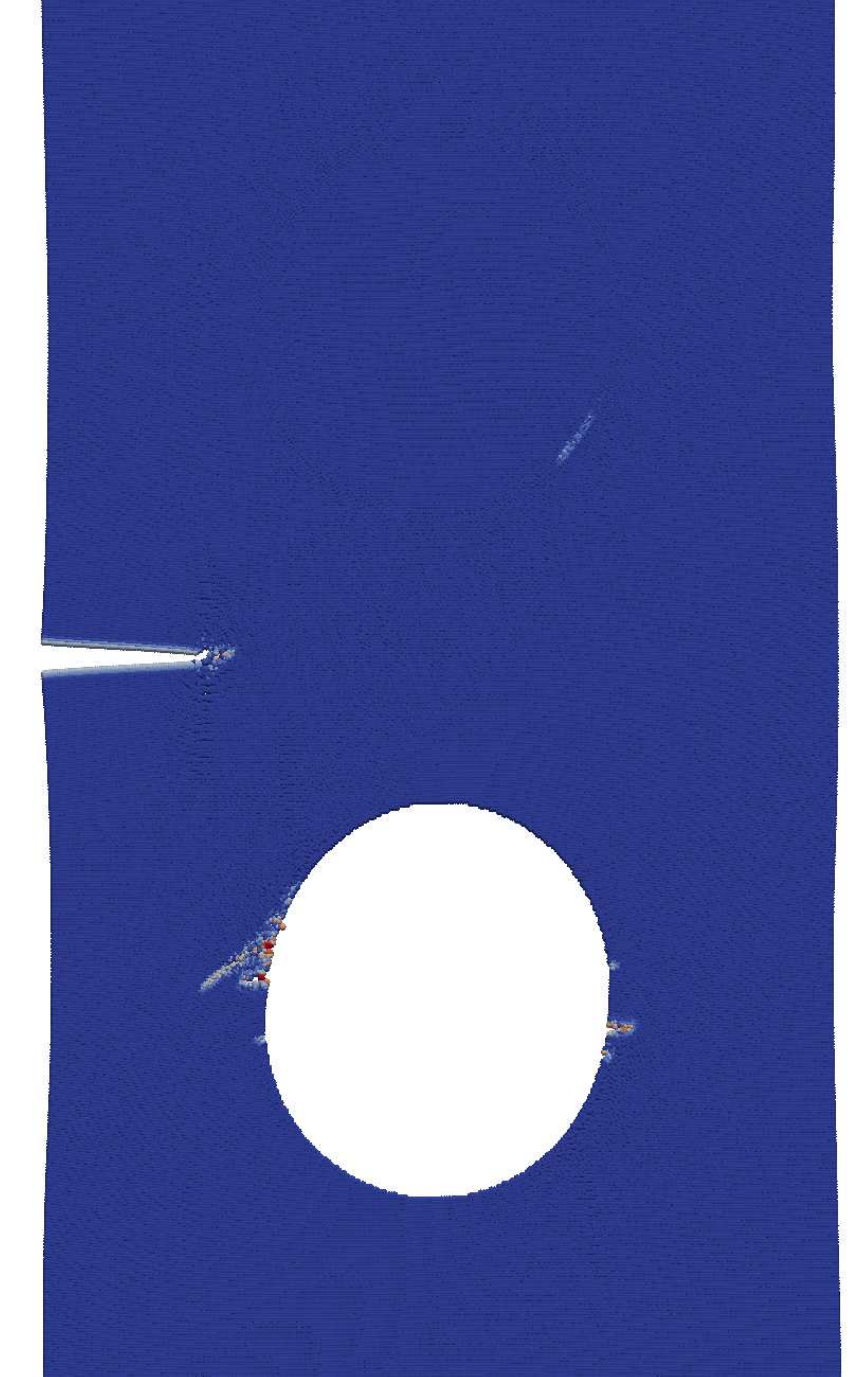} } 
\subfigure[$t = 4.25 \mu s$]{\includegraphics[scale=0.175]{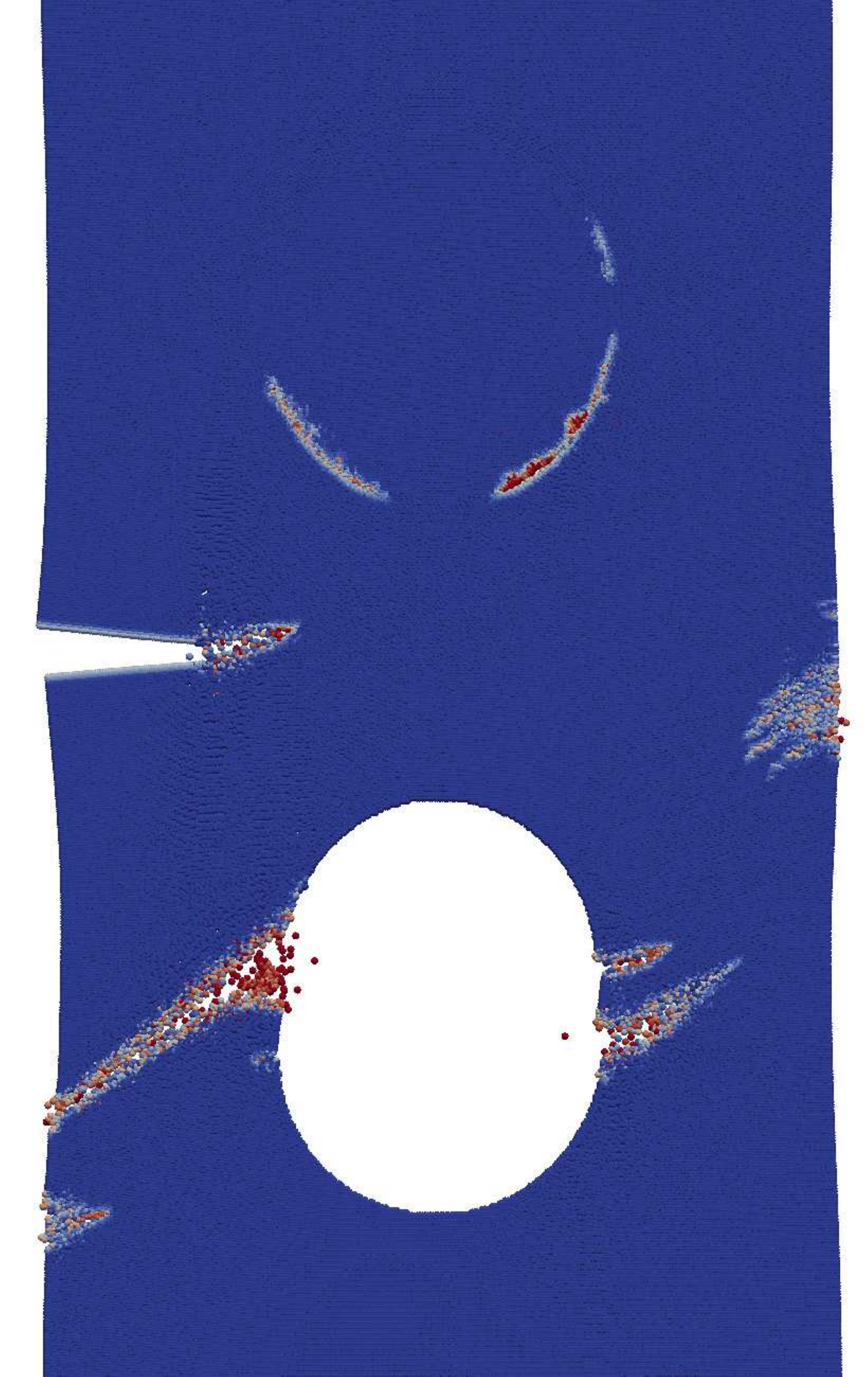} } 
\subfigure[$t = 6.13 \mu s$]{\includegraphics[scale=0.175]{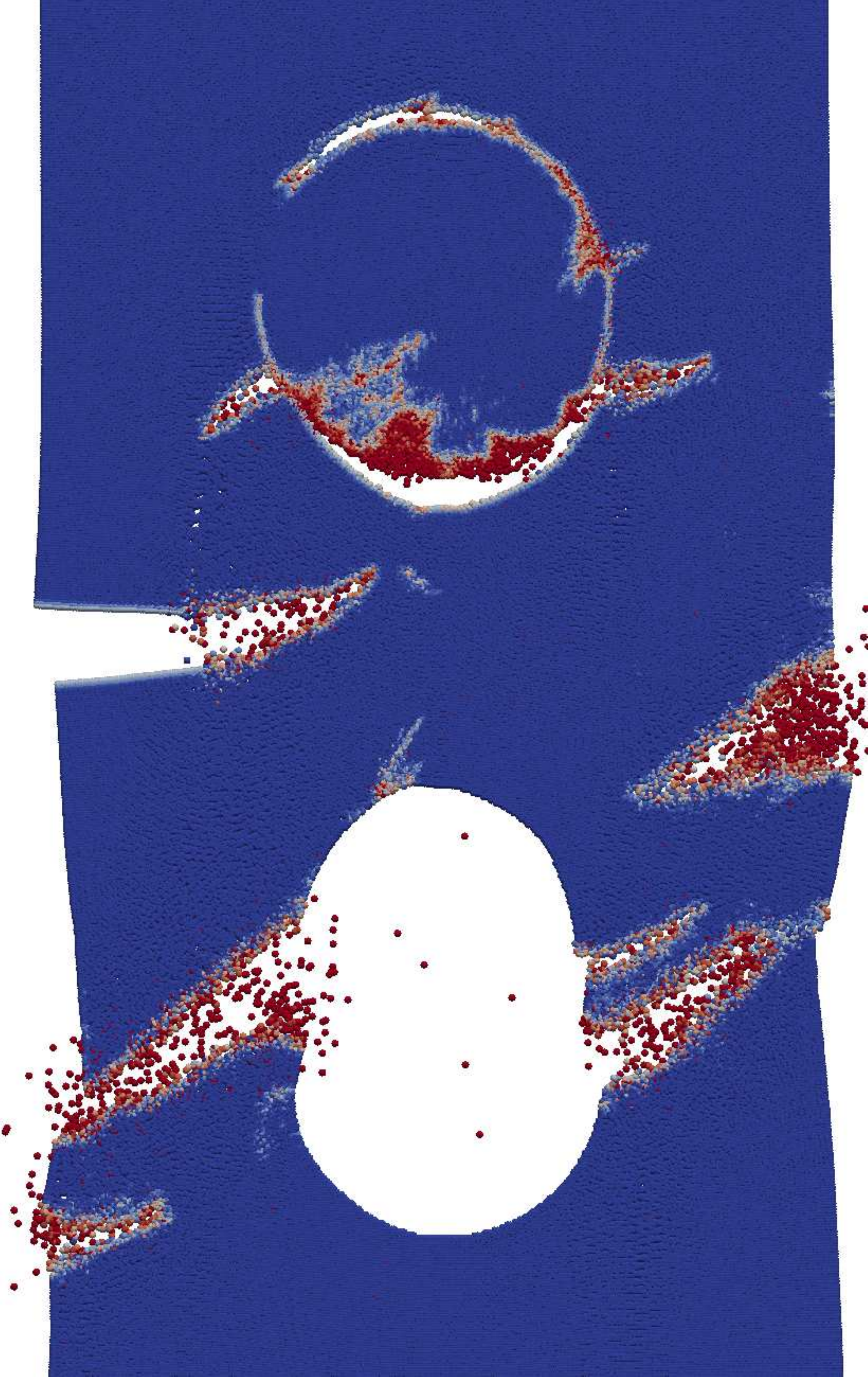} } 
\caption{Crack propagation for $\theta = 45^\circ$ - $v=50$ m/s - $n = 3$ - $300\times 600$ particles.}
\label{fig:theta45_inclusion_vel50}
\end{figure}

\begin{figure}[!htb]
\centering
\subfigure[$t = 2.04 \mu s$]{\includegraphics[scale=0.175]{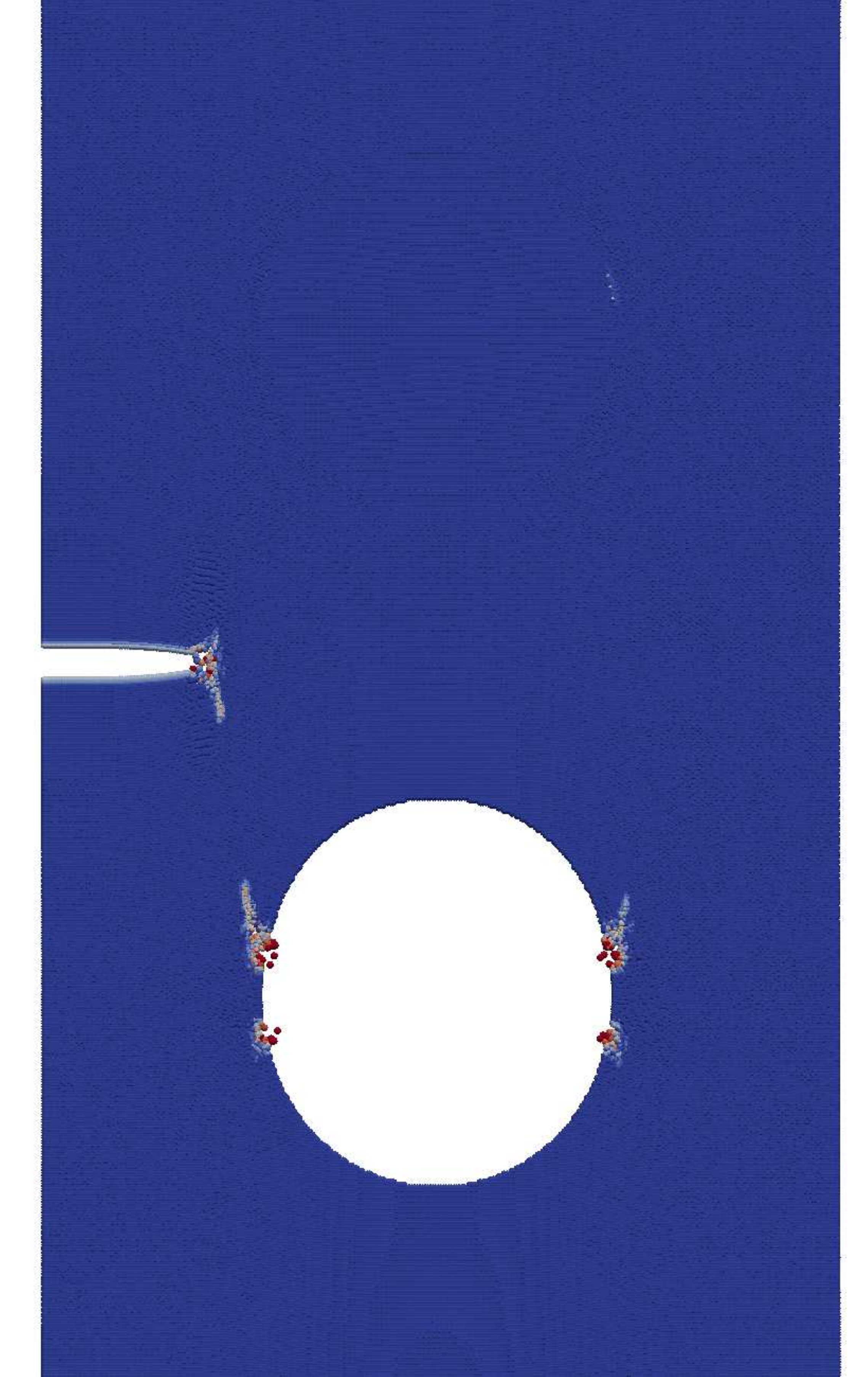} } 
\subfigure[$t = 3.40 \mu s$]{\includegraphics[scale=0.175]{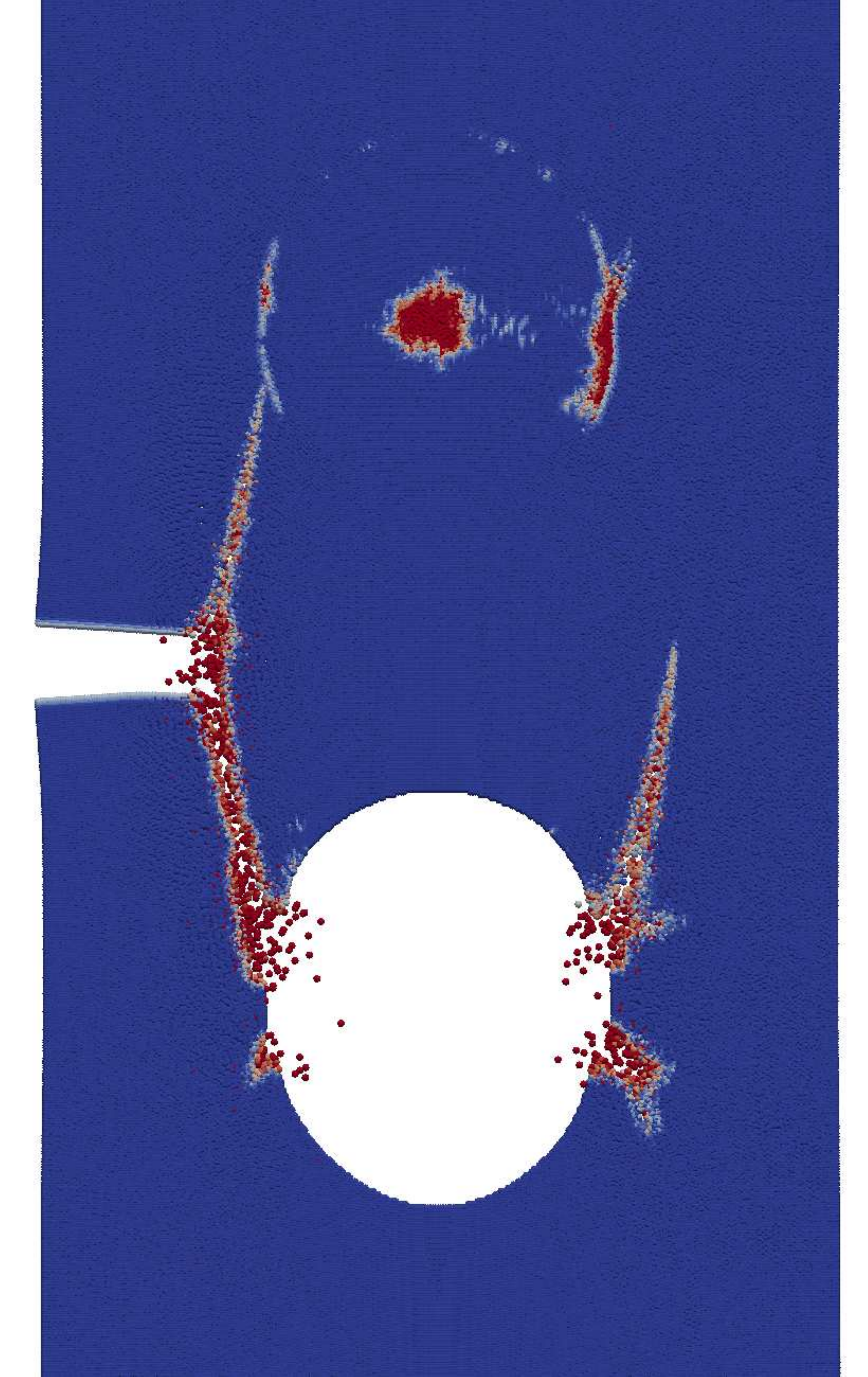} } 
\subfigure[$t = 5.44 \mu s$]{\includegraphics[scale=0.175]{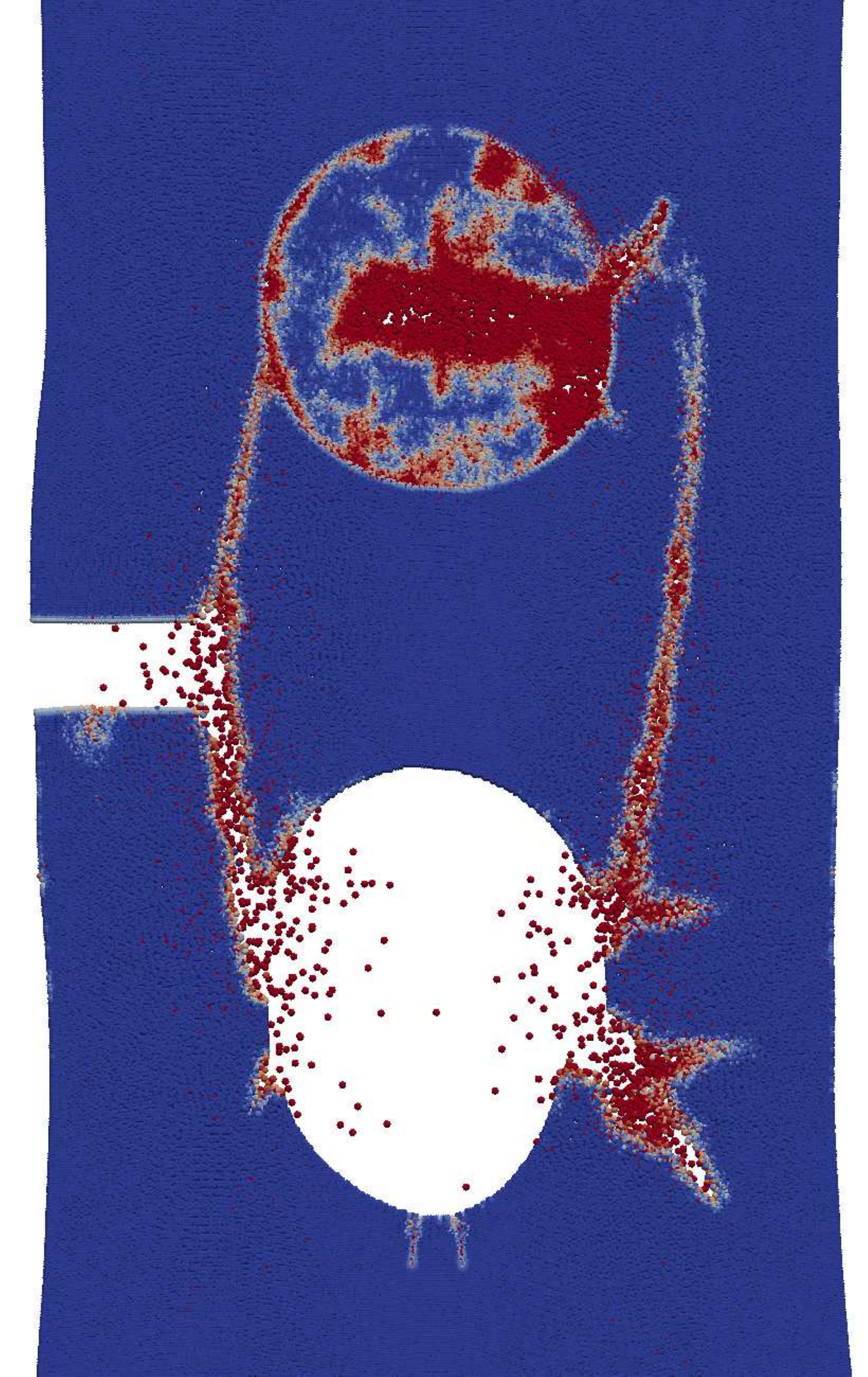} } 
\caption{Crack propagation for $\theta = 90^\circ$ - $v=50$ m/s - $n = 3$ - $300\times 600$ particles.}
\label{fig:theta90_inclusion_vel50}
\end{figure}

Now we increase the initial velocity to $v=50\frac{y}{2h}$ m/s and re-analyse the crack propagation for this example. The results are illustrated in Figures \ref{fig:theta0_inclusion_vel50}, \ref{fig:theta45_inclusion_vel50} and \ref{fig:theta90_inclusion_vel50}. It becomes clear that the crack propagation can change depending on the loading conditions. Figure \ref{fig:theta0_inclusion_vel50} now presents damage in the inclusion, and a double parallel crack originates from the hole. The crack propagation pattern in Figure \ref{fig:theta45_inclusion_vel50} is very similar to that shown in Figure \ref{fig:theta45_inclusion_vel25}, with the exception of the damage on the right side of the plate. Figure \ref{fig:theta90_inclusion_vel50} presents an almost vertical crack propagation, reaching both the inclusion and the void. Additionally, the applied velocity is sufficiently high that some damage is incurred in the inclusion.

Animations of this example for the initial velocity $v = 50$ m/s are available in the online version in the Supplemental Data section.

\section{Conclusions}

A generally anisotropic model for the non-ordinary state-based PD has been presented for the first time in the literature. The proposed formulation has been demonstrated for 2D materials assuming linear material behaviour and infinitesimal strains. The non-ordinary state-based framework has been used to model a composite anisotropic material. The PD formulation was validated against the FEM, and a very good agreement was achieved with both methods in the calculation of the dynamic stress intensity factors. The Tsai-Hill criterion for composite materials has been shown to provide good results for crack propagation problems. 

There are optimum values of the horizon size and grid spacing in order to reduce error and to obtain a reliable analysis. So far the horizon has been chosen empirically, but it becomes evident that it also depends on the material properties. 
The dynamic formulation has shown some interesting features for the crack propagation, where some oscillations have arisen at the edge of the plate when the crack is close to the edge.
Different loading may also lead to different crack propagation patterns, especially if heterogeneities are present in the material.

The formulation can be easily extended to 3D materials and different anisotropic materials, such as rocks. In this case, a damage criterion for this specific configuration would be necessary. Future work can include the assumption of large deformation in the model. Hence, the orientation of the material will evolve with deformation. An implicit formulation can also be implemented, allowing the use of larger time steps, leading to an enhanced computational efficiency.

\section*{Acknowledgements}\label{mypaper_sec:acknowledgements}

The first author acknowledges the Faculty of Science, Durham University, for his Postdoctoral Research Associate funding. Figures \ref{fig:theta0_cahill}, \ref{fig:theta45_cahill}, \ref{fig:theta60_cahill} and \ref{fig:theta90_cahill} have been reproduced with permission from Elsevier.

%\bibliographystyle{model2-names.bst}% \biboptions{authoryear}
%\biboptions{sort&compress}
%\bibliography{biblio_shale_corrected}

\begin{thebibliography}{51}
\expandafter\ifx\csname natexlab\endcsname\relax\def\natexlab#1{#1}\fi
\providecommand{\url}[1]{\texttt{#1}}
\providecommand{\href}[2]{#2}
\providecommand{\path}[1]{#1}
\providecommand{\DOIprefix}{doi:}
\providecommand{\ArXivprefix}{arXiv:}
\providecommand{\URLprefix}{URL: }
\providecommand{\Pubmedprefix}{pmid:}
\providecommand{\doi}[1]{\href{http://dx.doi.org/#1}{\path{#1}}}
\providecommand{\Pubmed}[1]{\href{pmid:#1}{\path{#1}}}
\providecommand{\bibinfo}[2]{#2}
\ifx\xfnm\relax \def\xfnm[#1]{\unskip,\space#1}\fi
%Type = Article
\bibitem[{Biner and Hu(2009)}]{biner2009simulation}
\bibinfo{author}{Biner, S.}, \bibinfo{author}{Hu, S.Y.}, \bibinfo{year}{2009}.
\newblock \bibinfo{title}{Simulation of damage evolution in composites: A
  phase-field model}.
\newblock \bibinfo{journal}{Acta Materialia} \bibinfo{volume}{57},
  \bibinfo{pages}{2088--2097}.
%Type = Article
\bibitem[{Bobaru and Hu(2012)}]{bobaru2012meaning}
\bibinfo{author}{Bobaru, F.}, \bibinfo{author}{Hu, W.}, \bibinfo{year}{2012}.
\newblock \bibinfo{title}{The meaning, selection, and use of the peridynamic
  horizon and its relation to crack branching in brittle materials}.
\newblock \bibinfo{journal}{International Journal of Fracture}
  \bibinfo{volume}{176}, \bibinfo{pages}{215--222}.
%Type = Article
\bibitem[{Bobaru et~al.(2009)Bobaru, Yang, Alves, Silling, Askari and
  Xu}]{bobaru2009convergence}
\bibinfo{author}{Bobaru, F.}, \bibinfo{author}{Yang, M.},
  \bibinfo{author}{Alves, L.F.}, \bibinfo{author}{Silling, S.A.},
  \bibinfo{author}{Askari, E.}, \bibinfo{author}{Xu, J.}, \bibinfo{year}{2009}.
\newblock \bibinfo{title}{Convergence, adaptive refinement, and scaling in {1D}
  peridynamics}.
\newblock \bibinfo{journal}{International Journal for Numerical Methods in
  Engineering} \bibinfo{volume}{77}, \bibinfo{pages}{852--877}.
%Type = Article
\bibitem[{Bouhala et~al.(2013)Bouhala, Makradi, Belouettar, Kiefer-Kamal and
  Fr{\'e}res}]{bouhala2013modelling}
\bibinfo{author}{Bouhala, L.}, \bibinfo{author}{Makradi, A.},
  \bibinfo{author}{Belouettar, S.}, \bibinfo{author}{Kiefer-Kamal, H.},
  \bibinfo{author}{Fr{\'e}res, P.}, \bibinfo{year}{2013}.
\newblock \bibinfo{title}{Modelling of failure in long fibres reinforced
  composites by {X-FEM} and cohesive zone model}.
\newblock \bibinfo{journal}{Composites Part B: Engineering}
  \bibinfo{volume}{55}, \bibinfo{pages}{352--361}.
%Type = Article
\bibitem[{Breitenfeld et~al.(2014)Breitenfeld, Geubelle, Weckner and
  Silling}]{breitenfeld2014non}
\bibinfo{author}{Breitenfeld, M.}, \bibinfo{author}{Geubelle, P.},
  \bibinfo{author}{Weckner, O.}, \bibinfo{author}{Silling, S.},
  \bibinfo{year}{2014}.
\newblock \bibinfo{title}{Non-ordinary state-based peridynamic analysis of
  stationary crack problems}.
\newblock \bibinfo{journal}{Computer Methods in Applied Mechanics and
  Engineering} \bibinfo{volume}{272}, \bibinfo{pages}{233--250}.
%Type = Article
\bibitem[{Cahill et~al.(2014)Cahill, Natarajan, Bordas, O'Higgins and
  McCarthy}]{cahill2014experimental}
\bibinfo{author}{Cahill, L.M.A.}, \bibinfo{author}{Natarajan, S.},
  \bibinfo{author}{Bordas, S.P.A.}, \bibinfo{author}{O'Higgins, R.M.},
  \bibinfo{author}{McCarthy, C.T.}, \bibinfo{year}{2014}.
\newblock \bibinfo{title}{An experimental/numerical investigation into the main
  driving force for crack propagation in uni-directional fibre-reinforced
  composite laminae}.
\newblock \bibinfo{journal}{Composite Structures} \bibinfo{volume}{107},
  \bibinfo{pages}{119 -- 130}.
%Type = Article
\bibitem[{{Courant} et~al.(1928){Courant}, {Friedrichs} and
  {Lewy}}]{courant1928uber}
\bibinfo{author}{{Courant}, R.}, \bibinfo{author}{{Friedrichs}, K.},
  \bibinfo{author}{{Lewy}, H.}, \bibinfo{year}{1928}.
\newblock \bibinfo{title}{{{\"U}ber die partiellen Differenzengleichungen der
  mathematischen Physik}}.
\newblock \bibinfo{journal}{Mathematische Annalen} \bibinfo{volume}{100},
  \bibinfo{pages}{32--74}.
%Type = Phdthesis
\bibitem[{Engwirda(2014)}]{mesh2d2014}
\bibinfo{author}{Engwirda, D.}, \bibinfo{year}{2014}.
\newblock \bibinfo{title}{Locally-optimal Delaunay-refinement and
  optimisation-based mesh generation}.
\newblock Ph.D. thesis.
%Type = Article
\bibitem[{Eringen and Edelen(1972)}]{eringen1972nonlocal}
\bibinfo{author}{Eringen, A.C.}, \bibinfo{author}{Edelen, D.},
  \bibinfo{year}{1972}.
\newblock \bibinfo{title}{On nonlocal elasticity}.
\newblock \bibinfo{journal}{International Journal of Engineering Science}
  \bibinfo{volume}{10}, \bibinfo{pages}{233--248}.
%Type = Article
\bibitem[{Garc\'ia-S\'anchez et~al.(2005)Garc\'ia-S\'anchez, S\'aez and
  Dom\'inguez}]{garcia2005anisotropic}
\bibinfo{author}{Garc\'ia-S\'anchez, F.}, \bibinfo{author}{S\'aez, A.},
  \bibinfo{author}{Dom\'inguez, J.}, \bibinfo{year}{2005}.
\newblock \bibinfo{title}{Anisotropic and piezoelectric materials fracture
  analysis by {BEM}}.
\newblock \bibinfo{journal}{Computers \& Structures} \bibinfo{volume}{83},
  \bibinfo{pages}{804--820}.
%Type = Article
\bibitem[{Gerstle et~al.(2007)Gerstle, Sau and
  Silling}]{gerstle2007peridynamic}
\bibinfo{author}{Gerstle, W.}, \bibinfo{author}{Sau, N.},
  \bibinfo{author}{Silling, S.}, \bibinfo{year}{2007}.
\newblock \bibinfo{title}{Peridynamic modeling of concrete structures}.
\newblock \bibinfo{journal}{Nuclear engineering and design}
  \bibinfo{volume}{237}, \bibinfo{pages}{1250--1258}.
%Type = Article
\bibitem[{Ghajari et~al.(2014)Ghajari, Iannucci and
  Curtis}]{ghajari2014peridynamic}
\bibinfo{author}{Ghajari, M.}, \bibinfo{author}{Iannucci, L.},
  \bibinfo{author}{Curtis, P.}, \bibinfo{year}{2014}.
\newblock \bibinfo{title}{A peridynamic material model for the analysis of
  dynamic crack propagation in orthotropic media}.
\newblock \bibinfo{journal}{Computer Methods in Applied Mechanics and
  Engineering} \bibinfo{volume}{276}, \bibinfo{pages}{431--452}.
%Type = Article
\bibitem[{Giordano et~al.(2017)Giordano, Zappal{\`a} and
  Kleiven}]{giordano2017anisotropic}
\bibinfo{author}{Giordano, C.}, \bibinfo{author}{Zappal{\`a}, S.},
  \bibinfo{author}{Kleiven, S.}, \bibinfo{year}{2017}.
\newblock \bibinfo{title}{Anisotropic finite element models for brain injury
  prediction: the sensitivity of axonal strain to white matter tract
  inter-subject variability}.
\newblock \bibinfo{journal}{Biomechanics and Modeling in Mechanobiology}
  \bibinfo{volume}{16}, \bibinfo{pages}{1269--1293}.
%Type = Article
\bibitem[{Giraud et~al.(2008)Giraud, Hoxha, Huynh, Do and
  Magnenet}]{giraud2008effective}
\bibinfo{author}{Giraud, A.}, \bibinfo{author}{Hoxha, D.},
  \bibinfo{author}{Huynh, Q.V.}, \bibinfo{author}{Do, D.P.},
  \bibinfo{author}{Magnenet, V.}, \bibinfo{year}{2008}.
\newblock \bibinfo{title}{Effective porothermoelastic properties of
  transversely isotropic rock-like composites}.
\newblock \bibinfo{journal}{International Journal of Engineering Science}
  \bibinfo{volume}{46}, \bibinfo{pages}{527--550}.
%Type = Article
\bibitem[{Griffith(1921)}]{griffith1921phenomena}
\bibinfo{author}{Griffith, A.A.}, \bibinfo{year}{1921}.
\newblock \bibinfo{title}{The phenomena of rupture and flow in solids}.
\newblock \bibinfo{journal}{Philosophical Transactions of the Royal Society of
  London A: Mathematical, Physical and Engineering Sciences}
  \bibinfo{volume}{221}, \bibinfo{pages}{163--198}.
%Type = Article
\bibitem[{Hattori et~al.(2017a)Hattori, Alatawi and
  Trevelyan}]{hattori2016extended}
\bibinfo{author}{Hattori, G.}, \bibinfo{author}{Alatawi, I.A.},
  \bibinfo{author}{Trevelyan, J.}, \bibinfo{year}{2017}a.
\newblock \bibinfo{title}{An extended boundary element method formulation for
  the direct calculation of the stress intensity factors in fully anisotropic
  materials}.
\newblock \bibinfo{journal}{International Journal for Numerical Methods in
  Engineering} \bibinfo{volume}{109}, \bibinfo{pages}{965--981}.
%Type = Article
\bibitem[{Hattori et~al.(2017b)Hattori, Trevelyan, Augarde, Cooms and
  Aplin}]{hattori2016numerical}
\bibinfo{author}{Hattori, G.}, \bibinfo{author}{Trevelyan, J.},
  \bibinfo{author}{Augarde, C.E.}, \bibinfo{author}{Cooms, W.M.},
  \bibinfo{author}{Aplin, A.C.}, \bibinfo{year}{2017}b.
\newblock \bibinfo{title}{Numerical simulation of fracking in shale rocks:
  Current state and future approaches}.
\newblock \bibinfo{journal}{Archives of Computational Methods in Engineering}
  \bibinfo{volume}{24}, \bibinfo{pages}{281--317}.
%Type = Article
\bibitem[{Henry(2008)}]{henry2008study}
\bibinfo{author}{Henry, H.}, \bibinfo{year}{2008}.
\newblock \bibinfo{title}{Study of the branching instability using a phase
  field model of inplane crack propagation}.
\newblock \bibinfo{journal}{EPL (Europhysics Letters)} \bibinfo{volume}{83},
  \bibinfo{pages}{16004}.
%Type = Book
\bibitem[{Holzapfel(2000)}]{holzapfel2000nonlinear}
\bibinfo{author}{Holzapfel, G.A.}, \bibinfo{year}{2000}.
\newblock \bibinfo{title}{Nonlinear solid mechanics}.
\newblock \bibinfo{publisher}{John Wiley \& Sons Ltd}.
%Type = Article
\bibitem[{Hu et~al.(2012)Hu, Ha and Bobaru}]{hu2012peridynamic}
\bibinfo{author}{Hu, W.}, \bibinfo{author}{Ha, Y.D.}, \bibinfo{author}{Bobaru,
  F.}, \bibinfo{year}{2012}.
\newblock \bibinfo{title}{Peridynamic model for dynamic fracture in
  unidirectional fiber-reinforced composites}.
\newblock \bibinfo{journal}{Computer Methods in Applied Mechanics and
  Engineering} \bibinfo{volume}{217}, \bibinfo{pages}{247--261}.
%Type = Article
\bibitem[{Hu et~al.(2014)Hu, Yu and Wang}]{hu2014peridynamic}
\bibinfo{author}{Hu, Y.l.}, \bibinfo{author}{Yu, Y.}, \bibinfo{author}{Wang,
  H.}, \bibinfo{year}{2014}.
\newblock \bibinfo{title}{Peridynamic analytical method for progressive damage
  in notched composite laminates}.
\newblock \bibinfo{journal}{Composite Structures} \bibinfo{volume}{108},
  \bibinfo{pages}{801--810}.
%Type = Article
\bibitem[{Lorentz and Andrieux(2003)}]{lorentz2003analysis}
\bibinfo{author}{Lorentz, E.}, \bibinfo{author}{Andrieux, S.},
  \bibinfo{year}{2003}.
\newblock \bibinfo{title}{Analysis of non-local models through energetic
  formulations}.
\newblock \bibinfo{journal}{International Journal of Solids and Structures}
  \bibinfo{volume}{40}, \bibinfo{pages}{2905--2936}.
%Type = Article
\bibitem[{Macek and Silling(2007)}]{macek2007peridynamics}
\bibinfo{author}{Macek, R.W.}, \bibinfo{author}{Silling, S.A.},
  \bibinfo{year}{2007}.
\newblock \bibinfo{title}{Peridynamics via finite element analysis}.
\newblock \bibinfo{journal}{Finite Elements in Analysis and Design}
  \bibinfo{volume}{43}, \bibinfo{pages}{1169--1178}.
%Type = Book
\bibitem[{Madenci and Oterkus(2014)}]{madenci2014peridynamic}
\bibinfo{author}{Madenci, E.}, \bibinfo{author}{Oterkus, E.},
  \bibinfo{year}{2014}.
\newblock \bibinfo{title}{Peridynamic theory and its applications}.
\newblock \bibinfo{publisher}{Springer}.
%Type = Article
\bibitem[{Milazzo(2012)}]{milazzo2012equivalent}
\bibinfo{author}{Milazzo, A.}, \bibinfo{year}{2012}.
\newblock \bibinfo{title}{An equivalent single-layer model for
  magnetoelectroelastic multilayered plate dynamics}.
\newblock \bibinfo{journal}{Composite Structures} \bibinfo{volume}{94},
  \bibinfo{pages}{2078--2086}.
%Type = Article
\bibitem[{Motamedi et~al.(2014)Motamedi, Milani, Komeili, Bureau, Thibault and
  Trudel-Boucher}]{motamedi2014stochastic}
\bibinfo{author}{Motamedi, D.}, \bibinfo{author}{Milani, A.S.},
  \bibinfo{author}{Komeili, M.}, \bibinfo{author}{Bureau, M.N.},
  \bibinfo{author}{Thibault, F.}, \bibinfo{author}{Trudel-Boucher, D.},
  \bibinfo{year}{2014}.
\newblock \bibinfo{title}{A stochastic {XFEM} model to study delamination in
  {PPS}/{G}lass {UD} composites: {E}ffect of uncertain fracture properties}.
\newblock \bibinfo{journal}{Applied Composite Materials} \bibinfo{volume}{21},
  \bibinfo{pages}{341--358}.
%Type = Article
\bibitem[{Motamedi and Mohammadi(2012)}]{motamedi2012fracture}
\bibinfo{author}{Motamedi, D.}, \bibinfo{author}{Mohammadi, S.},
  \bibinfo{year}{2012}.
\newblock \bibinfo{title}{Fracture analysis of composites by time independent
  moving-crack orthotropic {XFEM}}.
\newblock \bibinfo{journal}{International Journal of Mechanical Sciences}
  \bibinfo{volume}{54}, \bibinfo{pages}{20--37}.
%Type = Book
\bibitem[{Muskhelishvili(1953)}]{Muskhelishvili53}
\bibinfo{author}{Muskhelishvili, N.I.}, \bibinfo{year}{1953}.
\newblock \bibinfo{title}{Some basic problems of the mathematical theory of
  elasticity}.
\newblock \bibinfo{publisher}{Leiden: Noordhoff}.
%Type = Article
\bibitem[{Nobile and Carloni(2005)}]{Nobile05}
\bibinfo{author}{Nobile, L.}, \bibinfo{author}{Carloni, C.},
  \bibinfo{year}{2005}.
\newblock \bibinfo{title}{Fracture analysis for orthotropic cracked plates}.
\newblock \bibinfo{journal}{Composite Structures} \bibinfo{volume}{68},
  \bibinfo{pages}{285--293}.
%Type = Article
\bibitem[{Oterkus and Madenci(2012)}]{oterkus2012peridynamic}
\bibinfo{author}{Oterkus, E.}, \bibinfo{author}{Madenci, E.},
  \bibinfo{year}{2012}.
\newblock \bibinfo{title}{Peridynamic analysis of fiber-reinforced composite
  materials}.
\newblock \bibinfo{journal}{Journal of Mechanics of Materials and Structures}
  \bibinfo{volume}{7}, \bibinfo{pages}{45--84}.
%Type = Article
\bibitem[{Oterkus et~al.(2012)Oterkus, Madenci, Weckner, Silling, Bogert and
  Tessler}]{oterkus2012combined}
\bibinfo{author}{Oterkus, E.}, \bibinfo{author}{Madenci, E.},
  \bibinfo{author}{Weckner, O.}, \bibinfo{author}{Silling, S.},
  \bibinfo{author}{Bogert, P.}, \bibinfo{author}{Tessler, A.},
  \bibinfo{year}{2012}.
\newblock \bibinfo{title}{Combined finite element and peridynamic analyses for
  predicting failure in a stiffened composite curved panel with a central
  slot}.
\newblock \bibinfo{journal}{Composite Structures} \bibinfo{volume}{94},
  \bibinfo{pages}{839--850}.
%Type = Article
\bibitem[{Pens{\'e}e et~al.(2002)Pens{\'e}e, Kondo and
  Dormieux}]{pensee2002micromechanical}
\bibinfo{author}{Pens{\'e}e, V.}, \bibinfo{author}{Kondo, D.},
  \bibinfo{author}{Dormieux, L.}, \bibinfo{year}{2002}.
\newblock \bibinfo{title}{Micromechanical analysis of anisotropic damage in
  brittle materials}.
\newblock \bibinfo{journal}{Journal of Engineering Mechanics}
  \bibinfo{volume}{128}, \bibinfo{pages}{889--897}.
%Type = Article
\bibitem[{Queiruga and Moridis(2017)}]{queiruga2017numerical}
\bibinfo{author}{Queiruga, A.F.}, \bibinfo{author}{Moridis, G.},
  \bibinfo{year}{2017}.
\newblock \bibinfo{title}{Numerical experiments on the convergence properties
  of state-based peridynamic laws and influence functions in two-dimensional
  problems}.
\newblock \bibinfo{journal}{Computer Methods in Applied Mechanics and
  Engineering} \bibinfo{volume}{322}, \bibinfo{pages}{97--122}.
%Type = Article
\bibitem[{Santiuste et~al.(2014)Santiuste, Rodr\'iguez-Mill\'an, Giner and
  Migu\'elez}]{santiuste2014influence}
\bibinfo{author}{Santiuste, C.}, \bibinfo{author}{Rodr\'iguez-Mill\'an, M.},
  \bibinfo{author}{Giner, E.}, \bibinfo{author}{Migu\'elez, H.},
  \bibinfo{year}{2014}.
\newblock \bibinfo{title}{The influence of anisotropy in numerical modeling of
  orthogonal cutting of cortical bone}.
\newblock \bibinfo{journal}{Composite Structures} \bibinfo{volume}{116},
  \bibinfo{pages}{423 -- 431}.
%Type = Article
\bibitem[{Schl{\"u}ter et~al.(2014)Schl{\"u}ter, Willenb{\"u}cher, Kuhn and
  M{\"u}ller}]{schluter2014phase}
\bibinfo{author}{Schl{\"u}ter, A.}, \bibinfo{author}{Willenb{\"u}cher, A.},
  \bibinfo{author}{Kuhn, C.}, \bibinfo{author}{M{\"u}ller, R.},
  \bibinfo{year}{2014}.
\newblock \bibinfo{title}{Phase field approximation of dynamic brittle
  fracture}.
\newblock \bibinfo{journal}{Computational Mechanics} \bibinfo{volume}{54},
  \bibinfo{pages}{1141--1161}.
%Type = Book
\bibitem[{Sih(1991)}]{sih1991mechanics}
\bibinfo{author}{Sih, G.C.}, \bibinfo{year}{1991}.
\newblock \bibinfo{title}{Mechanics of fracture initiation and propagation}.
\newblock \bibinfo{publisher}{Springer}.
%Type = Article
\bibitem[{Sih et~al.(1965)Sih, Paris and Irwin}]{Sih65}
\bibinfo{author}{Sih, G.C.}, \bibinfo{author}{Paris, P.C.},
  \bibinfo{author}{Irwin, G.R.}, \bibinfo{year}{1965}.
\newblock \bibinfo{title}{On cracks in rectilinearly anisotropic bodies}.
\newblock \bibinfo{journal}{International Journal of Fracture}
  \bibinfo{volume}{1}, \bibinfo{pages}{189--203}.
%Type = Article
\bibitem[{Silling and Lehoucq(2010)}]{silling2010peridynamic}
\bibinfo{author}{Silling, S.}, \bibinfo{author}{Lehoucq, R.},
  \bibinfo{year}{2010}.
\newblock \bibinfo{title}{Peridynamic theory of solid mechanics}.
\newblock \bibinfo{journal}{Advances in Applied Mechanics}
  \bibinfo{volume}{44}, \bibinfo{pages}{73--166}.
%Type = Article
\bibitem[{Silling et~al.(2015)Silling, Littlewood and
  Seleson}]{silling2015variable}
\bibinfo{author}{Silling, S.}, \bibinfo{author}{Littlewood, D.},
  \bibinfo{author}{Seleson, P.}, \bibinfo{year}{2015}.
\newblock \bibinfo{title}{Variable horizon in a peridynamic medium}.
\newblock \bibinfo{journal}{Journal of Mechanics of Materials and Structures}
  \bibinfo{volume}{10}, \bibinfo{pages}{591--612}.
%Type = Article
\bibitem[{Silling(2000)}]{silling2000reformulation}
\bibinfo{author}{Silling, S.A.}, \bibinfo{year}{2000}.
\newblock \bibinfo{title}{Reformulation of elasticity theory for
  discontinuities and long-range forces}.
\newblock \bibinfo{journal}{Journal of the Mechanics and Physics of Solids}
  \bibinfo{volume}{48}, \bibinfo{pages}{175--209}.
%Type = Article
\bibitem[{Silling and Askari(2005)}]{silling2005meshfree}
\bibinfo{author}{Silling, S.A.}, \bibinfo{author}{Askari, E.},
  \bibinfo{year}{2005}.
\newblock \bibinfo{title}{A meshfree method based on the peridynamic model of
  solid mechanics}.
\newblock \bibinfo{journal}{Computers \& Structures} \bibinfo{volume}{83},
  \bibinfo{pages}{1526--1535}.
%Type = Article
\bibitem[{Silling et~al.(2007)Silling, Epton, Weckner, Xu and
  Askari}]{silling2007peridynamic}
\bibinfo{author}{Silling, S.A.}, \bibinfo{author}{Epton, M.},
  \bibinfo{author}{Weckner, O.}, \bibinfo{author}{Xu, J.},
  \bibinfo{author}{Askari, E.}, \bibinfo{year}{2007}.
\newblock \bibinfo{title}{Peridynamic states and constitutive modeling}.
\newblock \bibinfo{journal}{Journal of Elasticity} \bibinfo{volume}{88},
  \bibinfo{pages}{151--184}.
%Type = Article
\bibitem[{Silling and Lehoucq(2008)}]{silling2008convergence}
\bibinfo{author}{Silling, S.A.}, \bibinfo{author}{Lehoucq, R.B.},
  \bibinfo{year}{2008}.
\newblock \bibinfo{title}{Convergence of peridynamics to classical elasticity
  theory}.
\newblock \bibinfo{journal}{Journal of Elasticity} \bibinfo{volume}{93},
  \bibinfo{pages}{13--37}.
%Type = Book
\bibitem[{Ting(1996)}]{Ting96}
\bibinfo{author}{Ting, T.C.T.}, \bibinfo{year}{1996}.
\newblock \bibinfo{title}{Anisotropic Elasticity}.
\newblock \bibinfo{publisher}{Oxford University Press, New York}.
%Type = Article
\bibitem[{Wang et~al.(2016)Wang, Zhou and Xu}]{wang2016numerical}
\bibinfo{author}{Wang, Y.}, \bibinfo{author}{Zhou, X.}, \bibinfo{author}{Xu,
  X.}, \bibinfo{year}{2016}.
\newblock \bibinfo{title}{Numerical simulation of propagation and coalescence
  of flaws in rock materials under compressive loads using the extended
  non-ordinary state-based peridynamics}.
\newblock \bibinfo{journal}{Engineering Fracture Mechanics}
  \bibinfo{volume}{163}, \bibinfo{pages}{248 -- 273}.
%Type = Article
\bibitem[{Warren et~al.(2009)Warren, Silling, Askari, Weckner, Epton and
  Xu}]{warren2009non}
\bibinfo{author}{Warren, T.L.}, \bibinfo{author}{Silling, S.A.},
  \bibinfo{author}{Askari, A.}, \bibinfo{author}{Weckner, O.},
  \bibinfo{author}{Epton, M.A.}, \bibinfo{author}{Xu, J.},
  \bibinfo{year}{2009}.
\newblock \bibinfo{title}{A non-ordinary state-based peridynamic method to
  model solid material deformation and fracture}.
\newblock \bibinfo{journal}{International Journal of Solids and Structures}
  \bibinfo{volume}{46}, \bibinfo{pages}{1186--1195}.
%Type = Article
\bibitem[{Wu and Ren(2015)}]{wu2015stabilized}
\bibinfo{author}{Wu, C.}, \bibinfo{author}{Ren, B.}, \bibinfo{year}{2015}.
\newblock \bibinfo{title}{A stabilized non-ordinary state-based peridynamics
  for the nonlocal ductile material failure analysis in metal machining
  process}.
\newblock \bibinfo{journal}{Computer Methods in Applied Mechanics and
  Engineering} \bibinfo{volume}{291}, \bibinfo{pages}{197--215}.
%Type = Article
\bibitem[{Wu and Huang(2000)}]{Wu00}
\bibinfo{author}{Wu, T.L.}, \bibinfo{author}{Huang, J.H.},
  \bibinfo{year}{2000}.
\newblock \bibinfo{title}{Closed-form solutions for the magnetoelectric
  coupling coefficients in fibrous composites with piezoelectric and
  piezomagnetic phases}.
\newblock \bibinfo{journal}{International Journal of Solids and Structures}
  \bibinfo{volume}{37}, \bibinfo{pages}{2981--3009}.
%Type = Article
\bibitem[{W\"unsche et~al.(2011)W\"unsche, Zhang, Garc\'ia-S\'anchez, S\'aez,
  Sladek and Sladek}]{Wunsche11}
\bibinfo{author}{W\"unsche, M.}, \bibinfo{author}{Zhang, C.},
  \bibinfo{author}{Garc\'ia-S\'anchez, F.}, \bibinfo{author}{S\'aez, A.},
  \bibinfo{author}{Sladek, J.}, \bibinfo{author}{Sladek, V.},
  \bibinfo{year}{2011}.
\newblock \bibinfo{title}{Dynamic crack analysis in piezoelectric solids with
  non-linear electrical and mechanical boundary conditions by a time-domain
  {BEM}}.
\newblock \bibinfo{journal}{Computer Methods in Applied Mechanics and
  Engineering} \bibinfo{volume}{200}, \bibinfo{pages}{2848--2858}.
%Type = Article
\bibitem[{Yaghoobi and Chorzepa(2015)}]{yaghoobi2015meshless}
\bibinfo{author}{Yaghoobi, A.}, \bibinfo{author}{Chorzepa, M.G.},
  \bibinfo{year}{2015}.
\newblock \bibinfo{title}{Meshless modeling framework for fiber reinforced
  concrete structures}.
\newblock \bibinfo{journal}{Computers \& Structures} \bibinfo{volume}{161},
  \bibinfo{pages}{43 -- 54}.
%Type = Article
\bibitem[{Zhou and Wang(2016)}]{zhou2016numerical}
\bibinfo{author}{Zhou, X.P.}, \bibinfo{author}{Wang, Y.T.},
  \bibinfo{year}{2016}.
\newblock \bibinfo{title}{Numerical simulation of crack propagation and
  coalescence in pre-cracked rock-like brazilian disks using the non-ordinary
  state-based peridynamics}.
\newblock \bibinfo{journal}{International Journal of Rock Mechanics and Mining
  Sciences} \bibinfo{volume}{89}, \bibinfo{pages}{235 -- 249}.

\end{thebibliography}

\end{document}